\documentclass[a4paper,12pt,austrian,american]{book} 

\usepackage{makeidx}
\usepackage{acronym}
\usepackage{epsfig}
\usepackage{subfigure}

\usepackage{amsmath}
\usepackage{amssymb}
\usepackage{cite}
\usepackage{array}

\usepackage[austrian,american]{babel}
\usepackage[latin1]{inputenc}

\usepackage{fancyhdr} 
\fancypagestyle{plain}{%
  \fancyhf{}                       
  \cfoot{\thepage}                 
  
  }

\newcommand{\Ix}[1]{#1\index{#1}}

\newcommand{\beqs}{\begin{equation*}}
\newcommand{\beq}{\begin{equation}}

\newcommand{\eeqs}{\end{equation*}}
\newcommand{\eeq}{\end{equation}}

\newcommand{\beqas}{\begin{eqnarray*}}
\newcommand{\beqa}{\begin{eqnarray}}

\newcommand{\eeqas}{\end{eqnarray*}}
\newcommand{\eeqa}{\end{eqnarray}}

\newcommand{\seq}[5]{\newline 
\parbox{#1}{\begin{eqnarray*} #2 \end{eqnarray*}} \hfill
\parbox{#3}{\begin{eqnarray*} #4 \end{eqnarray*}} \hfill
\parbox{1cm}{\begin{equation} \label{#5} \end{equation}} 
\newline}

\newcommand{\eqs}[1]{\begin{equation} #1 \end{equation}}

\newcommand{\eq}[2]{\begin{equation} #1 \label{#2} \end{equation}}

\newcommand{\eqa}[2]{\begin{eqnarray} #1 \label{#2} \end{eqnarray}}

\newcommand{\meq}[2]{\begin{multline} #1 \label{#2} \end{multline}}

\newcommand{\eps}{\varepsilon}
\newcommand{\al}{\alpha}
\newcommand{\be}{\beta}
\newcommand{\ga}{\gamma}
\newcommand{\de}{\delta}
\newcommand{\om}{\omega}
\newcommand{\ka}{\kappa}
\newcommand{\la}{\lambda}
\newcommand{\si}{\sigma}

\newcommand{\De}{\Delta}
\newcommand{\Om}{\Omega}

\newcommand{\La}{\Lambda}
\newcommand{\Si}{\Sigma}
\newcommand{\Th}{\Theta}

\newcommand{\qb}{q_{\bot}}
\newcommand{\pb}{p_{\bot}}

\newcommand{\app}{\em}

\newcommand{\blist}{\begin{itemize}}

\newcommand{\elist}{\end{itemize}}

\providecommand{\href}[2]{#2}
\newcommand{\URL}[1]{{\href{#1}{{\tt{\small{#1}}}}}}

\newcommand{\myURL}{\href{http://stop.itp.tuwien.ac.at/~grumil/projects/myself/thesis/s4.nb}{{\tt{\small{http://stop.itp.tuwien.ac.at/$\sim$grumil/projects/myself/thesis/s4.nb}}}}}

\newcommand{\chapquote}[2]{{\it #1}\begin{flushright}#2\end{flushright}}

\newcommand{\twod}{$2d$}

\newcommand{\clearemptydoublepage}{\newpage{\pagestyle{empty}\cleardoublepage}}
\newcommand{\clearplaindoublepage}{\newpage{\pagestyle{plain}\cleardoublepage}}

\makeindex 

\begin{document}
\ifx\href\undefined\else\hypersetup{linktocpage=true}\fi


\frontmatter

\fancyhf{}%

\selectlanguage{american}

\begin{titlepage}

\thispagestyle{empty}

    \centering

    \vspace*{2\baselineskip}

    {\huge
      \textbf{DISSERTATION}}
        
    \vspace{3\baselineskip}

    {\Huge
      \textbf{Quantum Dilaton Gravity}\\
      \textbf{in Two Dimensions}\\
      \textbf{with Matter}\\}
    
    \vspace{4\baselineskip}

    A Thesis \\
    Presented to the Faculty of Science and Informatics \\
    Vienna University of Technology

    \vspace{1.5\baselineskip}
        
    Under the Supervision of Prof.\ Wolfgang Kummer \\
    Institute for Theoretical Physics

    \vspace{1.5\baselineskip}
        
    In Partial Fulfillment \\
    Of the Requirements for the Degree \\
    Doctor of Technical Sciences
    
    \vspace{1.5\baselineskip}
    
    By

    \vspace{1.5\baselineskip}
    
    \textbf{Dipl.-Ing.\ Daniel Grumiller} \\
    Kupelwiesergasse 45/5, A-1130 Vienna, Austria \\
    E-mail: \texttt{grumil@hep.itp.tuwien.ac.at}
  
    \vspace{\fill}
  
    \leftline{Vienna, May 4, 2001}

\end{titlepage}

\clearemptydoublepage

\begin{samepage}

\thispagestyle{plain}

\selectlanguage{austrian}
\frenchspacing


{\Huge\textbf{Kurzfassung}}

\bigskip\bigskip\bigskip

Eines der Hauptziele der Physik des 20. Jahrhunderts war die Quantisierung der
Gravitation. Trotz 70 Jahren Forschung erhielt man keine befriedigende
Quantengravitationstheorie. Es gibt mehrere Gr\"unde daf\"ur:
Gravitationstheorie ist nichtlinear und im Gegensatz zu anderen Feldtheorien,
die auf einer fixen Hintergrundmannigfaltigkeit definiert sind, wird die
Geometrie in der Allgemeinen Relativit\"atstheorie dynamisch. Im Unterschied 
zum Standard\-modell der Ele\-mentar\-teilchen\-physik ist sie
st\"orungs\-theo\-re\-tisch nicht\-renormierbar. Experimentelle Hinweise 
auf Quanten\-gravi\-tations\-effekte sind auf Grund ihrer Schw\"ache praktisch 
nicht vorhanden. Um konzeptionelle von technischen Problemen besser trennen zu 
k\"onnen wurden insbesondere die sogenannten zweidimensionalen 
Di\-la\-ton\-modelle untersucht. Leider fehlt den meisten dieser Theorien eine 
bestimmte Eigenschaft: Sie enthalten keine kontinuierlichen physikalischen 
Freiheitsgrade. Eine M\"oglichkeit dies zu beheben, ohne die Vorteile von 2 
Dimensionen aufzugeben, bietet die Kopplung an Materie.

In dieser Arbeit wird speziell das sph\"arisch reduzierte masselose
Klein-Gordon-Einstein-Modell betrachtet, wohl das wichtigste aller 
Dilatonmodelle mit Materie. Dabei wird ein Zugang erster Ordnung f\"ur die 
geometrischen Gr\"ossen verwendet. Nach einer hamiltonschen BRST Analyse wird 
die Pfad\-in\-te\-gral\-quantisierung 
in temporaler Eichung f\"ur die Cartanvariablen durchgef\"uhrt. R\"uckblickend
erweist sich der einfachere Faddeev-Popov-Zugang als ausreichend. Alle
unphysikalischen und geometrischen Freiheitsgrade werden eliminiert, was eine
nichtlokale und nichtpolynomiale Wirkung ergibt, die nur vom Skalarfeld und
von Integrationskonstanten, die durch entsprechende Randbedingungen an das
asymptotische effektive Linienelement fixiert werden, abh\"angt.

Danach werden die (zwei) Baumgraphen in niedrigster Ordnung St\"orungs\-theorie
berechnet, wobei implizit die G\"ultigkeit einer perturbativen Behandlung
angenommen wird. Jeder dieser Graphen enth\"alt f\"ur sich genommen einen 
divergenten Anteil, der \"uberraschenderweise in deren Summe wegf\"allt. 
Dadurch ergibt sich ein endliches $S$-Matrixelement. Wie die Betrachtung der 
(materieabh\"angigen) Metrik zeigt, ergibt sich wieder das Ph\"anomen eines 
``Virtuellen Schwarzen Loches'', das bereits im einfacheren Fall von in 2 
Dimensionen minimal gekoppelten Skalaren beobachtet wurde. Eine Diskussion der 
Streuamplitude f\"uhrt zu der Vorhersage vom Kugelwellenzerfall, einem neuen 
physikalischen Ph\"anomen. M\"ogliche Erweiterungen der hier besprochenen
Szenarien schlie{\ss}en die Dissertation ab.

\end{samepage}
\nonfrenchspacing
\selectlanguage{american}
\clearplaindoublepage

\begin{samepage} 

\thispagestyle{plain}


{\Huge\textbf{Acknowledgments}}

\bigskip\bigskip\bigskip


I am deeply indebted to my supervisor Prof. Wolfgang Kummer for his courage to
attack quantum gravity from a field theoretical point of view while being able 
of encouraging me to join the battle, and for recognizing good ideas while 
having them himself. Also the organization of the FWF project establishing the 
financial support of this thesis is due to him. He advised and assisted me as 
well in technical problems as in conceptual ones.

The stimulating discussions with my collaborators Dimitri Vassilevich, Daniel
Hofmann, ``Waldi'' Waltenberger, and Peter Fischer are gratefully 
acknowledged. We shared excitement and frustrations encountered in the 
calculations and in the attempts to interpret them.

Supplementary, the input and interest of the students who compiled a 
``Vor\-be\-reit\-ungs\-prak\-ti\-kum'' -- i.e. Robert Wimmer, ``Waldi'' 
Waltenberger, Daniel Hofmann, Maria H\"orndl, Peter Fischer, Manfred Herbst, 
Andreas Gerhold, and Joachim Wabnig -- was of great help.

I have profited from the numerous exchanges of views and e-mails with 
several experienced colleagues, in particular with Peter Aichelburg and 
\linebreak[4] Michael Katanaev.

It is a pleasure to express my gratitude to all the members of our institute 
for their everlasting willingness to dispute {\it Mathematica}l, physical, 
\TeX nical and philosophical (a.k.a. trivial) topics and for the excellent 
ambiente. In addition to the aforementioned, I render special thanks to 
Herbert Balasin, Martin Ertl, Dragica Kahlina, Alexander Kling, Maximilian 
Kreuzer, Anton Rebhan, Erwin Riegler, Paul Romatschke, Peter Schaller, Axel 
Schwarz and Dominik Schwarz.

Finally, I wish to send hugs and kisses to Wiltraud Grumiller, Heidrun Rieger
and Rosita Rieger for several private reasons, but particularly for their 
constant engagement with my children Laurin and Armin, which was a necessary 
condition for this dissertation being done.

This work was supported by the projects P 12.815-TPH and P 14.650-TPH of the
FWF (\"Osterreichischer Fonds zur F\"orderung der Wissen\-schaft\-lichen 
For\-schung). Additional sponsoring was granted by my parents Ingo 
Gru\-miller and Helga Weule. Part of my travel expenses were subsidized by 
the Deka\-nat der Fakult\"at f\"ur Tech\-nische Natur\-wis\-sen\-schaften 
und Infor\-matik.

\end{samepage}
\clearplaindoublepage

\begin{samepage}

\thispagestyle{plain}


{\Huge\textbf{Abstract}}

\bigskip\bigskip\bigskip


One of the main goals of 20$^{th}$ century physics was the quantization of 
gravity. Despite of 70 years of research a comprehensive theory fulfilling this
task could not be obtained. There are various explanations for this failure:
Gravity is a non-linear theory and as opposed to other field theories which
are defined on a fixed background manifold, geometry becomes dynamical in 
general relativity. It is perturbatively non-renormalizable in contrast to the 
Standard Model of particle physics. Experimental evidence for quantum gravity 
is scarce due to its sheer weakness. Therefore, physicists have considered 
various toy models -- among them the so-called dilaton models in two 
dimensions -- in order to separate technical problems from conceptual ones. 
Unfortunately, most of them lack a certain feature present in ordinary 
gravity: They contain no continuous physical degrees of freedom. One way
to overcome this without leaving the comfortable realm of two 
dimensions is the inclusion of matter.

In this thesis special emphasis is put on the spherically reduced 
Einstein-massless-Klein-Gordon model using a first order approach for 
geometric quantities, because phenomenologically it is probably the 
most relevant of all dilaton models with matter. After a Hamiltonian BRST 
analysis path integral quantization is performed using temporal gauge for the 
Cartan variables. Retrospectively, the simpler Faddeev-Popov approach turns 
out to be sufficient. It is possible to eliminate all unphysical and geometric 
quantities establishing a non-local and non-polynomial action depending solely 
on the scalar field and on some integration constants, fixed by suitable 
boundary conditions on the asymptotic effective line element.

Then, attention is turned to the evaluation of the (two) lowest 
order tree vertices, explicitly assuming a perturbative expansion in the 
scalar field being valid. Each of them diverges, but unexpected 
cancellations yield a finite $S$-matrix element when both contributions are 
summed. The phenomenon of a ``virtual black hole'' -- already encountered 
in the simpler case of minimally coupled scalars in two dimensions -- occurs, 
as the study of the (matter dependent) metric reveals. A discussion of the 
scattering amplitude leads to the prediction of gravitational decay of 
spherical waves, a novel physical phenomenon. Several possible extensions 
conclude this dissertation.

\end{samepage}

\clearplaindoublepage

\begin{samepage}

\thispagestyle{plain}


\chapter*{}

\chapquote{An architect's first work is apt to be spare and clean.  He knows
he doesn't know what he's doing, so he does it carefully and with great
restraint.

As he designs the first work, frill after frill and embellishment
after embellishment occur to him.  These get stored away to be used ``next
time.'' Sooner or later the first system is finished, and the architect,
with firm confidence and a demonstrated mastery of that class of systems,
is ready to build a second system.

This second is the most dangerous system a man ever designs.
When he does his third and later ones, his prior experiences will
confirm each other as to the general characteristics of such systems,
and their differences will identify those parts of his experience that
are particular and not generalizable.

The general tendency is to over-design the second system, using
all the ideas and frills that were cautiously sidetracked on the first
one.  The result, as Ovid says, is a ``big pile.''}
{Frederick Brooks, ``The Mythical Man Month''}

\end{samepage}

\clearplaindoublepage

\lhead[\thepage]{\slshape \contentsname}                       
\rhead[\slshape \contentsname]{\thepage}
\tableofcontents 

\clearemptydoublepage

\listoffigures 

\clearemptydoublepage


\mainmatter


\lhead[\thepage]{\slshape \leftmark}
\rhead[\slshape \rightmark]{\thepage}
\renewcommand{\chaptermark}[1]{
  \markboth{\chaptername\ \thechapter.\ #1}{}}
\renewcommand{\sectionmark}[1]{
  \markright{#1}}

\chapter{Introduction}

\chapquote{Everything starts somewhere, although many physicists disagree.\\
But people have always been dimly aware of the problem with the start of 
things. They wonder aloud how the snowplough driver gets to work, or how the 
makers of dictionaries look up the spelling of the words. Yet there is the 
constant desire to find some point in the twisting, knotting, ravelling nets of
space-time on which a metaphorical finger can be put to indicate that here, 
{\em here}, is the point where it all began...}
{Terry Pratchett, ``The Hogfather''}

\section{Historical remarks}

Statements of the form ``this theory started with the work of X'' are inexact, 
because they neglect existing previous work. For example, the statement ``(the 
theory of) electromagnetism was found in 1862 by Maxwell'' undoubtedly neglects
important previous work in this field by Coulomb, Amp\`ere, Faraday and 
others. 
Still, from a theoretical point of view, Maxwell´s work was the most important 
one, because it established a unified framework for the description of eletric,
magnetic and optic phenomena. Having this in mind, it is safe to say that 
general relativity -- the geometrical description of gravity -- was started in 
1915 with two celebrated papers of Einstein \cite{ein15a, ein15b} and quantum 
mechanics emerged in 1926 \cite{hei25, sch26}. The first paper devoted to a 
combination of these two rather distinct theories, quantum gravity, appeared 
already 1930 \cite{ros30}. For a brief historical overview of the developments 
of quantum gravity since 1930 I refer to a note of Rovelli and references 
therein \cite{rov00}. 

Because quantum gravity is still an evolving field, historical remarks are 
probably premature and of course always rather subjective. At the moment, 
there are two major lines of research trying to attack the quantization of 
gravity: string theory (for a brief review cf. e.g. \cite{sen98} and 
references therein) and loop quantum gravity (for a pedagogical introduction 
cf. e.g. \cite{gar99}). For a discussion of advantages and drawbacks of these 
two fields (and related, more exotic ones, like e.g. Penrose's ingenious 
twistor approach \cite{pen67}) cf. \cite{rov97}. The main disadvantage of 
(present day) string theory is its background dependence. On the other hand, 
it is not quite clear how to extract physical predictions of (present day) 
loop quantum gravity. In a certain sense, the situation of strings is inverse 
to the status of loops: In string theory, an enormous amount of technical work 
has been done during the last three decades, revealing beautiful mathematical 
structures, but conceptually it suffers still from the problem of background 
dependence. In loop quantum gravity a lot of emphasis has been put into 
conceptual problems and their solution, but apart from the counting of \ac{BH} 
microstates practically nothing has been calculated 
analytically until now.

Thus it is desirable to have an easier framework, where one can study 
analytically or at least perturbatively certain issues (like 
$S$-matrix elements) and at the same time avoid misconceptions (i.e. 
without introducing a fixed background). The hope is that results obtained in 
this way give valuable insights into the corresponding problems of the 
original theory. Therefore, physicists have considered models of quantum 
gravity in three dimensions \cite{djt82a,djt82b,djt84,dej84,wit88,car95} and 
generalized\footnote{Ordinary gravity in two dimensions yields just the 
Gauss-Bonnet term as the Einstein-Hilbert action. Therefore, 
various extensions of gravity have been proposed for twodimensional (\twod) 
models, the most important ones being the so-called ``generalized dilaton 
theories''.} gravity models in two dimensions -- cf. e.g. \cite{tei83,hfj83,
jac85,kvo86}. In this thesis I will concentrate on the latter.

The seminal work of Callan, Giddings, Harvey and Strominger (\acs{CGHS}) in 
1992 \cite{cgh92} rekindled the interest in {\twod} models and many 
important results have been obtained since that time:
\blist
\item A unified approach of all \ac{GDT} in the 
framework of Poisson-$\si$ models \cite{sst94b,kst96a}
\item A global classification of all \ac{GDT} \cite{kst96b}
\item The existence of a conserved quantity in all \ac{GDT} \cite{kus92,gkp92,
kuw94} -- also in the presence of static matter \cite{man93} or even 
dynamical one \cite{kut99} -- leading to a quasilocal energy definition 
\cite{kul97}
\item Conceptual discussions about the r{\^o}le of time in quantum gravity 
\cite{sst94a}
\item A clarifying investigation of the r{\^o}le of conformal transformations 
showing the global difference between conformally related theories 
\cite{kuv99,ghk00}
\item The derivation of the Hawking radiation of \ac{GDT} 
\cite{klv97b,klv98,kuv98}
\item The exact path integral quantization of all \ac{GDT} (without matter) in 
\ac{EF} gauge \cite{klv97a}
\elist
Of course this list is 
strongly biased. For example, it neglects important results that have been 
obtained in conformal gauge (cf. e.g. \cite{lgk94,gkl95,set96,bak96}). The 
reason for this omission is simple: The global classification and the 
quantization procedure are much easier in \ac{EF} gauge 
than in conformal gauge. For a selected example showing this difference I 
refer to the global discussion of the \ac{KV} model \cite{kvo86} in conformal 
gauge \cite{kat93} and in \ac{EF} gauge \cite{kst96b}. For a more 
comprehensive review cf. e.g. \cite{str99} and references therein. 

Due to these successes the {\em plafond} of what can be achieved with 
{\twod} models soon seemed to be reached: After all, these theories are 
only topological ones and thus no propagating (physical) modes exist. In 
particular, important phenomena such as scattering of gravity waves or matter 
waves coupled to gravity cannot be described by such models.

There are three possible ``next natural steps'': 
\blist
\item Neglect that ``essentially everything'' is known about \ac{GDT} and try 
to find some loopholes without changing the topological nature of the theory. 
Prototypes following this line of research can be found in the review article 
\cite{noo00}. However, the brave attempts of renormalizing {\twod} 
dilaton gravity perturbatively seem to be a misdirection of resources in view 
of the exact results obtained in \cite{klv97a}.
\item Leave ``flatland'' and investigate higher dimensional theories. In 
particular, there are interesting results obtained in $2+\eps$ dimensional
dilaton gravity \cite{wei79,gkt78,chd78,kan90,ntt94}. However, in this way one
looses many favorable features of {\twod} theories.
\item Add field degrees of freedom, making the theory nontopological. 
\elist
In this thesis I will follow the last route.

\section{Mathematical remarks}

I am using the ARARA\footnote{As rigorous as reasonably achievable.}-principle 
as a guideline. Most of the calculations are contained in the appendices:
\blist
\item Appendix A lists the notations and conventions (in particular signs and 
indices) I am using.
\item Appendix B recalls some important mathematical preliminaries about the 
Einstein-Cartan formulation of general relativity (including important 
simplifications in two dimensions) and Hamiltonian path integral quantization 
using the \ac{BRST} approach.
\item Appendix C is devoted to the spherical reduction of higher dimensional 
Einstein gravity to a {\twod} dilaton model and various formulations of 
the latter.
\item Appendix D shows the relation between second order and first order 
formalism and Poisson-$\si$ models. It contains the \ac{EOM} and a brief 
derivation of the absolute conservation law present in all such theories. The 
r{\^o}le of conformal transformations is discussed briefly and as an
example of an exactly integrable system static scalars are discussed.
\elist
The subsequent appendices contain essential new calculations not
included in the main text of the thesis in order to increase its readability:
\blist
\item Appendix E provides a Hamiltonian and \ac{BRST} analysis of first order 
dilaton gravity coupled to scalars and discusses useful sets of constraints, 
canonical transformations and gauge fixing conditions. 
\item Appendix F collects useful formulae and contains the calculation of the 
lowest order tree-graph scattering amplitude.
\elist

\section{Physical remarks}

String theory is a covariant attempt of building a quantized theory of 
gravity, while loop quantum gravity is a canonical one. But there exists (at 
least) a third possibility: The sum over histories approach. Invented by 
Feynman for ordinary quantum field theory \cite{fey48, fey51} it has been 
initiated for gravity already in 1957 by Misner \cite{mis57} and revived by 
Hawking in 1979 \cite{haw79}. 

This approach has one big advantage and an equally sized disadvantage: 
Perturbation theory can be applied most easily using the path integral 
formulation. On the other hand, there is a big danger that ``subtleties'' are 
swept under the rug because in general there exist difficulties with the 
path integral measure. In most cases, the path integral approach is justified 
only because one already knows the results from more complicated, but also 
more rigorous calculations. 

Thus, the optimal strategy seems to pick up the advantages of the sum over 
histories approach avoiding its pit traps. This is possible (at least in the 
geometric sector) within the first order formulation of \ac{GDT} where an 
exact path integration can be performed. In the matter 
sector one has to use perturbative techniques like in ordinary \ac{QFT}, but 
this is less dangerous than a (also conceptually dissatisfying) perturbation 
in the geometric sector (it is known since 1973 that gravity is a 
nonrenormalizable theory \cite{hoo73, hov74}).

The structure of this thesis is as follows:
\blist
\item Chapter 2 introduces \ac{SRG} in the first order formulation and shows 
three pedagogical examples of the formalism's power.
\item Chapter 3 is the main part of the thesis containing the quantization of 
the \ac{EMKG} model and the calculation of the lowest order tree graph 
scattering amplitude.
\item Chapter 4 generalizes the basic scenario providing an outlook to further 
theories of interest.
\item Chapter 5 contains a summary.
\elist

\clearplaindoublepage

\chapter[Classical {\twod} dilaton gravity]
{Classical $\boldsymbol{2d}$ dilaton gravity}

\chapquote{Physics is mathematical not because we know so much about the 
physical world, but because we know so little: It is only its mathematical 
properties that we can discover.}{Bertrand Russel, ``An outline of 
Philosophy''}

\section{From Einstein to first order gravity}

\subsection{Geometry}\index{geometry}

Given a Pseudo-Riemannian manifold $M^4$ in $d=4$ with a metric $g_{\mu\nu}$ 
possessing Minkowskian signature, it is possible to define the 
Ricci-scalar (or scalar curvature) $R$ by contraction of the 
Ricci-tensor \cite{wal84}. The \ac{EH} action for gravity is
\eq{
L = \int_{M^4} d^4x \sqrt{-g} R,
}{C1}
with $g = \det g_{\mu\nu}$. I am using natural units $8\pi G_N = c = 
\hbar = 1$ and the Bj{\o}rken-Drell convention sig $g = (+,-,-,-)$.

Imposing some symmetry on the solutions of the \ac{EOM} will in general lead 
to  the same result as imposing this symmetry in the action. However, if no
Killing vector exists dimensional reduction may yield a wrong action, like in 
the case of a general warped product metric \cite{kkk98}.
 
Fortunately, in the case of spherical symmetry it is possible to work 
with the ``spherically reduced'' action and to obtain the correct 
(classical) result \cite{tih84, haj84, lau96}. Therefore one can work with a 
{\twod} Lagrangian instead of the more complicated \ac{EH} 
Lagrangian in $d=4$.

Twodimensional models\index{twodimensional models} have many remarkable 
properties:
\blist
\item If there is no further scale, they are conformally invariant
 --  cf. e.g. \cite{efs00} and references therein.
\item The spin connection has just one independent 1-form component and is 
proportional to $\eps_{ab}$ -- cf. e.g. \cite{wal84}.
\item The Riemann tensor has just one independent scalar component 
\linebreak[4] (namely the scalar curvature) -- cf. e.g. \cite{wal84}.
\item There are no continuous physical degrees of freedom (in absence of 
matter); that means pure gravity in $d=2$ does not allow asymptotic 
``gravity-states'' (i.e. gravitational waves) -- cf. {\app appendix E}.
\item They are naturally related to string theories since the world-sheet 
dynamic is given by a {\twod} model -- cf. e.g. \cite{sen98}.
\item They are usually special cases of so-called Poisson-$\si$ models 
\cite{sst94b} -- cf. {\app appendix D} on p. \pageref{poisson sigma}.
\item  There exists always a conservation law \cite{kut99} -- cf. {\app 
appendix D} on p. \pageref{conservation law}.
\item In a certain sense (see below) they are a {\twod} analogon of 
\ac{JBD} theory or other scalar-tensor-theories
\cite{jor55, jor59, brd61}.
\item Some specific models can be successfully quantized (see next chapter).
\elist

Thus, many features which are not completely understood in gravity -- the
quantization procedure and non-trivial generalizations of \ac{EH} 
gravity -- can be studied in a simpler and friendlier ``ambiente''.

It is true that the quantization of a $d=2$ model may not lead to the same 
phenomenology as the quantization of a spherically symmetric $d=4$ model, 
because non-trivial fluctuations in ``angular'' directions may occur 
(see p. \pageref{dranomaly}).  
Nevertheless the knowledge of certain techniques in $d=2$ may serve as a
canonical example for $d=4$ models leading to a deeper understanding.

Imposing spherical symmetry in the metric
\eq{
g_{mn}(x_m) dx^m dx^n = g_{ab}(x_a)dx^a dx^b - \Phi^2(x_a) d \Om^2 ,
}{C2}
where I have used the notation explained in {\app appendix A} (on p. 
\pageref{conventions}), leads after a lengthy but straightforward calculation 
(which is performed in {\app appendix C}) to the {\twod} \Ix{dilaton action}:
\eq{
L_{\text{dil}} = \int_{M^2} d^{2}x \sqrt{-g} \left( \Phi^2R + 2\left(\nabla 
\Phi \right)^2 - 2 \right) .
}{C3} 
Note that the first term of (\ref{C3}) corresponds to a non-minimally coupled
\ac{EH} term in $d=2$ because it contains the {\twod} curvature
linearly. The ``\Ix{dilaton field}'' $\Phi^2$ is (from the $4d$ 
point of view) part of the geometry. From the {\twod} point of view it
is some ``quintessence'' field to which \ac{EH} gravity is coupled.
The situation is very similar to \ac{JBD} theory where the coupling
constant $G_{N}$ is replaced by some (space-time dependent) scalar field 
$\phi$ \cite{jor55, jor59, brd61}. In fact, one of the reasons why 
quintessence models are {\em en vogue} nowadays\footnote{Of ocurse, for
cosmologists the main motivation is the possibility to explain the otherwise
puzzling type Ia supernovae \cite{pal97,per98}, which eventually led to the
proposal of quintessence \cite{zws99}.} is precisely this 
correspondence between the dilaton field coupled to the \ac{EH}-term in the 
lowdimensional theory and a geometrical field in a higher dimensional 
theory\footnote{The simplest model leading to a $4d$ 
quintessence model is a five dimensional Kaluza-Klein theory with one 
compactified dimension \cite{kal21, kle26a, kle26b}.} (cf. e.g. \cite{brj99}).

Spherically symmetric \ac{EH} theories minimally coupled (in $d=4$) to matter 
will produce a {\em non-minimally} coupled matter Lagrangian in the dilaton 
action, since the dilaton field is part of the $4d$ metric. Therefore, 
confusion may arise when talking about ``(non)-minimally'' coupled matter. For
the rest of this work I will always refer to the {\twod} theory, and it
will be made explicit whenever this is not the case.

Although much simpler, eq. (\ref{C3}) is still not of the desired form. I
introduce Cartan variables instead of the metric because the
$\sqrt{-g}$-term gives rise to undesired non-linearities.

In {\app appendix D} (p. \pageref{Cartan formulation} ff.) the equivalence of 
the \Ix{first order formulation} (``Poisson-$\si$ models'' \cite{sst94a}) and 
the \Ix{dilaton formulation} for a \ac{GDT} is reviewed (cf. \ref{lag1}). Using
\eq{
D\wedge e^{\pm} = d \wedge e^{\pm} \pm \om \wedge e^{\pm},
}{C04} 
this \Ix{first order action} reads
\begin{eqnarray}
L_{\text{FO}} &=& \int_{M^2} \left( X^+ D\wedge e^- + X^- D\wedge e^+ + X 
d\wedge \omega \right. \nonumber \\
&& \hspace{1cm} \left. - e^- \wedge e^+ \left(V(X) + X^+X^- U(X)\right) 
\right),
\label{C4}
\end{eqnarray}
with
\eq{
\left. V(X) \right|_{\text{SRG}} = -\frac{\lambda^2}{4} = const., 
\hspace{0.5cm} \left. U(X) \right|_{\text{SRG}} = -\frac{1}{2X},
}{C4a}
for \ac{SRG}\index{spherically reduced Lagrangian}. The constant $\lambda$ 
appearing in (\ref{C4a}) has a length-dimension of $-1$ due to its definition 
(\ref{dilredef1}). $e^{\pm}$ and $\om$ are the 
Cartan-variables (vielbein 1-form and spin-connection 1-form\footnote{In two 
dimensions, the matrix valued 1-form $\om^a{}_b$ contains only one 
independent 1-form valued component, denoted by $\om$. The relation is 
$\om^a{}_b = \eps^a{}_b\om$ -- cf. {\app appendix B} on p. 
\pageref{simplifications}.}, respectively) and $X$ is the
dilaton field (redefined by eq. (\ref{dilredef2})). For \ac{SRG} $X$ has to be 
(semi-)definite and I chose it to be positive (apart from the singularity at 
the origin). $X^{\pm}$ are target space coordinates giving rise to torsion 
terms in the action. Other choices of the functions $V(X)$ and $U(X)$ lead to 
different models which are listed in a table on p. \pageref{table1}.

Note that it is always possible to eliminate the spin-connection and the target
space coordinates ending up with an action of the form (\ref{C3}) -- but this
would spoil the linearity and simplicity of the action and the \ac{EOM}
(cf. {\app appendix D}). By using the \Ix{first order action} 
(\ref{C4}) instead of (\ref{C3}) or (\ref{C2}) I follow a principle which has 
proven very useful in theoretical physics (especially in the Hamiltonian 
analysis and there especially in the construction of the \ac{BRST}
charge \cite{brs75, tyu75}): Enlarging the phase space properly (thus 
introducing more (unphysical) degrees of freedom) simplifies the dynamics of 
the complete system.

\subsection{Matter}\index{matter}

I restrict myself to scalar matter with arbitrary dilaton coupling in
\twod:
\eq{
L_{\text{m}} = \int_{M_2} F(X) \left(dS \wedge * dS - e^- \wedge e^+ f(S) 
\right).
}{Cmatter}
$S$ is the scalar field, $F(X)$ some arbitrary (smooth) coupling function and
$f(S)$ some arbitrary (smooth) self-interaction potential (it may contain e.g.
a $m^2S^2$ and a $\la S^4$ term). 

In {\app appendix C} on p. \pageref{matter part} it is shown that for \ac{SRG} 
the proper form is given by $F(X) = - X/2 $. The numerical factors are due to 
my conventions, but the linear appearance of $X$ is easy to understand: It
stems from the four dimensional measure $\sqrt{-g^{(4)}} = X \sqrt{-g}$. Thus,
minimal coupling in d=4 induces non-minimal coupling to the dilaton
in d=2. 

In the following section I repeat some well-known calculations in the
Cartan formulation in order to show the formalism's power.

\section{Some examples of the formalism's power}

\subsection{The search for \Ix{Killing-horizons}}

The large class of \ac{GDT} with one Killing-horizon has been
classified using the Cartan formulation \cite{kkl96}. I will perform an
analogous calculation for a phenomenologically relevant model which has up to
{\em three} Killing-horizons given by the effective action (\ref{C4}) with the
``potential''
\eq{
V(X) = -3A X-\frac{\lambda^2}{4} + \frac{B}{X}, \hspace{0.5cm}
U(X) = -\frac{1}{2X}.
}{C51}
The function $U(X)$ remains unchanged while in $V(X)$ there are two additional 
terms. The first one can be interpreted as a $(d=4)$ cosmological 
constant\footnote{The factor 3 has been introduced for later convenience.} and
the last one corresponds to some (electric) monopole charge (see below). For a 
complete discussion of the global properties of generic models cf. 
\cite{kst96a, kst96b}.

The conserved quantity (cf. {\app appendix D} on p. \pageref{conservation 
law}) for a model with a potential belonging to the following class
\eqs{
V(X) = \sum_{n=-\infty}^{\infty} c_n X^n, \hspace{0.5cm}U(X) = \frac{1}{bX}, 
}
is given by
\eq{
{\cal C}^{(g)} = X^{\frac{1}{b}}X^+X^- + \sum_{n=-\infty}^{\infty} \frac{c_n}
{n+1+\frac{1}{b}} X^{n+1+\frac{1}{b}}.
}{C6}

A very useful ansatz, which eventually will lead to an \ac{EF} 
gauge, is given by
\eq{
e^+ = X^+ g(X) df .
}{C5}
The Lorentz index is captured by the 0-form $X^+$, the form degree of freedom 
by the 1-form $df$. I have allowed for a further dependence on the (scalar) 
dilaton field via an arbitrary function $g(X)$. The \ac{EOM} imply immediately
\eq{
e^- = X^- g(X) df + \frac{dX}{X^+}, \hspace{0.5cm} g(X) = c X^{\frac{1}{b}},
c \in \mathbb{R}^* .
}{C7}
After the redefinition $\tilde{X} = k X^{1+1/b}$ and the identification
$c = k \left(1+1/b\right)$ (with $k \in \mathbb{R}^*$) the 
line-element in \ac{EF} gauge reads
$(ds)^2 = df \otimes (2 d\tilde{X} + K^2(\tilde{X}) df)$
with the Killing-norm
\eq{
K^2(\tilde{X}) = h(\tilde{X})\left({\cal C}^{(g)}-\sum_{n=-\infty}^{\infty}
\frac{c_n}{n+1+\frac{1}{b}}\left(\frac{\tilde{X}}{k}\right)^{\frac{(n+1)b+1}
{b+1}}\right),
}{C8}
where $h(\tilde{X}) = 2k^2 \left(1+1/b\right)^2 (\tilde{X}/k)^{1/(b+1)}$. I 
will use for the dilaton field the 
gauge (\ref{dilgauge}). Then the choice $k = (d-2)/\la$ allows the 
identification $\tilde{X} = r$, where $r$ is the ``radius''.

Zeros of the Killing-norm correspond to Killing horizons (cf. e.g. 
\cite{tow97}). For our model this reduces to finding all positive 
zeros\footnote{Negative zeros would give rise to negative radii and I 
disregard this possibility.} of the algebraic equation 
\eq{
A \tilde{X}^4 + \tilde{X}^2+\frac{4{\cal C}^{(g)}}{\la^3}\tilde{X}
+\frac{16 B}{\la^4} = 0.
}{C9}
Before investigating the most general case lets throw a short look at some 
important special cases:

\subsubsection{Schwarzschild horizon}

For $A=0=B$ one obtains the Schwarzschild solution \cite{sch16} with one 
Killing-horizon at
\eq{
\tilde{X} = - \frac{4{\cal C}^{(g)}}{\lambda^3}.
}{C10}
In Schwarzschild gauge the identity $\tilde{X} = r_h$ yields the 
desired relation $r_h \propto - {\cal C}^{(g)} \propto m$
provided that ${\cal C}^{(g)}$ is negative\footnote{It may seem unesthetic to
need a negative value for the conserved quantity in order to get some positive
\ac{BH} mass. This sign is due to the conventions I have used and could be
``repaired'' easily -- however, this would introduce (harmless but unwanted) 
signs in the quantization procedure and therefore I stick to this convention 
which is compatible with \cite{gru99, grk00}.}.

Using the \ac{ADM} mass definition derived in \linebreak[4] 
\cite{gru99, grk00} (and used in the \ac{RN} case below), 
$m = -2{\cal C}^{(g)}/\la^3$, one obtains the correct numerical factor 
$r = 2m$.

\subsubsection{Cosmological horizon}

For $B=0={\cal C}^{(g)}$ one has got an ``uncharged vacuum'' with 
non-vanishing cosmological constant. For negative values of $A$ -- the
\ac{dS} case -- one obtains a Killing horizon at 
$\tilde{X} = 1/\sqrt{-A}$ which (again in Schwarzschild gauge) yields the 
well-known relation \cite{tow97} $\Lambda \propto 1/r^2$
where I have substituted $\Lambda$ instead of $A$. 

\subsubsection{Reissner Nordstr\"om horizons}

For $A=0$ up to two Killing-horizons are obtained, depending on the size of 
$B$ and the conserved quantity at
\eq{
\tilde{X}_{12} = \left( m \pm \sqrt{m^2-\frac{16B}{\lambda^4}} \right). 
}{C14}
Thus, one has to identify the square of the charge of the \ac{BH} with 
$Q^2 = 16B/\lambda^4$ in order to get the \ac{RN} 
solution \cite{rei16, nor16}.

\subsubsection{The generic case}

\begin{figure}
\noindent
\begin{minipage}[b]{\linewidth}
  \centering
  \subfigure[There are three positive zeroes in the Killing-norm $\xi(r)$...]
            {\epsfig{file=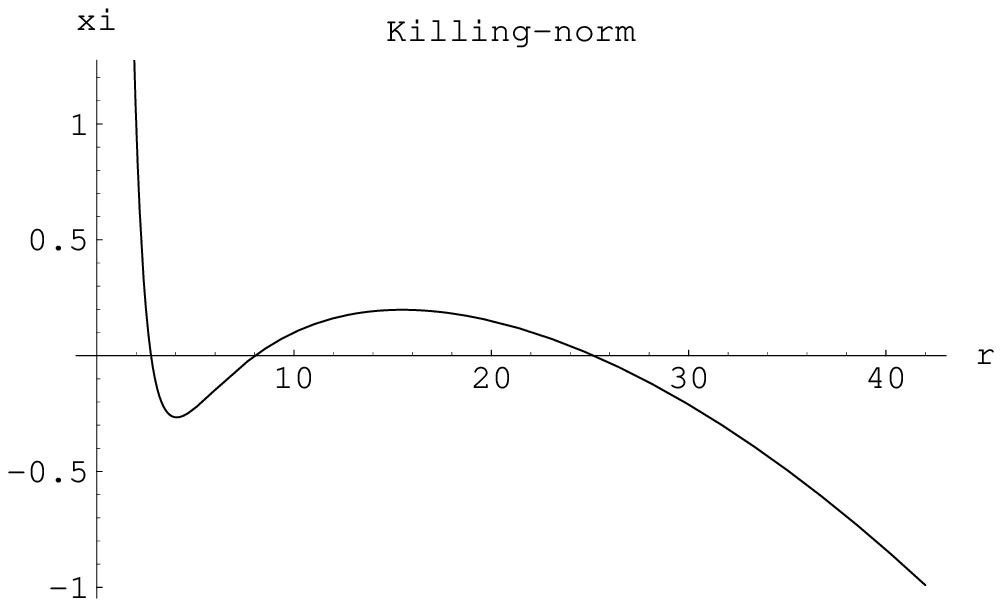,width=.47\linewidth}} \hfill
  \subfigure[...leading to three singularities in $\partial v_{(2)}/
            \partial r$.]{\epsfig{file=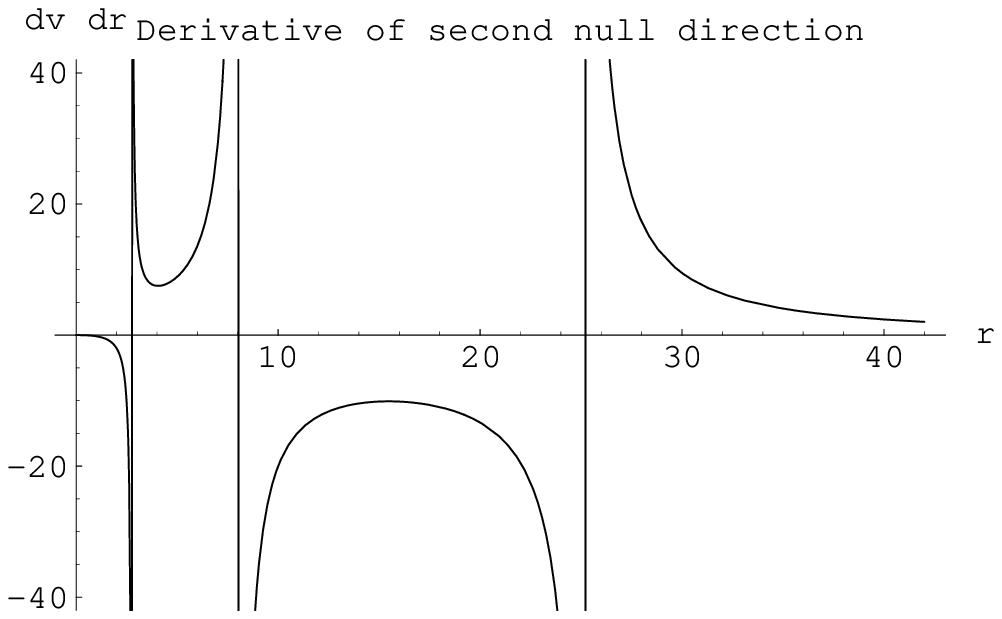,width=.47\linewidth}}
  \subfigure[The second null direction $v_{(2)}$ in the complete region...]
            {\epsfig{file=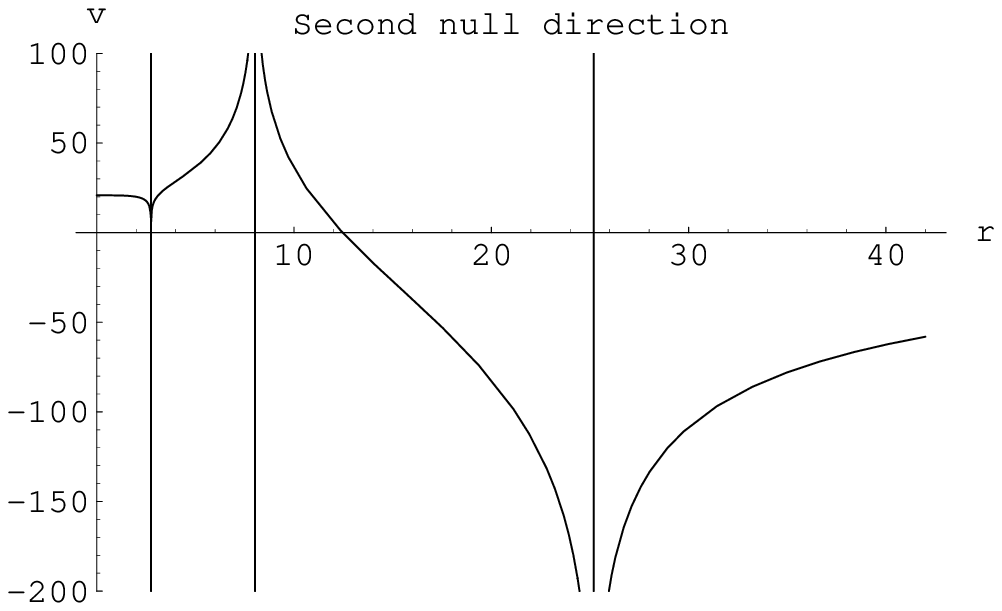,width=.47\linewidth}} \hfill
  \subfigure[...and in the ``\ac{RN}-region''.]
            {\epsfig{file=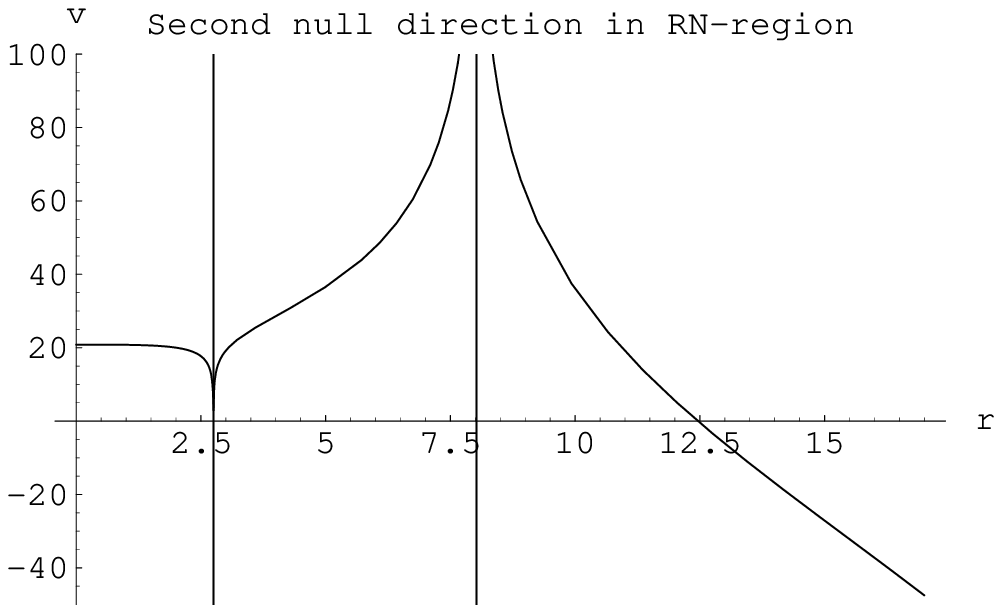,width=.47\linewidth}}
\end{minipage}
\caption[Killing norm and null directions]
{Numerical values have been chosen s.t. all horizons are 
clearly visible: $m = 5$, $Q^2=20$ and $\La = - 10^{-3}$ (the neccessary 
conditions $M^2 > Q^2$ and $\La < 0$ are obviously fulfilled but the values are
far from being realistic for a macroscopic \ac{BH}).
Note that in graph (d) the well-known ``asymptotic'' behavior $v_{(2)} = -2r$ 
is reproduced. For very small absolute values of the cosmological constant 
$\Lambda$ the second null-direction $v_{(2)}$ follows for a wide range of $r$ 
this pseudo-asymptotic line before approaching the outer (cosmological) 
horizon in the \ac{dS} case. This part of the graph reproduces the familiar
Reissner-Norstr{\"o}m patches in the Carter-Penrose diagram.}
\label{fig:nullgraphs}
\end{figure}

At first glance one might expect up to four horizons. However, it is easy to
see that the algebraic equation (\ref{C9}) has a maximum of three positive
real solutions. Therefore, there are always less than four horizons (cf.
graph (a) of figure \ref{fig:nullgraphs}).

For the (phenomenologically relevant) limit $\left|{\cal C}^{(g)}\right| >> 
\left| \Lambda \right|$ up to three horizons emerge, two of which are located 
near the origin and correspond to the
two \ac{RN} horizons, while the third one is far from the origin
resembling the cosmological horizon in the \ac{dS} case.

Lets construct now the Carter-Penrose diagram \cite{car69,pen69} for the 
generic case starting with the EF line-element
\eq{
(ds)^2 = du \otimes \left( 2dr+ \xi(r) du \right),
}{C17}
with $du = df$ and $dr = d\tilde{X}$ and obtain a familiar expression for the 
Killing-norm
\eq{
\xi(r) =  \frac{Q^2}{r^2} - \frac{2 m}{r} + 1 + \La r^2.
}{C50}

\begin{figure}
\begin{minipage}[b]{\linewidth}
   \centering\epsfig{file=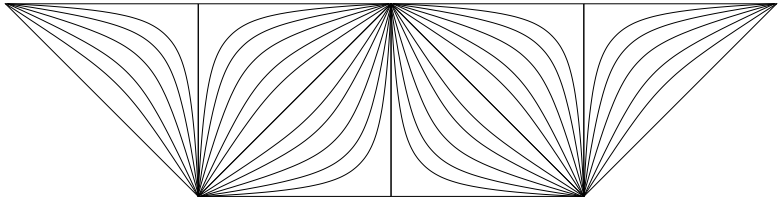,width=.75\linewidth}
\end{minipage}
\caption[Patch of {C}arter-{P}enrose diagram with 3 {K}illing horizons]
{One patch obtained from graph (c) in figure \ref{fig:nullgraphs}. The first 
three sub-patches on the left side are the \ac{RN} patches. The last one is 
due to the ``cosmological horizon'' symmetrizing the complete diagram to a 
more ``esthetic'' one. The diagonal line on the \acs{l.h.s.} and \acs{r.h.s.}
corresponds to the singularity at the origin and at $i_0$, respectively. All
vertical lines are Killing horizons. The curved lines are equi-radial.}
\label{fig:cppatch}
\end{figure}

\begin{figure}
\begin{minipage}[b]{\linewidth}
   \centering\epsfig{file=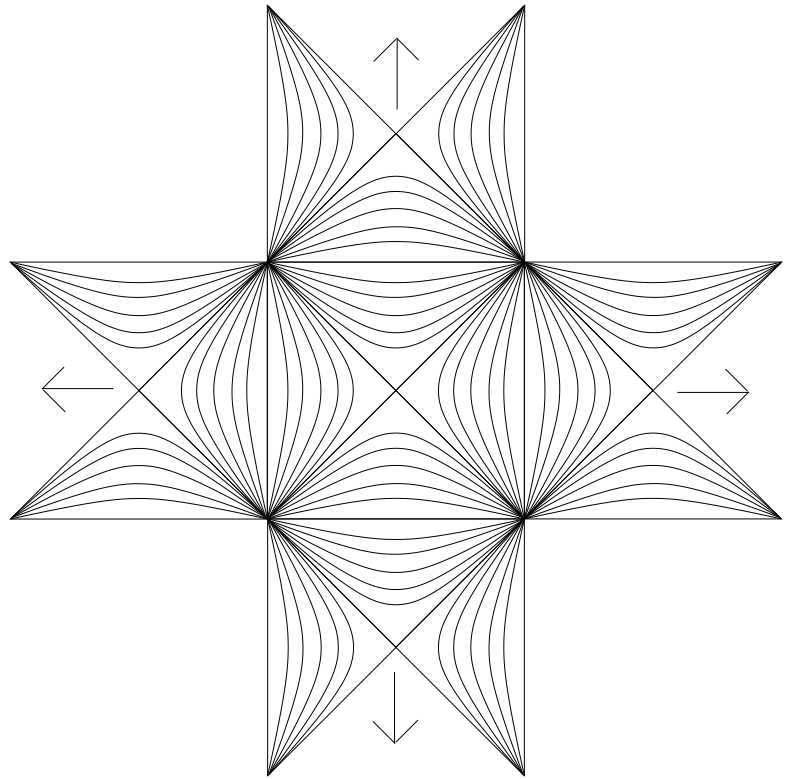,width=.75\linewidth}
\end{minipage}
\caption[Full {C}arter-{P}enrose diagram]
{The full Carter-Penrose diagram has to be continued 
on the top, bottom {\em and} on the left and right side leading to an almost 
complete tiling of the plane. It is identical to the Carter-Penrose diagram 
of the {\bf R5}-case of \cite{kst96b} leading to topologies of arbitrary 
genus. For further discussion (e.g. of geodesic (in-)completeness) I refer to 
that paper. Note that apart from the additional triangular patches at the left 
and right side of the diagram it is completely equivalent to the 
Carter-Penrose diagram of the \ac{RNBH}.}
\label{fig:cpdiagram}
\end{figure}

By analogy with \cite{kst96a, kst96b} I investigate the null directions (cf. 
the graphs in figure \ref{fig:nullgraphs})
\eq{
v_{(1)} = u = const. ,\hspace{1cm} v_{(2)} = - \int\limits_{r_0}^{r} 
\frac{2}{\xi(r')} dr',
}{C19}
and introduce the conformally compactified metric via the line element
\eq{
\left(d\tilde{s}\right)^2 = d\tilde{U}d\tilde{V} = \frac{d v_{(1)}}
{1+v_{(1)}^2} \frac{dv_{(2)}}{1+v_{(2)}^2}, 
}{C20}
thus obtaining one patch of the Carter-Penrose diagram (cf. 
figure \ref{fig:cppatch}). By continuation the complete one is obtained 
(cf. figure \ref{fig:cpdiagram}). Of course, there exist ``extremal'' black 
holes, where two (or even three) Killing-horizons coincide allowing one to 
``play around'' with lots of special cases -- but their discussion is  
analogous to the extremal \ac{RNBH}.

In the notation of \cite{kst96b} the generic case with three horizons is the 
{\bf R5} solution of $R^2$-gravity. It is remarkable, that our model  -- which 
is nothing else but a \ac{RNBH} in \ac{dS} background -- has the same global
structure as a {\twod} $R^2$-model with an (also \twod) cosmological constant. 
Further it is noteworthy that arbitrary higher genus solutions exist in our 
model as opposed to the \ac{SSBH} or \ac{RNBH}.

It is amusing that asymptotically (i.e. for very large or very small $r$) one
gets in the \ac{dS}-case a duality relation of the form
\eq{
r \leftrightarrow \frac{1}{r} ,\hspace{1cm} \Lambda \leftrightarrow -Q^2, 
}{C21}
reflected also in the symmetric form of the Carter-Penrose diagram (fig. 
\ref{fig:cpdiagram}) or the corresponding $X \leftrightarrow 1/X$ duality in 
the ``potential'' (\ref{C51}).

\subsubsection{Summary}

I have reviewed basic steps for the construction of a Carter-Penrose diagram 
and shown the location of (up to three phenomenologically relevant) 
Killing-horizons in the Cartan formulation of \ac{SRG} for a \ac{RNBH} in 
\ac{dS} background with a ``potential'' given by
\eq{
V(X) = -3 \La X-\frac{\la^2}{4} + \frac{\la^4}{16} \frac{Q^2}{X}, 
\hspace{0.5cm}U(X) = -\frac{1}{2X} .
}{C35}
Note that the full potential ${\cal V} = U(X)X^+X^-+V(X)$
scales linearly with the square of the inverse length scale $\la^2$ provided 
the gauge fixing (\ref{dilatongauge}) is chosen. Thus also the total action 
(\ref{C4}) scales linearly with $\la^2$ and the resulting \ac{EOM} are independent 
of it.

\subsection{Critical collapse}\index{critical collapse}

Since the pioneering work of Choptuik on (spherically symmetric) critical 
collapse \cite{cho93} there have been many efforts in this field of 
research (for the reader who is not familiar with this topic I refer to a 
review of Gundlach \cite{gun98}). Although in many of these works spherical 
symmetry is assumed, to the best of my knowledge none of them contains some 
calculations in the Cartan-formulation. This section will fill this gap at
least partially.

The \ac{EOM} for the \ac{EMKG} model 
have been derived elsewhere \cite{grk00}, therefore I concentrate here 
on the critical  solution, which has two remarkable features: It is discretely 
self-similar\footnote{I.e. there exists a discrete diffeomorphism $\Phi$ and a
real constant $\Delta$ s.t. \[ \Phi^* g_{ab} = e^{2\Delta}g_{ab}, \] where 
$\Phi^*g_{ab}$ is the pull-back of $g_{ab}$ under the diffeomorphism $\Phi$.} 
(\acs{DSS}) and it leads to a universal\footnote{i.e. independent of 
the specific choice of the one-parameter family of initial data} mass-scaling 
near the critical point, justifying the word ``critical'' in this context.

I will use the metric defined via the ($4d$) line element
\eq{
(ds)^2 = e^{2\tau} \left( Ad\tau^2+2Bd\tau dx+Cdx^2 - R^2 d\Om^2 \right),
}{C24}
with $A, B, C, R$ being periodic functions of $\tau$ with the same period 
$\De$. These coordinates are usually called ``adapted coordinates'', since they
reflect the \ac{DSS} directly. I use the latter as an input in this ansatz.

In terms of Cartan-variables one has four neccessary conditions
\seq{3cm}{
X &=& e^{2\tau} R^2(\tau,x), \\
2e_0^-e_0^+ &=& e^{2\tau} A (\tau,x), 
}{3cm}{
e_0^-e_1^++e_1^-e_0^+ &=& e^{2\tau} B (\tau,x), \\
2e_1^-e_1^+ &=& e^{2\tau} C (\tau,x),
}{C25}
restricting the possible gauge fixings compatible with (\ref{C24}).

I impose the \ac{SB} gauge conditions
\seq{3cm}{
&& e_0^- = 1, \\
&& e_0^+ = 0,
}{3cm}{
&& e_1^- = -\frac{h}{2}, \\
&& e_1^+ = (e), \\
&& X = e^{2\tau} R^2 (\tau),
}{C26}
with $(e) = e^{2\tau}f_e$ and $-h = f_h$,
where again $R^2, f_e$ and $f_h$ are periodic in $\tau$ with period $\De$. 
$f_e$ and $f_h$ are additionally functions of $x$. In this section I will 
denote periodic functions in $\tau$ with the period $\De$ by $f_{\dots}$ with 
some (easy interpretable) index. For positive values of the Killing-norm $h$ 
the coordinate $\tau$ is timelike and $x$ is lightlike. The only residual 
gauge freedom left after the fixing (\ref{C26}) is $x$-rescaling.

For the scalar field I make the ansatz
\eq{
S = \ka \tau + f_s (x, \tau),
}{C28}
guaranteeing the periodicity of $\partial_{\tau} S$.

Plugging (\ref{C26}-\ref{C28}) into the \ac{EOM} (cf. p. 
\pageref{equations of motion}) leads to the conserved quantity (and thus
following the lines of \cite{gru99, grk00} to the mass-aspect function)
\eq{
m (\tau, x) = e^{\tau}f_m , \hspace{1cm} \left.\left<m (\tau, x)\right>
\right|_{\text{1 period}} = m_0 \cdot e^{\tau},
}{C29}
which is directly related to the \ac{BH} mass (cf. next section). This 
relation (\ref{C29}) is absolutely neccessary for the mass-scaling law which 
can be derived perturbatively introducing a fiducial time \cite{gun98}
\eq{
\left. \frac{\partial C_{*00}}{\partial p} \right|_{p_*} \cdot (p-p_*)
e^{-\la_{*0} \tau_p} = \eps,
}{C32}
where $p$ is the free parameter of the one-parameter family of initial data, 
$p_*$  is the critical parameter at the threshold of black-hole formation, 
$\la_{*0}$ is some (universal) positive real number, $\eps$ is some arbitrary, 
but fixed small positive real number, $\tau_p$ is the fiducial time which is 
defined by the relation above and the first term contains the family-dependent 
information near the critical point\footnote{I do not bother about the 
technical details (breakdown of the linear ansatz or possibility to neglect the
stable modes) involved in the proper definition and range of $\tau_p$. I simply
assume that it is always possible to find such a fiducial time, provided the
difference $p-p_*$ is small enough.}.

From (\ref{C29}) and (\ref{C32}) follows immediately the famous scaling 
law\footnote{Usually a dimensional argument is used in this context; it seems
that this is not neccessary, since I could {\em derive} (\ref{C29}) rather
than {\em assume} it. The crucial point is that not only the mass-aspect 
function but also the (eventual) final state \ac{BH} scales with $e^{\tau}$ due 
to the relations between the mass-aspect function, the \acs{ADM}-,
Bondi- and final state \ac{BH} mass presented in the next section. The 
drawback of our method is the assumption of \ac{DSS} which holds only for the 
critical solution. So in fact our ``derivation'' is equivalent to the 
assumption that \ac{DSS} holds ``approximately'' which can only
be true very close to the critical point.}
\eq{
m_{b.h.} = C (p-p_*)^{\ga} , \hspace{1cm} \ga = \frac{1}{\la_{*0}} .
}{C33} 

For sake of completeness I list here some explicit expressions encountered in
the derivation of (\ref{C29}) (using the notation of {\app appendix D})
\seq{3cm}{
&& S^+ = - e^{2\tau} f_e \partial_1 f_s, \label{C45} \\
&& S^- = \ka + \partial_0 f_s - \frac{1}{2} f_h \partial_1 f_s, \\
&& X^+ = e^{2\tau} 2R \left( R + \frac{\partial R}{\partial \tau} \right), \\
&& X^- = \frac{Rf_h}{f_e} \left( R + \frac{\partial R}{\partial \tau} \right),
}{3cm}{
&& \om_0 = -1-\partial_0 \ln\left((e)R\right), \\
&& \om_1 = \frac{f_h}{2} \partial_0 \left(\ln(e)\right)-\frac{1}{2}\partial_0 
f_h, \\
&& W_0 = \frac{\sqrt{X}}{(e)}\partial_1 f_s \partial_0 X \cdot W,
}{C30}
\eq{
W_1 = \frac{\sqrt{X}}{(e)}\partial_0 X \left[ \left(\ka+\partial_0f_s
\right) \cdot W + \left(W+\frac{1}{2}f_h\partial_1f_s\right)\partial_1f_s 
\right],
}{C83}
with $W = \ka+\partial_0f_s-f_h\partial_1f_s.$
An important observation is the definite $\tau$-scaling of all left-hand-side
quantities above (\ref{C30}-\ref{C83}) -- so  all these quantities $Q$ are of 
the generic form
\eq{
Q (\tau, x) = e^{n \cdot \tau}\cdot f_Q, \hspace{1cm} n \in \left\{
0, 1, 2 \right\} .
}{C46}
Especially the scaling of $W_0 \simeq e^{\tau} \simeq W_1$ is crucial for the
derivation of the scaling-property of the final state \ac{BH} mass.

The \ac{EMKG} equation simplifies to a \ac{PDE}
\eq{
\partial_0f_s - \frac{1}{2}f_h\partial_1f_s + \left(1+\partial_0 \left(\ln R
\right)\right)f_s = C(\tau),
}{C31}
which in the case of \ac{CSS} further reduces to a linear 
first order \ac{ODE} because $f_s$, $R$ and $C$ become $\tau$-independent.

It would lead beyond the scope of this thesis if I would redo all the 
calculations encountered in spherically symmetric critical collapse -- but
I would like to emphasize again the simplicity of the first order approach as 
compared with the $4d$ approach \cite{mac96} and the appearance of a conserved 
quantity related to the \ac{BH} mass.

\subsection{\acs{ADM}, Bondi and \ac{BH} mass}

For the \ac{EMKG} model a remarkably simple relation 
between \ac{ADM}-mass\index{ADM mass} \cite{adm62}, \Ix{Bondi-mass} 
\cite{bon62, sac62} and the conservation law (\ref{EMKGcons}) using a \ac{SB} 
gauge has been established \cite{gru99, grk00}. {\em Per definitionem} the 
Bondi-mass at 
${\cal I}^-$ is given by the limit
\eq{
m_-(v) = \lim_{r \to \infty} m(r, v) = m (v, \infty; \infty, 0),
}{C22}  
where $m(r, v)$ is the so-called ``\Ix{mass-aspect function}'' being equivalent
to the integrated conserved quantity (cf. {\app appendix D} on p. 
\pageref{conservation law}).

The \ac{ADM}-mass is simply the limit
\eq{
m_{\text{ADM}} = \lim_{v \to \infty} m_-(v) .
}{C23}
An analogous calculation yields the Bondi-mass at ${\cal I}^+$ and its relation
to the \ac{ADM} mass:
\beqa
&& m_+(u) = \lim_{r \to -\infty}m(r, u) = m(u,\infty;-\infty,0), \label{C42} \\
&& m_{\text{ADM}} = \lim_{u \to -\infty} m_+(u).
\eeqa 

The initial state \ac{BH} mass\footnote{Supposing the spacetime being 
stationary at $i_-$ and $i_+$ -- in general the quantity given in eq. 
(\ref{C40}) is the complete ingoing timelike matter flux and eq. (\ref{C41}) 
yields the total outgoing timelike matter flux.}\index{black hole mass} is the 
limit
\eq{
m_{i.b.h.} = \lim_{v \to -\infty} m_-(v) = m_{\text{ADM}} 
- \int\limits_{-\infty}^{\infty} A_{\text{in}}(v', \infty) dv',
}{C40}
with $A$ being defined by the zero component of the matter part of the 
conserved quantity (i.e. it is proportional to $W_0$ defined in eq. 
(\ref{WEMKG}) for the \ac{EMKG} model in (ingoing) \ac{SB} gauge).

The final state \ac{BH} mass is given by
\eq{
m_{f.b.h.} = \lim_{u \to \infty} m_+(u) = m_{\text{ADM}} 
- \int\limits_{-\infty}^{\infty} A_{\text{out}}(u', \infty) du' .
}{C41}

Thus, with very little computational effort I could derive rather simple 
formulae (\ref{C22}-\ref{C41}), establishing the importance of the conserved 
quantity for energy considerations and (eventual) \ac{BH} masses.

\clearplaindoublepage

\chapter[Quantization of {\twod} gravity]
{Quantization of $\boldsymbol{2d}$ gravity}

\chapquote{Another great Dane has made free\\
With a question of Be or Not Be.\\
Now might Schr\"odinger's puss,\\
In descending by schuss,\\
Leave one track on each side of a tree?}{Peter J. Price}

\section{A word of warning}

For the reader who feels insulted by the combination of many words and few 
formulae I recommend to skip the following section. I have often considered to
delete it and would it have been a paper rather than a thesis I surely would 
have done it. Finally I have concluded that the advantages of keeping it (it 
is readable for non-specialists, it gives the thesis a personal note and it 
provides some interpretation of the formulae below) seem to balance
the disadvantages (in contrast to formulae which are either correct or wrong 
words are often fuzzy, part of the interpretation may seem trivial to some and 
incorrect to others and some statements are a little bit polemic).

The next two sections after these remarks merely review the simpler cases of 
matterless \ac{GDT} and minimally coupled scalars to point out and discuss 
certain weak points which exist in that derivations. The main section of this 
chapter (and the thesis) treats the case of nonminimally coupled matter 
(p. \pageref{nonminimally coupled matter} ff.) and is self-contained, but uses 
intensively all appendices, particularly \app{appendices E} and \app{F}. It 
includes as the most prominent special case the \ac{EMKG}.

\section{Some \Ix{conceptual remarks}}

I recommend the reader to consult the section above before starting to read 
this one. In the following, I would like to give my personal answer to some 
conceptual questions. Or rather, the answer that certain physicists have given
to those questions. Or rather, the answer that certain physicists have given to
those questions which coincide with my personal point of view.

\subsection[What is meant by ``quantization of $2d$ gravity''?]
{What is meant by ``quantization of $\boldsymbol{2d}$ gravity''?}

There exist various attempts to quantize gravity. The current research on this 
topic can essentially be divided into two classes: $M$-theory (including 
superstring theories) -- cf. e.g. \cite{sen98} and canonical quantum gravity 
 --  cf. e.g. \cite{ash99}. We will focus on the latter.

Canonical quantum gravity may not be the ``true'' theory of quantum gravity 
 --  and it surely seems not to be a very inspired way of trying to quantize 
gravity since at first glance only well-known methods (of quantum field 
theory) are used. However, it turns out that the conceptual problems in this 
formulation are rather awkward leading e.g. to the problem of 
time \cite{kie99}.  

Thus, one can hope to get some insight how to cope with these fundamental 
issues in easier models than $4d$ quantum gravity -- e.g. in {\twod} effective
action models which can mimic spherically symmetric $4d$ models.

Depending on the choice of canonical variables, the canonical theory can be 
subdivided into the following approaches \cite{kie99}:
\blist
\item {\em Quantum geometrodynamics}: The traditional approach using the 
three-dimensional metric as its configuration variable.
\item {\em Quantum loop dynamics}: The configuration variable is the trace of a
holonomy with respect to a loop, analogous to a Wilson loop.
\item {\em Quantum connection dynamics}: The configuration variables are the 
Cartan variables having many similarities to gauge theories. 
\elist

We are going to use the last set of configuration variables treating
gravity as a field theory of Cartan-variables, target space coordinates 
and (if present) matter fields -- thus we can use many of the 
techniques frequently used in \ac{QFT}. 
Of course, many of these calculations rely on perturbation 
theory, an approach which does not seem useful in the context of quantum 
gravity, since fluctuations in the geometric fields lead to ``fluctuations of
the manifold'' defined by that quantities, giving rise to the well-known 
conceptual problems of background (in)dependence (cf. e.g. the discussion in 
\cite{wal84} or \cite{kie99}). Our opinion is that quantum gravity should be 
considered exactly or not at all -- so-called ``semi-classical'' calculations 
may give hints what new scales or effects may appear, but the situation is 
similar to the one in the early days of quantum physics: Using Bohr's atom 
model for the description of the hydrogen atom instead of solving 
Schr\"odinger's equation (or Dirac's equation) gives some correct answers, but 
the theoretical picture is completely unsatisfactory.

Apart from this conceptual reason there is another hint that quantum gravity
should be considered non-perturbatively at least in the geometric sector: It is
well-known that the naive self-energy of a charged point particle is UV 
divergent. However, including gravitational effects leads to a finite 
(UV-regularized) expression for the self-energy\footnote{This argument can be 
found in many textbooks and reviews -- cf. e.g. \cite{ash91}.} (for clarity I
have undone the convention $8\pi G_N = 1$):
\eq{
m(r) = \frac{r}{2G_N}\left(\sqrt{1+\frac{4G_Nm_0}{r}+\frac{4G_Ne^2}{r^2}} - 
1\right)
}{selfenergy}
If we first expand in terms of the gravitational coupling constant and then let
$r \to 0$ each term in this series diverges in the UV-regime, but the 
(non-perturbative) sum is finite everywhere.
 
In our approach the answer to the question stated above is: ``The same way as
we would quantize any other \ac{QFT}''. We ``just'' have to ensure, that 
gravity is treated exactly and no perturbative arguments are used for the 
geometric part of the action. Of course, this task is non-trivial and the main 
part of this thesis.

Another question one may be tempted to ask is:

\subsection{``If the geometric fields are treated as quantum fields, to what 
background manifold do they refer?''}
 
The answer we would give is ``to $\mathbb{R}^n$ (with $n=2$ in our case)''. So 
the metrical and topological properties of the ``effective manifold'' that 
matter-fields interacting with the geometric fields feel is completely 
determined by the dynamics of the latter together with the interaction 
(``backreaction'') between gravity and matter. The topology and metric emerges 
through the dynamics of the system, which is consistent with the idea, that 
topology-changing quantum fluctuations may exist. Note, however, that in our 
approach we have to impose certain boundary conditions on some of the 
geometric fields, thus restricting the topological structure of the 
``effective manifold''.

Finally -- since we are using Hamiltonian formalism -- one may inquire:

\subsection{``What about time? Is it not some delicate coordinate favored 
by the Hamiltonian formalism?''} 

This question is answered in great detail in an 
article of Rovelli \cite{rov99}, boiling down to a simple ``No, it isn't.''. 
Even for the notion of ``motion'' time is not needed explicitly -- according 
to Henneaux and Teitelboim it emerges due to the ``unfolding'' of a gauge 
transformation \cite{het92}. Finally, even the concept of ``phase space'' does
not need a ``time-evolution'' explicitly if the phase space is defined as the
space of solutions of the classical \ac{EOM} rather than the space of states of the
dynamical system ``at a given time''. In the language of quantum mechanics this
viewpoint corresponds to the Heisenberg picture rather than the Schr\"odinger
picture\footnote{In what was perhaps his last public seminar, in Sicily, Dirac
used just a single transparency with just one sentence: ``The Heisenberg
picture is the right one'' \cite{rov99}.}.

A related question -- almost a ``never ending story'' -- having concerned (and
sometimes also confused) more than three generations of physicists is the 
``quantum enigma'' of the ``correct'' interpretation of quantum mechanics -- 
especially the phenomenon of macroscopic decoherence\footnote{This phenomenon 
is often characterized as the ``reduction of the wave packet''.}. My viewpoint 
is the FAPP-interpretation ({\bf F}or {\bf A}ll {\bf P}ractical 
{\bf P}urposes) which is similar to the FAPP-interpretation in thermodynamics: 
It is not impossible, that the pieces of a broken glass elevate from the floor 
by lowering its temperature and join together to form again the glass it has 
been before it fell from a table -- but it is extremely unlikely, because the 
Poincar{\'e} recurrence time is orders of magnitude larger than the estimated 
age of the universe. Analogically, it is not impossible that some human being 
is scattered as it walks through some door or some poor cat is in a 
superposition state of being alive and dead -- but due to its interaction with 
the environment it is extremely unlikely. Unless some decisive experiment is 
performed, Occam's razor cuts out all other interpretations, although they may 
have more ``sex-appeal''\footnote{E.g. Penrose's idea of a ``more even-handed 
marriage between quantum mechanics and gravity'' using what he calls 
``Objective Reduction'' \cite{pen89}.}. For a recent review on the decoherence 
problem and the emergence of the classical world cf. e.g. \cite{gjk96}.

Before treating a model with matter, we would like to review briefly the 
quantization procedure for the matterless case using some recent results 
obtained by path integral quantization \cite{klv97}.

\section{The \Ix{matterless case}}

Since the complete Hamiltonian analysis of the \ac{EMKG}-model is performed in
{\app appendix E} containing as a special case minimal coupled matter and the
matterless case, we restrict ourselves to quoting the main results of
\cite{klv97}.

As expected, we obtain no continuous physical degrees of freedom and the path
integral reduces to
\eqa{
W &=& \int \left({\cal D} X \right)\left({\cal D} X^+ \right)\left({\cal D} 
X^- \right)\de_{(1)}\de_{(2)}\de_{(3)} \det F \nonumber \\
&& \times \exp{i \int \left(J^+X^- + J^-X^+ + JX\right)}, 
}{Q1}
where the three $\de$-functions can be used to integrate out the remaining 
three fields in (\ref{Q1}). It is important, that non-trivial homogeneous
solutions may exist, because the arguments of the $\de$-functions contain some
differential operators. They are also neccessary, because otherwise the 
generating functional for connected Green functions would be identically zero
for vanishing ``momenta-sources'' $J, J^{\pm}$. In this sense, the paper 
\cite{klv97} is only correct up to eq. (33), because the homogeneous solution 
has been skipped there.

\section{Minimally coupled matter}\index{minimally coupled matter}

This special case for massless scalars has been investigated in \cite{klv99}. 
In the geometric part, the treatment is more general than our treatment, since 
we are going to restrict ourselves to \ac{SRG} (nevertheless, the complete 
Hamiltonian analysis performed in {\app appendix E} is still valid for the 
most general case).
However, in the matter part minimal coupling is imposed, unfortunately 
disregarding the phenomenologically interesting \ac{EMKG} model. Still, \cite{klv99}
serves in a certain way as a ``canonical'' example for more general cases,
which is why we review some crucial steps, generalizing to massive scalars with
arbitrary (non-derivative) self interactions.

The first order Lagrangian in terms of Cartan variables (see {\app appendix D})
\eq{
{\cal{L}}_{\text{FO}} = X^+ D\wedge e^- + X^- D\wedge e^+ + 
X d\wedge \omega - e^- \wedge e^+ {\cal V} + \ka dS \wedge *dS,
}{Q2}
yields together with\footnote{Although intuitively clear when looking at the
corresponding terms in the Lagrangian (\ref{Q2}), it is not 
obvious to identify the target space coordinates and the dilaton field with
the canonical conjugate momenta of the Cartan variables' one-components.
Indeed, in a more general ansatz one could treat {\em all} fields as 
independent canonical coordinates, obtaining new constraints (this time second
class). After the introduction of the Dirac bracket one can 
see that this identification is a justified shortcut reducing the Dirac-bracket
to the ordinary Poisson-bracket \cite{vasxx}.}
\seq{4cm}{
&& \bar{q}_i = \left( \omega_0, e^-_0, e^+_0 \right), \\
&& q_i = \left( \omega_1, e^-_1, e^+_1 \right),
}{4cm}{
&& p^i = \left( X, X^+, X^- \right), \\
&& P = \frac{\partial \cal{L}}{\partial \partial_0 S} = 
\frac{\partial {\cal{L}}^{\left( m \right)}}{\partial \partial_0 S},
}{Q3d}
after a Hamiltonian analysis performed in {\app appendix E} three primary and 
three secondary constraints, all of them being first class:
\eq{
\bar{p}_i \approx 0, 
}{Q3}
\eqa{
&& G_1 = -X^+ e^-_1 + X^- e^+_1 + \partial_1 X \approx 0, \label{Q3a} \\
&& G_2 = \partial_1 X^+ + \omega_1 X^+ - e^+_1 {\cal \tilde{V}}  + \frac{\ka}
{4e^-_1} \left[ \left( \partial_1 S \right) - \frac{P}{F(X)} \right]^2 
\approx 0, \label{Q3b} \\
&& G_3 = \partial_1 X^- - \omega_1 X^- + e^-_1 {\cal \tilde{V}} - \frac{\ka}
{4e^+_1} \left[ \left( \partial_1 S \right) + \frac{P}{F(X)} \right]^2 
\approx 0.
}{Q3c}
The symbol $\approx$ means ``weakly equal to zero'' or -- 
equivalently -- ``zero on the surface of constraints''. 

The algebra of the secondary constraints with themselves is rather simple
\begin{eqnarray}
&& \left\{ G_1, {G'}_2 \right\} = - G_2 \de(x-x'), \label{Q6} \\
&& \left\{ G_1, {G'}_3 \right\} = G_3 \de(x-x'), \label{Q7} \\
&& \left\{ G_2, {G'}_3 \right\} = - \left[ \frac{d {\cal V}}{d X} G_1
+ \frac{d {\cal V}}{d X^+}G_2 + \frac{d {\cal V}}{d X^-}G_3 
\right] \de(x-x'), \label{Q8}
\end{eqnarray}
although more general than in the Yang-Mills case, because the 
``structure-constants'' are no longer constants, but functions of the canonical
variables. Nevertheless, the cohomological series does not yield any higher
order structure functions, which is the generic case for Poisson-$\si$ models
(for this general statement cf. \cite{sst94a}; for an 
explicit calculation cf. {\app appendix E} on p. \pageref{BRST} ff.). A 
comparison with the more general algebra (\ref{algebra1}-\ref{algebra3})
shows that only in the third Poisson-bracket arises an additional term due to
the non-minimal coupling function $F(X) \neq \text{const}$.

The extended Hamiltonian can be written as a sum over constraints
\eq{
H_{ext} = - \bar{q}^i G_i + \la^i \bar{p}_i + \mu^i G_i,
}{Q4}
leading to an interpretation of the coordinates $\bar{q}_i$ as Lagrange 
multipliers\footnote{That is why we lifted the index of $\bar{q}$ in 
(\ref{Q4}).} analogously to the r{\^o}le of the zero component $A_0$ of the 
vector potential in \ac{QED}. 

Now comes the first subtle point in \cite{klv99}: Although not a canonical
gauge, temporal gauge is used for simplifying calculations. In the next
section we will discuss this issue to some extent. It suffices to say one
has to bother about boundary conditions, which are not treated there. Further,
only 3 gauge fixing relations are used like in Lagrangian path integral
quantization (instead of 6, which is the number of first class constraints and 
the number of gauge fixing conditions which have to be imposed in the 
Hamiltonian quantization procedure). Thus, at first glance, it seems that half 
of our constraints (and thus the corresponding classical \ac{EOM}) are 
``lost''. In {\app appendix E} on p. \pageref{lost equations} we provide a 
sketch how the ``lost constraints'' can be re-obtained.

Integrating out the ghosts, ghost-momenta, antighosts, 
anti\-ghost-\-mo\-menta, Lagrange-multipliers and their momenta results in a 
generating functional
\eqa{
W &=& \int \left({\cal D} p^i \right)\left({\cal D} q_i \right)
\left({\cal D} S \right)\left({\cal D} P \right) \det F \nonumber \\
&& \times \exp{i \int \left({\cal L}^{(eff)}+\text{sources}\right)},
}{Q5}
the explicit form of which will be given in the next section when we discuss
the non-minimal coupled model which includes minimal coupling as a special case
 --  cf. eqs. (\ref{Q33}, \ref{Q34}). 

The next subtle point is a change of the measure introduced by hand. It is 
argued in \cite{klv99} that -- since the correct diffeomorphism invariant 
measure for a scalar field on a curved background is ${\cal D}\left((-g)^{1/4}
S\right) = {\cal D}(\sqrt{(e)}S)$ -- one should perform this 
change, thus introducing a factor $\sqrt{q_3}$ in the path integral (because 
in that gauge $(e)=q_3$). A second factor coming out of the blue looking 
rather similar is $1/\sqrt{q_2}$ which cancels a term emerging from the 
scalar momentum integration. Since any measure is possible in principle, a 
convenient one is better, which is the reason for our choice.

Afterwards an integration over the coordinates leads to a path integral similar
to (\ref{Q1}) with the three $\de$-functions
\beqa
&&\de \left( - \nabla_0 \left( p_1 - \hat{B}_1 \right) \right), \\
&&\de \left( - \nabla_0 \left( p_2 - \hat{B}_2 \right) \right), \\
&&\de \left( - e^{-\hat{T}}\nabla_0e^{\hat{T}} \left( p_3 - \hat{B}_3 
\right) \right),
\eeqa 
with the operator $\hat{T}$ defined by
\eq{
\hat{T} := \nabla_0^{-1} \left( U(\hat{B}_1)\hat{B}_2 \right) .
}{Q9}

The operators $\hat{B}_i$ are given explicitly up to $\mathcal{O}(S^2)$. For 
the first time a perturbative expansion is used, which is 
harmless insofar as the neglected terms do not contain any geometric 
quantities. Fortunately, there is no conceptual problem in using perturbation 
theory for the matter fields, {\em after} quantum gravity has been treated 
exactly (i.e. non-perturbatively)\footnote{It amounts to first determining the
geometry of the ``background-manifold'' by an exact treatment of the 
geometrical degrees of freedom and afterwards using a traditional perturbation 
theory for the matter fields \'a la \ac{QFT} given on that background.}. By 
virtue of these $\de$-functions all geometric degrees of freedom may be 
integrated out yielding a generating functional containing only the scalar 
field and the sources.

Finally, another subtlety sneaks into the derivation of the generating 
functional: An additional term is added to the Lagrangian depending on an
arbitrary function of $x$ which is traced back to an ambiguity in the
definition of momentum-sources. This observation is backed by the calculations
performed in \cite{hak94} where the ``traditional'' order of 
integrations\footnote{By ``traditional'' we mean first the momentum 
integrations and {\em afterwards} the coordinate integration.} has been used. 
Only this term survives in the matterless case when the external 
momentum-sources are 
put to zero promoting it to the ``generator'' of the quantum \ac{EOM} of the
coordinates $q_i$. The residual gauge transformations investigated in {\app
appendix E} on p. \pageref{residual gauge transformations} sheds some light
on the obscure origin of this ambiguity.

The last step is the integration of the scalars which is performed only 
perturbatively and (although technically challenging) does not contain any
open points as opposed to the integration of the geometric part. Because of 
our introduction of a mass-term there is an additional 
${\mathcal O}(S^2)$ contribution to $\hat{B}_3$ (and in the case of 
self-interactions of course also higher order terms are present) which changes 
most of the formulae of chapter 4 of \cite{klv97}. 

In the next section the \ac{EMKG} model will be quantized in the same way and 
some of the aforementioned subtle issues will be clarified.

\section{Nonminimally coupled matter}\label{nonminimally coupled 
matter}\index{nonminimally coupled matter}

\subsection{The classical theory}

A review of the most important classical properties (e.g. the conservation law)
is outlined in {\app appendix D} and part of {\app appendix E}. Our 
starting point is the first order Lagrangian in 
terms of Cartan variables $e^{\pm}$ and $\om$, the dilaton field $X$, the 
target space coordinates $X^{\pm}$ and some massless scalar field $S$
\beqa
{\cal{L}}_1 &=& X^+ D\wedge e^- + X^- D\wedge e^+ + 
X d\wedge \omega - e^- \wedge e^+ {\cal V} \nonumber \\
&&+ F(X) \left[ dS \wedge *dS - e^- \wedge e^+ f(S) \right] 
\label{Q20} \\
&=:& {\cal{L}}^{\left(g\right)}+{\cal{L}}^{\left(m\right)} .
\eeqa
The interaction-function $f(S)$ can contain a matter term and arbitrary 
self-interaction terms. $F(X)$ is some smooth (coupling-)function of $X$ which 
will later be specified for the case of \ac{SRG} 
($F(X)=-1/2 X$). The splitting into a ``geometric part'' 
${\cal{L}}^{\left(g\right)}$ given by the first line of (\ref{Q20}) and a 
``matter part'' ${\cal{L}}^{\left(m\right)}$ given by the second is for later 
convenience.

\subsection{\Ix{Hamiltonian analysis}}

An extensive treatment is given in {\app appendix E}. The algebra
of the secondary constraints differs slightly from (\ref{Q6}-\ref{Q8}):
\begin{eqnarray}
&& \left\{ G_1, G'_2 \right\} = - G_2 \de(x-x') \label{Q21} \\
&& \left\{ G_1, G'_3 \right\} = G_3 \de(x-x') \label{Q22} \\
&& \left\{ G_2, G'_3 \right\} = - \left[ \frac{d {\cal V}}{d X} G_1
+\frac{d {\cal V}}{d X^+}G_2+\frac{d {\cal V}}{d X^-}G_3 \right] \de(x-x') 
\nonumber \\ 
&& \quad\quad\quad\quad\quad\quad + \frac{F'(X)}{(e)F(X)}{\cal{L}}^{(m)} G_1 
\de(x-x') \label{Q23}
\end{eqnarray}

The extended action reads
\eq{
L_{ext} = \int \left( \dot{q}_i p^i + \dot{S} P + \dot{\bar{q}}^i \bar{p}_i  
- H_0 \right) ,
}{Q24}
with $H_0 = -\bar{q}^i G_i$. The \Ix{\ac{BRST}-charge} is given by
\eq{
\Omega = c^i G_i + \frac{1}{2}c^i c^j C_{ij}{}^k p^c_k + b^i \bar{p}_i 
}{Q25}
introducing the ghosts $c^i$, the ghost momenta $p^c_i$ the antighosts $p^b_i$
and the anti-ghost momenta $b^i$. The structure functions $C_{ij}{}^k$
can be read off from (\ref{Q21}-\ref{Q23}). Note that the \ac{BRST}-charge 
still does not contain higher order structure functions which is a non-trivial 
result in the presence of nonminimally coupled matter.

The \Ix{gauge fixing fermion} $\Psi = p^{b}_i \chi^i + p^c_i \bar{q}^i$
contains the gauge fixing functions $\chi$ which will be chosen s.t. they
produce a temporal gauge. 

This implies the \ac{BRST}-extended gauge-fixed action
\eq{
L_{BRST} = \int \left[ \dot{q}_i p^i + \dot{S} P + \dot{\bar{q}}^i \bar{p}_i + 
\dot{c}^i p^c_i + \dot{b}^i p^k_i - H_{gf} \right] ,
}{Q28}
with the \ac{BRST}-extended gauge-fixed Hamiltonian being the 
\ac{BRST}-exact\footnote{Since the physical Hamiltonian is zero on the surface 
of constraints there is no non-exact contribution to the \ac{BRST}-closed 
Hamiltonian. Thus, the \ac{BRST}-extended Hamiltonian vanishes and only a 
gauge fixing term appears in (\ref{Q29}).} quantity
\eq{
H_{gf} = \left\{ \Psi, \Om \right\} .
}{Q29}

\subsection{Temporal gauge and \Ix{boundary conditions}}

By ``\Ix{temporal gauge}'' we define a class of gauge fixings of the generic
form $\bar{q}^i = a^i \in \mathbb{R}^3$. By ``non-singular temporal 
gauge'' we define a temporal gauge satisfying the condition 
$det g_{\al\be} \neq 0$. An example for a singular temporal gauge is the 
choice $a^i = 0$ with $i = 1, 2, 3$. 

We will be using the (reasonable) choice $a^1=a^3=0$ and $a^2=1$ leading to an
identification of $e_1^+$ with the determinant $(e)$ and to a proportionality 
between $e_1^-$ and the Killing-norm, thus promoting the zeroes of $e_1^-$ to
\Ix{Killing-horizons}. Note that this choice is equivalent to an outgoing 
\Ix{\ac{SB} gauge} of the metric if we identify $x_0 = r$ and $x_1 = u$
were $r$ and $u$ are ordinary outgoing \ac{EF} coordinates or to 
an ingoing \ac{SB} gauge with $x_0 = r$ and $x_1 = - v$.

Temporal gauge is not a canonical gauge because Lagrange multipliers are 
involved in the gauge fixing relations. Nevertheless it is possible to achieve
this gauge in the (extended) Hamiltonian formalism treating the Lagrange 
multipliers as canonical variables (cf. eq. (\ref{Q24})). 

The proper choice for the gauge fixing functions $\chi$ leading to a temporal
gauge is
\eq{
\chi = \frac{1}{\eps} \left(\bar{q}^i-a^i\right), \hspace{0.5cm}a^i = (0,1,0), 
}{Q32}  
with some (small) constant $\eps$. We intend to take the limit $\eps \to 0$ -- 
a trick which is well-known and can be found e.g. in a review of Henneaux 
\cite{hen85}.

Having integrated out all unphysical variables we obtain the path integral
\begin{eqnarray}
W_{\varepsilon=0} &=& \int \left({\cal D}q_i\right)\left({\cal D}p^i\right)
\left({\cal D}S\right)\left({\cal D}P\right) \left( \det \partial_0 
\right)^2 \det \left( \partial_0 + X^+U(X) \right) \nonumber \\
&&\times \exp \left[ i \int \left( {\cal L}_{\text{eff}}^{\varepsilon = 0} + 
J_i p^i + j_i q_i + Q S \right) d^2x \right], \label{Q33}
\end{eqnarray}
with the \Ix{effective Lagrangian}
\begin{equation}
{\cal L}_{\text{eff}}^{\varepsilon = 0} = p^i \dot{q}_i + P \dot{S} + G_2 .
\label{Q34}
\end{equation}

In {\app appendix E} on p. \pageref{lost equations} ff. it is shown that one
has to impose certain \Ix{boundary conditions} on the fields. We are going to 
assume asymptotic flatness for the ``\Ix{effective manifold}'' but postpone
the fixing of boundary values till we have integrated out the geometric 
variables. This will lead to some residual gauge degrees of freedom which are
fixed afterwards.

\subsection{Path integral quantization}\index{path integral quantization}

We are choosing a rather peculiar order of integration: First, we integrate out
the scalar field momentum $P$, but afterwards we eliminate the geometric 
{\em coordinates} before integrating the momenta. This is not the traditional
order of integrations, but it turns out to be a very convenient route of 
calculations, because all coordinates appear only {\em linearly} in the action.
Therefore, we will obtain $\de$-functions of the momenta and the scalar field
as a result which can be used to perform the geometric momentum integrations.
After adding the ambiguous term we are able to present a (nonlocal and 
non-polynomial) action containing only the scalar field.

\subsubsection{The result of the $\boldsymbol{P}$-integration}

The action yields
\eq{
{\cal L} = - \dot{p}_i q_i + \partial_1 p_2 + q_1 p_2 - q_3 {\cal V} + 
F(p_1) \left[ \left(\partial_0 S\right) \left(\partial_1 S\right) - q_2 
\left(\partial_0 S\right)^2 \right] .
}{Q100}
The measure contains an additional factor $\sqrt{\det{F(p_1)}}$ as compared to
the minimal coupled case which is just the proper one for s-waves. Again we
have to change the measure by hand if we want the proper Polyakov factor 
${\cal D}(S \det g^{1/4})$ \cite{pol81}.

\subsubsection{The result of the $\boldsymbol{q_i}$-integrations}

Since all coordinates appear now linearly, we obtain just three 
$\de$-functions:
\eqa{
&& \de\left(\dot{p}_1 - p_2 - j_1 \right) \label{Q101} \\
&& \de\left(\dot{p}_2 + F(p_1) \left(\partial_0 S\right)^2 -j_2 \right) 
\label{Q102} \\
&& \de\left(\dot{p}_3 + p_2p_3 U(p_1) + V(p_1) + F(p_1) f(S)- j_3 \right) .
}{Q103}

\subsubsection{The result of the $\boldsymbol{p_i}$-integrations}

The Faddeev-Popov determinant is canceled by these integrations. They 
are non-trivial in the sense, that the argument of
the $\de$-functions contain (a linear partial) differential operator. 
Therefore, homogeneous solutions and integration functions (which will be 
called constants in the following despite of their $x_1$-dependency) exist 
which must be added to the particular solutions of (\ref{Q101}-\ref{Q103}). 

Here we encounter a severe qualitative change in the behavior as an inspection
of the minimally coupled case reveals: The equations (\ref{Q101}) and 
(\ref{Q102}) are coupled due to the appearance of the coupling function 
$F(p_1)$ in the latter. Hence, one has to use a perturbative expansion in $S$
already at this step, i.e. one assumes that $S$ is a small perturbation (weak
matter approximation) and solves the system (\ref{Q101}-\ref{Q102}) iteratively
up to a given order in $S$. Afterwards, the solution for $p_1, p_2$ can be 
plugged into (\ref{Q103}) yielding a result for $p_3$. If $F(p_1)$ is linear 
in $p_1$ (as it is the case for \ac{SRG}) our sub-system of \ac{PDE} 
(\ref{Q101}-\ref{Q102}) becomes linear.

As an example we provide a perturbative solution for the simple case 
$F(p_1) = p_1$ up to quartic order in the scalar field
\eqa{
&& \hat{p}_1 = {\cal J} - \nabla_0^{-2} \left(\left(\partial_0 S\right)^2 
\left( {\cal J} - \nabla_0^{-2} \left( \left(\partial_0 S\right)^2 {\cal J} 
\right)\right)\right), \label{Q104} \\
&& \hat{p}_2 = \nabla_0^{-1} j_2 + \bar{p}_2 - \nabla_0^{-1}\left(\left(
\partial_0 S\right)^2 \left({\cal J}-\nabla_0^{-2} \left( \left(\partial_0 S
\right)^2 {\cal J}\right)\right)\right), \label{Q105} \\
&& \hat{p}_3 = e^{-\hat{T}} \left[\nabla_0^{-1} e^{\hat{T}}\left(j_3 - 
V(\hat{p}_1) - F(\hat{p}_1) f(S) \right)+\bar{p}_3\right],
}{Q106}
with the homogeneous contributions
\eq{
\nabla_0 \bar{p}_i = 0,
}{Q107}
and the same formal definitions as in \cite{klv98}
\eq{
\hat{F} = e^{-\hat{T}} \nabla_0^{-1} e^{\hat{T}}, \hspace{0.5cm}\hat{T} = 
\nabla_0^{-1}\left(\hat{U}\hat{p}_2\right), \hspace{0.5cm}\hat{U} = 
U(\hat{p}_1),
}{Q108a}
as well as a new one
\eq{
{\cal J} = \nabla_0^{-1} j_1 + \nabla_0^{-2} j_2 + \bar{p}_1 + \nabla_0^{-1} 
\bar{p}_2 .
}{Q108}
The solution (\ref{Q106}) is also valid for the most general case, provided
that one is able to obtain a solution for $\hat{p}_1$ and $\hat{p}_2$.

Of course, these ``inverse derivatives'' have to be taken with a grain of salt:
Either one bothers about proper regularization of this integral-like operators
by analogy with \cite{klv99} or one uses a certain short-cut for the 
evaluation of scattering amplitudes like in \cite{gkv00}. Since our main 
interest are $S$-matrix elements we will follow the last route.

\subsubsection{The \Ix{ambiguous terms}}

The resulting action
\eq{
{\cal L} = J_i \hat{p}_i + SQ + F\left(\hat{p}_1\right) \left(\partial_0 S
\right) \left(\partial_1 S\right)
}{Q109}
must be supplemented by additional terms surviving in the limit of vanishing 
sources coming from the following ambiguity of the $J_i \hat{p}_i$-terms 
(cf. {\app appendix E} on p. \pageref{ambigous terms}):
\eqa{
J_1 \hat{p}_1 &\to& J_1 \hat{p}_1 + \hat{g}_1 K_1(\nabla_0^{-1}, \left(
\partial_0 S\right)^2, j_1, j_2), \label{Q110} \\
J_2 \hat{p}_2 &\to& J_2 \hat{p}_2 + \hat{g}_2 K_2(\nabla_0^{-1}, \left(
\partial_0 S\right)^2, j_1, j_2), \label{Q111} \\
J_3 \hat{p}_3 &\to& J_3 \hat{p}_3 + \hat{g}_3 e^{\hat{T}} \left(j_3 - 
V(\hat{B_1}) - F(\hat{p}_1) f(S) \right), 
}{Q112}
where the $\hat{g}_i$ are homogeneous solutions of $\nabla_0 \hat{g}_i = 0$ and
$K_i$ can be obtained perturbatively for the simple case shown in the example
above (\ref{Q104}-\ref{Q105}).
They are again slightly different from the ambiguity terms obtained in
\cite{klv99}\footnote{There the matter contributions of $\hat{p}_1$ and 
$\hat{p}_2$ have been omitted, while the third ambiguous term is essentially
equivalent to (\ref{Q112}).}. Note that in the limit of vanishing sources $J_i$
{\em and} $j_i$ only (\ref{Q112}) yields a non-vanishing contribution. Thus
for the calculation of scalar vertices only (\ref{Q112}) is relevant.
The constants $\hat{g}_1$ and $\hat{g}_2$ can be absorbed into integration 
constants which will be fixed by asymptotic conditions below.

Since the three ambiguous terms are closely related to $q_i$ 
(cf. (\ref{qEOM})), one could say that the only nontrivial effective 
interaction of the scalar field with geometry arises due to its coupling to 
the determinant $(e) = q_3$. In fact, the term 
$\hat{g}_3 e^{\hat{T}}|_{j_i=0,J_i=0}$ appearing in the only nontrivial 
ambiguous contribution is identical to $q_3$ which further fortifies this 
observation.

\subsubsection{The result}

After the smoke clears we obtain a path integral containing only the scalar 
field $S$ and various sources as well as integration constants:
\eq{
W = \int \left({\cal D}S\right) \sqrt{\det{\hat{q}_3}} 
\sqrt{\det{F(\hat{p}_1)}} \exp i \int {\cal L}_S,
}{Q113}
with
\eq{
{\cal L}_S = J_i \hat{p}_i + SQ + \hat{p}_1 \left(\partial_0 S\right) \left(
\partial_1 S\right) + \hat{g}_3 e^{\hat{T}} \left(j_3 - V(\hat{p_1}) - 
F(\hat{p}_1) f(S) \right) .
}{Q114}
The non-polynomiality is partly due to the ambiguous term, but also due to the
non-minimal coupling $\hat{p}_1$ (which contains the scalar field in a 
non-polynomial way as can be seen from the solution of (\ref{Q101})) in front 
of the propagator term which is a novel feature.

At first glance, this result was obtained without restricting the potentials
$U(p_1)$, $V(p_1)$, $f(S)$ or the coupling function $F(p_1)$. However, we have
only checked for the simple case of $F(p_1) = c_0 + c_1 p_1$ with 
$c_i \in \mathbb{R}$ explicitly that the ambiguous contributions proportional 
to $g_1$ and $g_2$ can be absorbed into redefinitions of integration constants.
It seems likely that this property holds also in general, but this has not 
been proved yet.

In principle, one can use (\ref{Q113}-\ref{Q114}) as a starting point for a 
perturbation theory in the scalar field $S$. However, in order to obtain
mathematically meaningful results one would have to take care about a proper 
definition of the ``inverse derivatives'' appearing in the solutions 
(\ref{Q104}-\ref{Q106}). Therefore, we will follow a simpler strategy which has
proven very useful already in the minimally coupled case \cite{gkv00}. 

\addtocontents{toc}{\protect\newpage}
\subsection{The 4-point \Ix{vertices}}

From now on we will specialize to the \ac{EMKG}, i.e.
\seq{3cm}{
&& U(p_1) = -\frac{1}{2p_1}, \\
&& V(p_1) = -\frac{\la^2}{4},
}{3cm}{
&& F(p_1) = -\frac{p_1}{2}, \\
&& f(S) = 0.
}{Q137}

\addcontentsline{toc}{subsection}{\protect\hspace*{1.33truecm}3.5.5.1 \, 
The trick}
\subsubsection{The trick}

It is sufficient to assume the following second order combinations of the 
scalar field being localized at a single point\footnote{Actually, the 
sources should be localized at different points, but for the lowest order 
tree-graphs -- which are our main goal -- this makes no difference.}
\eq{
S_0 := \frac{1}{2}\left(\partial_0 S\right)^2 = c_0 \de (x-y), \hspace{0.5cm} 
S_1 := \frac{1}{2}\left(\partial_0 S\right)\left(\partial_1 S\right) = c_1 
\de(x-y),
}{Q115}
and solve the classical \ac{EOM} following from the gauge 
fixed action (\ref{Q114}) 
\seq{2.5cm}{
&& \partial_0 p_1 = p_2, \\
&& \partial_0 p_2 = p_1 S_0, \\
&& \partial_0 p_3 = 2 + \frac{p_2p_3}{2p_1},
}{4cm}{
&& \partial_0 q_1 = \frac{q_3p_2p_3}{2p_1^2} + S_1 - q_2 S_0, \\
&& \partial_0 q_2 = - q_1 - \frac{q_3p_3}{2p_1}, \\
&& \partial_0 q_3 = - \frac{q_3p_2}{2p_1},
}{Q116}
up to linear order in the ``sources'' $c_0$ or
$c_1$. Then the solutions have to be substituted back into the interaction 
terms in (\ref{Q114}). Higher orders in $c_0, c_1$ would yield either loop 
contributions or vertices with at least 6 outer legs. We emphasize again that 
we are using perturbative methods in the matter sector only. Thus no a priori 
split into background- and fluctuation-metric occurs in our approach. 
As already mentioned before, this is a perturbative approach and justified 
only if the scalar field is ``small'', i.e. its energy is negligible as 
compared to Planck's mass. 

Using the ``causal'' prescription for the boundary values at $x_0 \to \infty$ 
introduced in \cite{klv99} as well as the useful on-shell relation 
$\partial_0 = p_2 \partial/\partial p_1$ we obtain up to linear order in $c_0$ 
and $c_1$:
\eqa{
\hspace{-1cm} && p_1 (x) = x_0 + (x_0 - y_0) c_0 y_0 h(x,y), \label{Q117} \\
\hspace{-1cm} && p_2 (x) = 1 + c_0 y_0 h(x,y), \label{Q118} \\
\hspace{-1cm} && q_2 (x) = 4 \sqrt{p_1} + \left(2c_0y_0^{3/2}-c_1y_0 + 
(c_1-6c_0y_0^{1/2})p_1 \right) h(x,y), \label{Q119} \\
\hspace{-1cm} && q_3 (x) = \frac{1}{\sqrt{p_1}}, 
}{Q120}
where $h(x,y) := \theta (y_0 - x_0) \de(x_1-y_1)$.
We have used as matching conditions at $x_0=y_0$ the following continuity
properties: $p_1, q_2$ and $q_3$ are $C^0$ and $\partial_0 q_2(y_0+0) - 
\partial_0 q_2(y_0-0) = \left(c_1 - q_2(y_0) c_0\right) \de(x_1-y_1)$. 
Integration constants which would produce an asymptotic (i.e. for $x_0 \to 
\infty$) Schwarzschild term and a Rindler term have been fixed to zero. Thus, 
a black hole may appear only at an intermediate stage (the ``virtual black 
hole'', see below), but should not act asymptotically. Due to the infinite 
range of gravity this is necessary for a proper $S$-matrix element, if
we want to use as asymptotic states spherical waves for the incoming and 
outgoing scalar particles. 

\addcontentsline{toc}{subsection}{\protect\hspace*{1.33truecm}3.5.5.2 \, 
The effective line element}
\subsubsection{The \Ix{effective line element}}

The solutions (\ref{Q119}) and (\ref{Q120}) yield the outgoing \ac{SB} 
line element
\eq{
(ds)^2 = 2 (dr) (du) + K(r,u) (du)^2,
}{Q123} 
if we identify\footnote{Note the somewhat unusual r{\^o}le of the indices 0 and 1:
$x_0$ is asymptotically proportional to $r^2$, thus our Hamiltonian evolution 
is with respect to a ``radius'' as ``time''-parameter.}
\eq{
u = 2\sqrt{2} x_1, \hspace{0.5cm} r = \sqrt{p_1(x_0)/2} .
}{Q132}
Since $p_1$ is the dilaton field it really makes sense to call this coordinate 
``radius''. The Killing norm
\eq{
\left. K (r,u) \right|_{x_0 < y_0} = \left(1 - \frac{2m_u}{r} - a_u 
r\right)\left(1+ {\cal O}(c_0) \right),
}{Q124}
with $m_u = \de(x_1(u)-y_1)(-c_1 y_0 - 2 c_0 y_0^{3/2})/2^{7/2}$ and 
$a_u = \de(x_1(u)-y_1)(c_1-6c_0y_0^{1/2})/2^{5/2}$, has two zeros located 
approximately at $r = 2m$ and $r = 1/a$ corresponding to a Schwarzschild- and 
a Rindler-horizon if they are positive. In the asymptotic region the 
Killing norm is constant by construction: 
$\left. K (r,u)\right|_{x_0 > y_0} = 1$. These boundary conditions fix all 
integration constants and residual gauge transformations.

We would like to recall that (\ref{Q123}) is not only a result of the temporal
gauge, but also a result of choice: We have set the asymptotic mass- and
Rindler-term equal to zero thus fixing the residual gauge freedom and hence all
integration constants. This choice was triggered not so much by physical 
considerations, but by simplicity. Of course, other choices are possible and
interesting. However, we do not know how to treat the asymptotic states of the
scalar field in the general case. Therefore, we restrict ourselves to the 
simplest possibility. But we will nevertheless obtain a non-trivial result 
for the $S$-matrix.

\addcontentsline{toc}{subsection}{\protect\hspace*{1.33truecm}3.5.5.4 \, 
The virtual black hole}
\subsubsection{The \Ix{virtual black hole}}

As in \cite{gkv00} we turn next to the conserved quantity, which exists in all 
two dimensional generalized dilaton theories \cite{kus92}, even in the 
presence of matter \cite{kut99}. For SRG its geometric part reads
\eq{
{\cal C}^{(g)} = \frac{p_2p_3}{\sqrt{p_1}} - 4 \sqrt{p_1} 
}{conserved}
and by assumption it vanishes in the asymptotic region $x_0 > y_0$. A simple 
argument shows that ${\cal C}^{(g)}$ is discontinuous: $p_1$ and
$p_3$ are continuous, but $p_2$ jumps at $x_0 = y_0$. This phenomenon
has been called \acf{VBH} in \cite{gkv00}. It is generic 
rather than an artifact of our special choice of asymptotic conditions.

The solutions (\ref{Q117}) and (\ref{Q118}) establish
\eq{
\left. {\cal C}^{(g)} \right|_{x_0 < y_0} = 4 c_0 y_0^{3/2} \propto - m_{VBH}. 
}{vbh}
Thus, $c_1$ only enters the Rindler term in the Killing norm, but not 
the VBH mass (\ref{vbh}). 

\addcontentsline{toc}{subsection}{\protect\hspace*{1.33truecm}3.5.5.3 \, 
The result}
\subsubsection{The vertices}

\begin{figure}
\centering
\epsfig{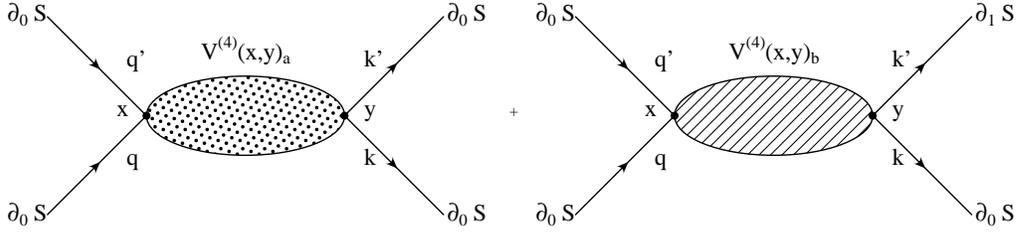}
\caption{Total $V^{(4)}$-vertex with outer legs}
\label{fig:s4treegraphs}
\end{figure}
For the vertices we obtain (independently of the prescription)
\meq{
V^{(4)}_a = \int_x\int_y S_0(x) S_0(y) \left. \left( \frac{d q_2}{d c_0} p_1 + 
q_2 \frac{d p_1}{d c_0} \right)\right|_{c_0 = 0 = c_1} \\
= \int_x\int_y S_0(x) S_0(y) \left| \sqrt{y_0}-\sqrt{x_0} \right| 
\sqrt{x_0y_0} \left( 3x_0+3y_0+2\sqrt{x_0y_0} \right) \de(x_1-y_1),
}{Q121}
and
\meq{
V^{(4)}_b = \int_x\int_y \left. \left(S_0(y) S_1(x) \frac{d p_1}{d c_0} - 
S_0(x) S_1(y) \frac{d q_2}{d c_1} p_1 \right)\right|_{c_0 = 0 = c_1} \\ 
= \int_x\int_yS_0(x) S_1(y) \left|x_0-y_0 \right| x_0 \de(x_1 - y_1),
}{Q122}
with $\int_x := \int\limits_0^\infty dx^0\int\limits_{-\infty}^\infty dx^1$.

\addcontentsline{toc}{subsection}{\protect\hspace*{1.33truecm}3.5.5.5 \, 
The asymptotics}
\subsubsection{The \Ix{asymptotics}}

With $t := r + u$ the scalar field satisfies asymptotically the spherical 
wave equation. For proper s-waves only the spherical Bessel function (cf. e.g.
\cite{lal86})
\eq{
R_{k0} (r) = 2k \frac{\sin (kr)}{kr}
}{Q126}
survives in the mode decomposition of $S(r,t)$:
\eq{
S(r,t) = \frac{1}{\sqrt{2\pi}} \int\limits_0^{\infty} \frac{dk}{\sqrt{2k}} 
R_{k0} \left[a^+_k e^{ikt} + a^-_k e^{-ikt}\right] 
}{Q127}
This mode decomposition will be used to define asymptotic states and to build 
the Fock space. The normalization factor is chosen such that the Hamiltonian 
reads
\eq{
H = \frac{1}{2} \int\limits\limits_0^{\infty} \left[ (\partial_t S)^2 + 
(\partial_r S)^2 \right] r^2 dr = \int\limits_0^{\infty} dk k a^+_k a^-_k .
}{Q128}
Note that in \cite{fgk01} different normalization has been chosen, which will
eventually lead to slightly different pre-factors in the scattering amplitude.

\subsection{Scattering amplitudes}\index{scattering amplitudes}

\addcontentsline{toc}{subsection}{\protect\hspace*{1.33truecm}3.5.6.1 \, 
Preliminaries}
\subsubsection{Preliminaries}

\begin{figure}
\centering
\epsfig{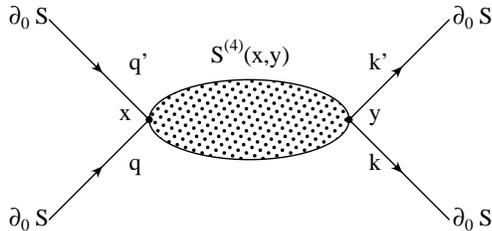}
\caption{Nonlocal $S^{(4)}_{min}$-vertex with outer legs}
\label{fig:s4min}
\end{figure}

Before presenting details of the calculation is is worthwhile to collect the 
results obtained in the minimally coupled case. A first result for the 
(nonlocal) $V^{(4)}$-vertex was obtained in \cite{klv97}. It was used in 
\cite{gkv00} to calculate the scattering amplitudes for this vertex (cf. 
figure \ref{fig:s4min} for the massless case).

In the massless case we could show that we either obtain a singular amplitude 
or (by imposing regularizing boundary conditions) a zero amplitude. Because no
loop-integrations were involved this was interpreted as following: Either
always a \ac{BH} forms (independently how ``strong'' the initial conditions
may be) or no scattering occurs at all (the \ac{BH} has been ``plugged'' by
suitable boundary conditions). 

\begin{figure}
\centering
\epsfig{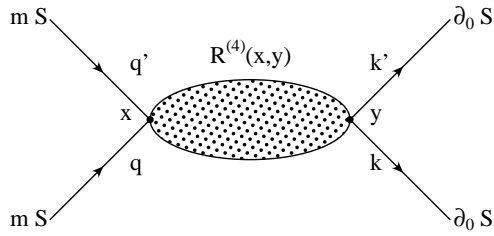}
\caption{Nonlocal $R^{(4)}_{min}$-vertex with outer legs}
\label{fig:r4min}
\end{figure}

In the massive case, however, a non-trivial finite result could be obtained
due to the appearance of an additional vertex (cf. figure \ref{fig:r4min}) 
which we called $R^{(4)}$ in \cite{gkv00}. For energies $E$ fulfilling 
$m \ll E \ll m_{planck}$, where $m$ is the mass of the scattered scalars, we
could distillate a simple ``asymptotic'' behavior for the complete vertex:
\eq{
S^{(4)}+R^{(4)} \approx R^{(4)} \approx c m^6 \frac{\ln E}{E^5},
}{Q40} 
with $c$ being a dimensionless constant near unity and the 
$S^{(4)}$-contribution being suppressed by a further factor of $m^2/E^2$.

From the (classical) numerical results of spherical symmetric collapse 
\cite{cho93} it is to be expected, that the amplitudes extracted from the 
(classical) vertex must somehow reproduce threshold behavior of \ac{BH}
formation. Since this behavior could not be seen in the minimal coupled case
we expect essential differences to the non-minimal coupled one.

Indeed, the nonminimal case provides an interesting result already in the 
simple case of massless scalars. At first glance it may be surprising that the 
innocent looking factor $X$ in front of the matter Lagrangian induces such 
tremendous changes in the qualitative behavior. On the other hand, already the
constraint algebra differs essentially (\ref{Q23}) and the \ac{PDE} 
(\ref{Q101}-\ref{Q102}) become coupled as opposed to the minimal case giving 
rise to an additional vertex ($V^{(4)}_b$). Thus, this qualitative change in 
the scattering amplitude (see figure \ref{fig:s4treegraphs})
\eq{
T(q, q'; k, k') = \frac{1}{2}\left< 0 \left| a^-_ka^-_{k'} \left(V^{(4)}_a 
+ V^{(4)}_b \right) a^+_qa^+_{q'}\right| 0 \right> 
}{Q129}
is not so surprising after all.

\addcontentsline{toc}{subsection}{\protect\hspace*{1.33truecm}3.5.6.2 \, 
The final result}
\subsubsection{The \Ix{final result}}

After some long and tedious calculations (for details see {\app appendix F}) 
we obtain
\eq{
T(q, q'; k, k') = \frac{i\pi\de\left(k+k'-q-q'\right)}{4\sqrt{kk'qq'}} E^3
\tilde{T},
}{RESULT}
with
\meq{
\tilde{T} (q, q'; k, k') := \frac{1}{E^3} {\Bigg [}\Pi \ln{\Pi^2} + \frac{1}
{\Pi} \sum_{p \in \left\{k,k',q,q'\right\}}p^2 \ln{p^2} \\
\cdot {\Bigg (}3 kk'qq'-\frac{1}{2}\sum_{r\neq p} \sum_{s \neq r,p}
\left(r^2s^2\right){\Bigg )} {\Bigg ]},
}{feynman}
and $\Pi = (k+k')(k-q)(k'-q)$. It is fair to say that (\ref{RESULT}) is the 
main result of this thesis. In \cite{fgk01} essentially the same equation is 
presented with slightly different normalization factors. Since that result
is easier to comprehend from a fourdimensional point of view, we mention it as
well:
\eq{
T_4(q, q'; k, k') = -\frac{i\ka\de\left(k+k'-q-q'\right)}{2(4\pi)^4
|kk'qq'|^{3/2}} E^3 \tilde{T},
}{RESULT4}
having restored the $\ka$-dependence to enable dimensional analysis. 

\addcontentsline{toc}{subsection}{\protect\hspace*{1.33truecm}3.5.6.3 \, 
Discussion of the final result}
\subsubsection{Discussion of the final result}

The simplicity of (\ref{feynman}) is quite surprising, in view of the fact that
the two individual contributions (cf. figure \ref{fig:r4min}) are not only 
vastly more complicated, but also divergent (cf. (\ref{F6n})). This precise 
cancellation urgently asks for some deeper explanation.
The fact that a particular subset of graphs to a given order in perturbation
theory may be gauge dependent and even divergent, while the sum over all such
subsets should yield some finite, gauge-independent $S$-matrix is well known 
from gauge theory in particle physics (cf. e.g. \cite{kul73}).
However, it seems that only in the temporal gauge (\ref{Q32}) 
one is able to integrate out the geometric degrees of freedom 
successfully. Also that gauge is free from coordinate singularities which we 
believe to be a prerequisite for a dynamical study extending across the 
horizon\footnote{%
Other gauges of this class, e.g. the Painlev{\'e}-Gullstrand gauge 
\cite{pai21} seem to be too complicated to allow an application of our 
present approach.}.

The only possible singularities occur if an outgoing momentum equals an ingoing
one (forward scattering). Near such a pole we obtain with $k=q+\eps$ and
$q\neq q'$:
\eq{
\tilde{T}(q,q';\eps) = \frac{2(qq')^2}{\eps}\ln{\left(\frac{q}{q'}\right)} + 
{\cal O}(1).
}{forward}

The nonlocality of the vertex prevents the calculation of the usual $s$-wave 
cross section. However, an analogous quantity can be defined by squaring 
(\ref{RESULT}) and dividing by the spacetime integral over the product of the 
densities of the incoming waves $\bigl( \rho=(2\pi)^{-3} \sin^2(qr)/(qr)^2 
\bigr)$: $I = \int Dr dt \rho(q)\rho(q')$, $\sigma=I^{-1}
\int_0^{\infty} DkDk' |T|^2$. Together with the introduction of 
dimensionless kinematic variables $k=E \alpha, k'=E(1-\alpha), q=E \beta, 
q'=E(1-\beta), \alpha, \beta \in [ 0,1 ]$ this yields
\begin{equation}
 \frac{d\sigma}{d\alpha}=\frac{1}{4(4\pi)^3}\frac{\kappa^2 E^2 |\tilde{T}
(\alpha, \beta)|^2}{(1-|2\beta-1|)(1-\alpha)(1-\beta)\alpha\beta}.
\label{crosssection}
\end{equation}

Our result also allows the definition of a decay rate $d^3 \Gamma/(DqDkDk')$ 
of an $s$-wave with ingoing momentum $q$ decaying (!) into three outgoing 
ones with momenta $k,k',-q'$. Clearly, lifetimes calculated in this manner
will crucially depend on assumed distributions for the momenta.

Finally, we stress that in the more general four dimensional setup of
gravitational particle scattering combinations of non-spherical modes could 
contribute to the $s$-wave matrix element. Hence, our result (\ref{RESULT}) 
does not include the full ($4d$) classical information. Nonetheless, as the 
previous discussion shows, its physical content is highly nontrivial. We
emphasize especially the decay of $s$-waves, which is a new phenomenon. Note 
that it cannot be triggered by graviton interaction, since there are no 
spherically symmetric gravitons. Still, it is caused by gravity, i.e. by 
gravitational self interaction encoded in our non local vertices.

\clearplaindoublepage

\chapter{Possible Extensions}


\chapquote{It is possible by ingenuity and at the expense of clarity... \{to 
do almost anything in any language\}.  However, the fact that it is possible 
to push a pea up a mountain with your nose does not mean that this is a 
sensible way of getting it there.  Each of these techniques of language 
extension should be used in its proper place.}{Christopher Strachey}

\section{Loop calculations}\index{loop calculations}

Up to now only (the lowest order) tree graphs have been considered. A natural
generalization would be the calculation of one-loop diagrams. To this end, we
should step back and reconsider the minimally coupled case first for 
simplicity. Efforts in this direction already exist \cite{vasxx}, but it is 
too early to judge whether sensible results can be obtained for this case. It 
is to be expected, that even in the (initially) massless case one-loop 
diagrams yield through the renormalization procedure an effective mass term, 
since there is no symmetry protecting vanishing rest mass of the scalars.

\section{Dimensional reduction anomaly}\label{dranomaly}\index{dimensional 
reduction anomaly}

The fact that quantization and dimensional reduction may not commute has been
called ``dimensional reduction anomaly'' \cite{fsz00}. This is of particular
interest for spherically symmetric spacetimes like the spherically reduced 
\ac{EMKG} \cite{sut00}. 
Fortunately, it will not affect our tree graph calculations, since no 
renormalization was needed. But as soon as loop calculations are considered,
this anomaly should be taken into account.

\section{Nonperturbative calculations}

With our perturbative solution (\ref{Q104},\ref{Q105}) we were able to describe
the limit for a small (virtual) BH. On the other hand, semi-classical 
approaches are typically valid in the limit of a large BH -- the most 
prominent is e.g. the derivation of Hawking radiation\footnote{Since the 
Hawking temperature behaves like $T \propto 1/M$ the limit $M \to 0$ is 
obviously physically meaningless.} \cite{haw75}. It would be very interesting,
if we could obtain a nonperturbative solution of these equations.

Formally, this ``solution'' is of the form (cf. (\ref{Q104}))
\eqa{
&& p_1 = \frac{1}{1+\nabla_0^{-2} \left(\partial_0 S\right)^2} {\cal J}, 
\label{nonper1} \\
&& p_2 = \nabla_0 p_1.
}{nonper2}
However, this function of the integeral operator $\nabla_0^{-2}$ is only 
defined by its (perturbative) power series, so one cannot take (\ref{nonper1}) 
at face value. Nevertheless it is amusing to speculate a bit
using this formal ``solution'': For ``small'' $\nabla_0^{-2} (\partial_0 S)^2$ 
we re-obtain our perturbative result, i.e. the limit for a vanishing BH. For 
``large'' $\nabla_0^{-2} (\partial_0 S)^2$ the ``solution'' is regular, 
although in the expansion each term is large and the series diverges, in 
analogy to the example (\ref{selfenergy}). 

Unfortunately, prospects to give (\ref{nonper1}) some 
rigorous mathematical meaning beyond perturbation theory do not seem 
good\footnote{An alternative formulation of (\ref{nonper1}) -- which by 
``conservation of misery'' does not provide an exact solution apart from very 
special cases --  is given by the second order \acs{PDE} 
$(\nabla_0^2+(\partial_0 S)^2)p_1=0$. Note that we are using $F(p_1)=p_1$ as 
in (\ref{Q104}), so for \ac{SRG} the normalization factors have to be adapted
accordingly.}.

\section[$d$-dimensional spherical reduction]{$\boldsymbol{d}$-dimensional 
spherical reduction}\index{$d$-dimensional spherical reduction}

In principle only minor changes occur for $d$-dimensional spherical reduction
as compared to the special case $d=4$ -- c.f. (\ref{familiar}-\ref{dilgauge}).
The quantization procedure is completely equivalent, but a small complication
is induced by the fact that the potential $V \neq$ constant. One arrives again
at equation (\ref{Q113}-\ref{Q114}) but of course (\ref{Q137}) has to be 
modified accordingly.

In this case, the asymptotic solution\footnote{Assuming again the vanishing of
Rindler- and Schwarzschild-terms.} is given by
\eq{
\left. q_3 \right|_{as} = p_1^{-\frac{d-3}{d-2}}, \hspace{0.5cm} \left. q_2 
\right|_{as} = 4p_1^{\frac{d-3}{d-2}},
}{G1}
while for $p_1$ and $p_2$ one obtains the same solution as before (\ref{Q117}-
\ref{Q118}). With the redefinitions $r = p_1^{1/(d-2)}/\sqrt{2}$
and $u = \sqrt{2}(d-2)$ again the line element (\ref{Q123}) is produced.
The 4-point vertices will differ. It would be an interesting 
task to investigate the dimensional dependence of the scattering amplitude
(particularly the stability of finiteness of the scattering amplitude against 
small deviations from $d=4$).

\section{Non-vanishing \Ix{cosmological constant}}

A non-vanishing cosmological constant -- either at the $d=2$ level or at the
$d=4$ level -- induces just a change in the potential $V$. Thus, essentially 
the same remarks are valid as in the previous discussion of $d$-dimensional 
spherical reduction. The modification of $V$ is given by (\ref{C35}) with 
$Q=0$. Obviously, the cosmological constant will spoil asymptotic flatness. 
Instead, one has to assume an asymptotically (Anti-)deSitter spacetime
\eq{
(ds)^2 = 2(dr)(du)+K_{\La}(u,r)(du)^2, 
}{G2}
with $\lim_{r\to\infty}K_{\La}(u,r) = 1 + \La r^2$ provided that again the
Schwarzschild- and Rindler-terms vanish asymptotically. The main problem in 
this setup is of course that one cannot use Minkowski modes anymore.

\section{Non-vanishing \Ix{Schwarzschild term}}

If the boundary conditions are chosen s.t. only the Rindler term vanishes, but 
not the Schwarzschild term, then instead of Minkowski modes (i.e. spherical
waves) Schwarzschild modes have to be used \cite{eli87}. Unfortunately, there 
exists no general exact solution for this case, but in principle this problem 
could be solved numerically. This way, scattering on a real BH (rather than 
``just'' on the VBH) can be described.

\section{Non-linear \Ix{torsion terms}}

In contrast to previous modifications this one changes the theory essentially.
The reason is that I have been relying heavily on the standard form of the
potential ${\cal V} = U(X)X^+X^-+V(X)$. In non-linear torsion theories this
is not the generic form (the only exception is the \ac{KV}-model which is the 
special case of quadratic torsion). One would have to reconsider the whole 
quantization procedure from the beginning for this case. Fortunately, all 
prominent \ac{GDT} have a standard form potential (cf. table \ref{table1}). Of
course, this does not imply that there are no interesting models of this type
which could be investigated. But such an investigation goes beyond a simple
generalization of the scenario treated in this thesis. Therefore, I will not 
discuss it explicitly.

\section{\acs{JBD} and related theories}\index{Jordan-Brans-Dicke}

Such a generalization within the theoretical framework of first order formalism
has already been treated in \cite{hof99}. It suffices to say
that the introduction of a second ``dilaton field'' (the Dicke field) together
with dimensional reduction leads to an action the first order pendant of which
is very similar to (\ref{EMKG}):
\meq{
S_E = \int_{M_2} d^2x \sqrt{-g} \left[ X R + V_1^E (X,Y) \nabla_{\al} X 
\nabla^{\al} X + V_2^E (X, Y) \nabla_{\al} Y \nabla^{\al} Y \right. \\
\left. + V_3^E (X,Y) \nabla_{\al} X \nabla^{\al} Y + V_4^E (X,Y) \right].
}{G3}
The first dilaton ($X$) plays the r{\^o}le of ``the'' dilaton while the second 
one ($Y$) behaves like a scalar matter field. This point has already been 
investigated to a certain extend in the framework of ``two-dilaton theories'' 
\cite{ghk01}, a class of theories which includes the spherically reduced 
\ac{EMKG}, the polarized Gowdy model, spherically reduced Kaluze-Klein theory 
and spherically reduced scalar-tensor theories (and thus implicitly spherically
reduced nonlinear gravity theories \cite{mas94}). Especially scalar-tensor
theories have attracted attention recently in the context of quintessence 
cosmology \cite{zws99}.

As a demonstration I will pick out a very special model, the polarized Gowdy 
model \cite{gow71}. It turns out, that the result is completely equivalent to 
\ac{SRG}, apart from the fact that $V=0$ instead of being a non-vanishing 
constant. This small change has tremendous consequences: Setting the 
Schwarzschild- and Rindler-terms to zero yields $q_2 = 0$. This implies in 
particular a vanishing Killing norm at $i_0$. The total 4-vertex reduces to
\eq{
S^{(4)} = S_0(x)S_1(y)\frac{dp_1}{dc_0} = S_0(x)S_1(y)(y_0-x_0)x_0\Th(x_0-y_0)
\de(x_1-y_1),
}{G4}
and is similar to the non-symmetric vertex. In fact, it corresponds to the 
negative part of the signum function present in (\ref{Q122}). Due to 
symmetrization in the outer legs one concludes that the result for the 
polarized Gowdy model is equivalent to the non-symmetric part of the \ac{EMKG} 
result (i.e. containing dilogarithms and divergencies).

A more realistic choice of boundary conditions (leading to 
$\left.q_2\right|_{i_0} \neq 0$) could nevertheless lead to some finite result
for the scattering amplitude, but it will be accessible only numerically.

\section{Adding \Ix{fermions}}

Instead of (or in addition to) the scalar field one could use fermionic matter
degrees of freedom, given by the action\footnote{In $d=2$ the spin-connection 
does not enter the fermionic matter Lagrangian, which is a very nice feature.
The notation $\overleftrightarrow{d}$ is analogic to the more common
$\overleftrightarrow{\partial_{\mu}}$, meaning that $a\overleftrightarrow{d}b 
:= a \overrightarrow{d} b - a \overleftarrow{d} b$. The $\gamma^a$ are the 
(constant) $\gamma$-matrices (e.g. in light cone gauge).}
\eq{
{\cal L}^{(f)} = (e) F(X) e_a \left( \bar{\chi} \gamma^a 
\overleftrightarrow{d} \chi \right).
}{G5}
The resulting constraint algebra is actually simpler, despite the appearance of
second class constraints \cite{kum92, wal01}. The minimally coupled case 
provides again an example of a finite W-algebra \cite{kum92}. Since the 
calculations seem actually easier using fermions it will be interesting to 
quantize them analogical to the \ac{EMKG}. Note however, that a proper 
spherical reduction of $4d$ spinors will yield an action differing from 
(\ref{G5}), since additional (dilaton dependent mass-like) terms
will arise as well as terms coupled to the auxiliary fields $X^\pm$ 
\cite{penroserindler}.

\section{Adding \Ix{gauge fields}}

In the matterless case the introduction of gauge fields does not add any 
continuous physical degrees of freedom. Moreover, it is possible to eliminate the
Yang-Mills field completely and replace it by an effective term in the 
geometric potential. However, a new conserved quantity corresponding to the 
Yang-Mills charge appears \cite{kst96a}, denoted by $Q$. 
The request for a non-trivial bulk dynamics implies again the necessity of an 
introduction of matter degrees of freedom. I do not foresee any problems in 
this case and as a starting point for further work I will just state the new 
terms appearing in the potential (cf. eq. (\ref{C35})):
\eq{
{\cal V}(X^+X^-,X) \to {\cal V}(X^+X^-,X) + \frac{\la^2}{16}\frac{Q^2}{X}.
}{G6}
The asymptotic behavior ($X \to \infty$) is not changed to such an important 
degree as in the (A)dS case, but at least additional contributions in the 
vertex arise, caused by $d V(X)/d X \neq 0$ as opposed to (\ref{Q137}). 
Moreover, the asymptotic line element will look like the RN metric, thus 
spoiling our ansatz using Minkowski modes.

\section{Treatment of \Ix{boundary terms}}

In this thesis I have concentrated on the bulk dynamics, which is nontrivial
only in the presence of matter degrees of freedom. It is instructive to
investigate also the boundary dynamics, which can be nontrivial already in the
purely topological case (i.e. without matter). Since this is not the main
goal of this thesis I will only sketch its treatment.

In the matterless case all boundary contributions are trivial if one neglects
subtleties with large gauge transformations (i.e. finite diffeomorphisms)
\cite{gks97}. Unfortunately, one needs precisely these large gauge 
transformation if the topology is nontrivial (e.g. $S^1\times\mathbb{R}$ for
the \ac{KV} model \cite{sst94a}). They seem to be accessible only in 
Casimir-Darboux coordinates, but these coordinates are valid only in a certain 
patch. Moreover, if these boundary terms are used as an edge state action in 
order to count microstates contributing to the black hole entropy (for the 
spherically reduced case as well as for the \ac{CGHS} case) the result diverges 
\cite{gks97}. 

Thus it seems that despite the success in particular models \cite{sst94a} a
general satisfying treatment of the boundary dynamics is still not accessible
today. Insofar there is still work to be done even in the matterless case.

\clearplaindoublepage

\chapter{Conclusions}

\chapquote{A conclusion is simply the place where someone got tired of 
thinking.}{\tt /usr/share/games/fortunes/science} 

The first chapter was devoted to historical, physical and mathematical remarks,
providing a brief overview of general dilaton theories in two dimensions. We
argued that the coupling of matter to a \ac{GDT} could be interesting, because 
one could still work in two dimensions, but the theory had now continuous
physical degrees of freedom. Thus, either genuine matter degrees of freedom 
(as in the \ac{EMKG}) or gravitational degrees of freedom (as in the case 
of the polarized Gowdy model) could be treated in addition to the topological
ones. This introduction has to be read in conjunction with {\app appendix A} 
where the notation used for this work is listed, and with {\app appendix B} 
where very briefly some aspects of Einstein-Cartan gravity and quantization in 
the presence of constraints are reviewed.

In chapter two the relation between the fourdimensional \ac{EH} action and the
spherically reduced dilaton action was recapitulated (using the results of 
{\app appendix C} and {\app appendix D}). The first order formulation
of the latter was presented for the special case of the \ac{EMKG}. The reason,
why we focused on this model is twofold: On the one hand, it is still simple 
enough, to allow an analytic treatment (albeit perturbatively in the matter 
sector). On the other hand, it is dynamically nontrivial, as the seminal 
numerical work of Choptuik \cite{cho93} has revealed.
In order to demonstrate the power of the first order formalism three examples 
were given: A discussion of Killing-horizons for an \ac{AdS}-\ac{RNBH}, a 
presentation of the \ac{DSS} solutions of critical collapse and the relations 
between the \ac{ADM} mass, the Bondi mass, the effective \ac{BH} mass and the 
conserved quantity, which exists in all \ac{GDT}, even in the presence of 
matter. Further examples are contained in {\app appendix D}, e.g. the exactly
soluble case of static matter in section \ref{static solutions}. 

The main new contribution of this thesis, chapter three, was devoted to quantum
aspects and classical scattering amplitudes of the \ac{EMKG}. Starting with
some conceptual remarks concerning the quantization of gravity, a concise 
presentation of the quantization procedure for the matterless case and the (in 
$d=2$) minimally coupled case followed. Section \ref{nonminimally coupled 
matter} treated the quantization of nonminimally coupled matter, again focusing
on the \ac{EMKG}. We were heavily referring to {\app appendix E}, where the
Hamiltonian analysis and BRST quantization for the general case were performed.
It turned out, that the constraint algebra was changed essentially by the
nonminimal coupling function (cf. $C_{23}{}^1$ in (\ref{structure})), although 
in the Virasoro representation (\ref{virasoroalgebra}) this is not noticed. We
discussed a further set of constraints (``energetic constraints'') closely 
related to the conserved quantity providing an abelianization of the constraint
algebra in the matterless case. Returning to the ``natural'' constraints 
following directly from the Hamiltonian analysis, the use of a particular 
temporal gauge for the Cartan variables (cf. (\ref{Q32}) with (\ref{Q3d})) 
proved to be very convenient for two reasons: This gauge choice yielded
regular (\ac{SB}) coordinates at Killing horizons and was simple enough to
allow an exact treatment of the path integral for the geometric part of the
action. The order of integrations turned out to be crucial. 
We arrived at a nonlocal and nonpolynomial action consisting solely of the 
matter field and sources (\ref{Q109}). It had to be supplemented by an 
ambiguous term, eq. (\ref{Q112}), coming from residual gauge degrees of 
freedom. Our final (nonlocal and nonpolynomial) result of path integral 
quantization will be repeated here for convenience of the reader:
\eq{
W = \int \left({\cal D}S\right) \sqrt{\det{\hat{q}_3}} 
\sqrt{\det{F(\hat{p}_1)}} \exp i \int {\cal L}_S,
}{con1}
with
\eq{
{\cal L}_S = J_i \hat{p}_i + SQ + \hat{p}_1 \left(\partial_0 S\right) \left(
\partial_1 S\right) + \hat{g}_3 e^{\hat{T}} \left(j_3 - V(\hat{p_1}) - 
F(\hat{p}_1) f(S) \right),
}{con2}
where geometric quantities like $\hat{q}_i$, $\hat{p}_i$ or functions thereof 
(like $\hat{T}$) after {\em exact} path integrals only depend on the scalar 
field $S$ and sources $j_i$, $J_i$. The ambiguous term proportional to 
$\hat{g}_3$ had to be added as a homogeneous solution to a first order PDE, as
discussed in {\app appendix E}. The ambiguity was found to be the result of
residual gauge degrees of freedom which could be fixed by appropriate boundary
conditions on some of the canonical variables.

In the second part of this section, we treated the (two) lowest order 
tree graphs following from perturbation theory applied to (\ref{Q114}). We
had to assume the ``smallness'' of the scalar field in natural units, i.e. its
energy had to be small as compared to Planck energy. Using the same trick as 
in \cite{klv99,gkv00} we obtained the (matter dependent) effective line 
element (\ref{Q123}), the occurrence of a ``virtual black hole'' (\ref{vbh}) 
and the sought tree vertices. Using proper $s$-waves (\ref{Q126}) as 
asymptotic states we could derive the scattering amplitude (\ref{RESULT}) and 
(\ref{feynman}). All calculations were performed in {\app appendix F} and 
partly in a {\em Mathematica} notepad available at the URL \\
\myURL \\
This was necessary due to the sheer amount of terms present in intermediate
results. However, ``miraculous'' cancellations between the two amplitudes
(among them the divergent contributions) condensed the result to a final simple
form\footnote{We have restored the $\ka$-dependence and taken into account the
normalization factors used in \cite{fgk01}.}:
\eq{
T_4(q, q'; k, k') = -\frac{i\ka\de\left(k+k'-q-q'\right)}{2(4\pi)^4
|kk'qq'|^{3/2}} E^3 \tilde{T},
}{con3}
with the scale independent part
\meq{
\tilde{T} (q, q'; k, k') := \frac{1}{E^3} {\Bigg [}\Pi \ln{\Pi^2} + \frac{1}
{\Pi} \sum_{p \in \left\{k,k',q,q'\right\}}p^2 \ln{p^2} \\
\cdot {\Bigg (}3 kk'qq'-\frac{1}{2}\sum_{r\neq p} \sum_{s \neq r,p}
\left(r^2s^2\right){\Bigg )} {\Bigg ]}.
}{con4}
We explained these cancellations by gauge independence of the $S$-matrix and 
noted, that a different choice of gauge in principle could yield finite (and 
simple) intermediate results. However, we did not find such a gauge which still
has the two aforementioned virtues of temporal gauge\footnote{{\em We are 
indeed a blind race and the next generation, blind to its own blindness, will 
be amazed at ours.} L.L. Whyte, ``Accent on Form''}. Thus, we left this issue
as an open problem. A physical discussion of forward scattering, a 
cross section like quantity and the rate of an ingoing $s$-wave decaying (!)
into three outgoing ones concluded this chapter.

The final chapter four included an outlook to relevant possible extensions:
\blist
\item {\em Loop calculations}: They are an obvious next step, but for sake of
simplicity it seems convenient to discuss the (one) loop case first for 
minimally coupled matter. Note that due to the properties of our action 
``nonlocal loops'' will appear. Preliminary calculations in this direction
look quite promising \cite{vasxx}.
\item {\em Dimensional reduction anomaly}: The fact that
quantization and dimensional reduction do not commute in general has to be 
taken into account for a proper spherically reduced quantum theory 
\cite{fsz00,sut00}.
\item {\em Nonperturbative calculations}: Of course it would be extremely
interesting to obtain some exact solutions apart from the known static limit
(Fisher solution \cite{fis48}) and the \ac{CSS} case (Roberts solution 
\cite{rob89}). Maybe the simple form of (\ref{nonper1}) in temporal gauge, or
 its differential counterpart $(\nabla_0^2+(\partial_0 S)^2)p_1=0$, can help
to achieve this.
\item {\em $d$-dimensional spherical reduction}: This is a straightforward
application of our results. It would be interesting to investigate the 
dimensional dependence of the vertices and the scattering amplitude, since this
could give a clue to the dimensional reduction anomaly contribution, using
e.g. $d=4-\eps$ instead of $d=4$ and letting $\eps \to 0$.
\item {\em Nonvanishing cosmological constant} and {\em Schwarzschild case}: 
Both cases are extremely relevant for phenomenology, but the main difficulty 
is to find analytically manageable asymptotic modes. Possibly, a numerical 
treatment could be considered for these cases, but then again one might just 
stick to numerics in $d=4$ in the first place.
\item {\em Non-linear torsion terms}: It is possible to extend all results to
theories with non-linear torsion, but practically all known \ac{GDT} 
(cf. table \ref{table1}) have the standard form potential 
${\cal V} = U(X)X^+X^-+V(X)$.
\item {\em JBD and related theories}: Because of its impact on cosmology,
the treatment of scalar tensor theories in our framework would be very
interesting. A discussion heading in this direction can be found in 
\cite{ghk01}.
\item {\em Fermions}: A natural extension is the inclusion of fermions.
Preliminary results seem to imply that actually the calculations are 
{\em easier} than in the scalar case \cite{wal01}. Moreover, the relation to
W-algebras (cf. e.g. \cite{bht96}) in this context seems intriguing.
\item {\em Gauge fields}: The inclusion of such fields is already interesting 
in the purely topological scenario, since a second conserved quantity (the 
monopole charge) arises. As in the Schwarzschild case, the problem is the 
definition of asymptotic states.
\item {\em Boundary contributions}: The investigation of the r{\^o}le of 
large gauge transformations seems tailor made for the matterless case, because
no mixing between continuous and discrete physical degrees of freedom occur.
Still, it would be interesting to generalize e.g. the results of 
\cite{sst94a}.
\item {\em Supersymmetry}: For a comprehensive treatment of supersymmetric
dilaton theories we refer to the dissertation of M. Ertl \cite{ert01} and 
references therein. The extension of that work to models with matter (or, 
equivalently, the extension of the present work to models with supersymmetry) 
provides a huge field of applications of the methods and results developed in 
these theses.
\elist

\clearplaindoublepage


\begin{appendix}

\renewcommand{\chaptermark}[1]{
  \markboth{\appendixname\ \thechapter.\ #1}{}}

\chapter{Notations}
\label{conventions}

\section{Signs}\index{signs}

\blist
\item We use the ``West-Coast'' convention (or Bj{\o}rken-Drell convention) 
for the \Ix{signature} of the metric:
\begin{equation}
\text{sig}_4 = (+, -, -, -) \text{ and } \text{sig}_2 = (+, -). 
\label{signature}
\end{equation}
\item By $\varepsilon_{\mu \nu}$ we always denote the \Ix{Levi-Civit\'a symbol}
and not the corresponding tensor density. Thus
\begin{equation}
\varepsilon^{01} = 1 = - \varepsilon^{10} = - \varepsilon_{01} = 
\varepsilon_{10}
\end{equation}
is just a number.
\item We have the same sign convention for $\varepsilon^{ab}$ where $a, b$ are
target space light cone indices: $\eps^{+-} = 1$. But for sake of compatibility
with previous work \cite{kus92,kum92,kuw94,kuw95,klv97,klv97a,klv97b,klv98,
kuv98,klv99,kut99,kuv99,grk00,gkv00} we will nevertheless assume that the 
minus-component is ``the first one'', i.e. $d^2x\sqrt{-g} = e^-\wedge e^+$. In 
this sense, our sign convention is reversed: 
\eq{
\eps^{-+} = -1 .
}{epsilon}
\item In the Hamiltonian analysis we use the standard sign convention for the
fundamental Poisson bracket
\eq{
\left\{ q, p \right\} = + 1 .
}{PBsign}
\item  We are using the left-derivative giving rise to possible sign-changes 
in the \ac{EOM} whenever the right-derivative is used instead. By left-derivative 
we mean
\eq{
\frac{\de}{\de a} (a \wedge b) = b, \hspace{0.5cm}\frac{\de}{\de a} (b \wedge 
a) = (-)^{f_af_b}b,
}{leftderivative}
where $f_x$ is the form degree of $x$.
\item The conserved quantity (\ref{conserv}) is defined s.t. in \ac{SRG} it has to 
be negative for a positive \ac{BH} mass.
\item The secondary constraints are defined s.t. the Hamiltonian (\ref{h13}) 
is given by
\eq{
H_0 = - \bar{q}^i G_i .
}{hamilsign}
\item The sign definition for the gauge fixing fermion (\ref{fixfer}) is s.t. 
the gauge fixing part of the Hamiltonian reads
\eq{
H_{gf} = + \left\{ \Psi, \Om \right\} .
}{gfhamilsign}
\elist

\section{Indices}\index{indices}

We are using the Einstein-summation except if the corresponding indices are in 
brackets. Latin indices denote anholonomic indices while Greek indices are 
holonomic ones. \twod-tensor indices are always from the beginning of the 
alphabet ($a, b, \dots$; $\al, \be, \dots$). ``Angular'' tensor indices 
appearing in {\app appendix C} are denoted by $i, j, \dots$ while $4d$-tensor 
indices are from the middle of the alphabet ($n, m, \dots$; 
$\mu, \nu, \dots$). Light-cone indices are always denoted by (upper) $\pm$.

\section{Brackets}\index{brackets}

We are using the classical mechanics notation for the Poisson bracket like in
\cite{arn78}. Therefore, we always denote Poisson-brackets or anti-brackets by 
$\left\{\dots, \dots\right\}$. (Anti-)commutators will be denoted by 
$\left[\dots, \dots\right]$.

\section{Coordinates}\index{coordinates}

In the Hamiltonian analysis coordinates, constraints, ghosts and anti-ghost 
momenta get always lower (counting) indices $i, j, ...$ while momenta, 
Lagrange multipliers, ghost momenta and anti-ghosts get upper ones. The
(continuous) ``space-time'' indices are denoted by $x_0$ and $x_1$.

Special gauges leading to well-known coordinate systems are
\blist
\item the diagonal/Schwarzschild gauge with coordinates $t, r$
\item the outgoing \acf{SB}/\acf{EF} gauge with coordinates 
$u, r$
\item the ingoing \ac{SB}/\ac{EF} gauge with coordinates
$v, r$
\elist
In all of these cases we are using the standard notation for the coordinates,
e.g. for the null directions $u=t-r$ and $v=t+r$.

\section{\Ix{Cartan variables}}

In the text we are using the name ``\Ix{vielbein}'' also for the dual basis of 
one-forms. We are aware of possible confusion that may arise due to this 
sloppy ``convention''; still, we believe that it is more convenient to use 
``vielbein'' in our context since the expression ``dual basis of 1-forms'' is 
unwieldy.

\section{Physical \Ix{units}}

We use Planck units, i.e. $8\pi G_N = c = \hbar  = 1$. The proper 
coupling-factor of the matter Lagrangian is discussed in 
(\ref{matter}-\ref{EMKG}).

\section{Branch cuts}\index{branch cuts}

We will always assume that the branch cut of the logarithm is located at the
negative real axis. Thus we have
\eq{
\lim_{\eps \to 0} \ln{(x\pm i\eps)} = \ln{x} \pm i\pi \theta(-x) .
}{logcut}

\clearplaindoublepage

\chapter{Mathematical preliminaries}

\section[(Pseudo-)Riemannian manifolds and vielbein formalism]{(Pseudo-)Riemannian manifolds and \linebreak[4]vielbein formalism}

\subsection{\Ix{Cartan's structure equations}}

The basic definitions and relations used in classical general relativity are
extensively summarized in \cite{wal84}. For the Cartan formulation we refer
to \cite{nak90}.

The {\em first Cartan's structure equation} describes 
the torsion as a 2-form in terms of the vielbein basis: 
\begin{equation}
T^{a}=de^{a}+\omega^a{}_b\wedge e^{b} .
\label{cartan1}
\end{equation}

The metric compatibility reads $\omega _{cb}=-\omega _{bc}$. This important 
condition together with $T^{a}=0$ uniquely determines the connection 1-form 
$\omega_{\,\,\,\,b}^{a}$ which is then nothing else than the Levi-Civit\'a 
connection in the vielbein formalism. Nevertheless it turns out to be better 
to keep connection and vielbein as independent variables in the action, 
because the \ac{EOM} simplify. 

The {\em second Cartan's structure equation}
\begin{equation}
R^a{}_b = d\omega^a{}_b + \omega^a{}_c\wedge\omega^c{}_b
\label{cartan2}
\end{equation}
allows to compute the matrix valued curvature 2-form using the connection.

Note that by applying the covariant differential 
$D^a{}_b := \de^a{}_bd+\om^a{}_b$ several times to the structure equations we 
obtain always relations which contain only the \Ix{Cartan variables}. 
Therefore, classical gravity can be described completely in terms of the 
Cartan variables.

\subsection[Simplifications in $d=2$]{Simplifications in $\boldsymbol{d=2}$}
\label{simplifications}\index{simplifications in $d=2$}
\index{twodimensional models}

Counting some important independent degrees of freedom we conclude for $d=2$:
\begin{eqnarray}
&& \# \text{Riemann tensor components} = 1 \\
&& \# \text{Lorentz angles} = 1 \\
&& \# \text{Diffeomorphisms} = 2 \\
&& \# \text{vielbein components} = 4 \\
&& \# \text{connection components} = 2 \\
&& \# \text{continous physical degrees of freedom} = 0 
\end{eqnarray}
The last relation is a consequence of the topological nature of \ac{GDT} in 
$d=2$ and can be shown e.g. by a Hamiltonian analysis (cf. section 
\ref{gauge invariance} on p. \pageref{gauge invariance}).

Introducing light cone components for the target space coordinates bearing 
anholonomic indices
\begin{equation}
X^{\pm} := \frac{1}{\sqrt{2}} \left( X^0 \pm X^1 \right)
\end{equation}
simplifies some calculations: Scalar products (which contain second powers of 
each component) are transformed into bilinear terms
\begin{eqnarray}
&& X^i X_i = \eta_{ij} X^i X^j = \left(X^0\right)^2 - \left(X^1\right)^2 
\nonumber \\
&& \quad\quad\quad\quad = X^a X_a = \eta_{ab} X^a X^b = 2X^+X^-,    
\end{eqnarray}
which proves useful in the path integral quantization procedure. Note that the
Minkowski-metric in light-cone components is given by
\eq{
\eta_{ab} = \left( \begin{array}{cc}
0 & 1 \\
1 & 0
\end{array} \right) .
}{minkowski}

The connection 1-form in light-cone components reads (remember that the first 
component is always the minus-component)
\begin{equation}
\omega^a{}_b = \left( \begin{array}{cc}
-\omega & 0 \\
0 & \omega
\end{array} \right) = \eps^a{}_b \om. \label{omega2}
\end{equation}

The quadratic term in $\omega$ in the second Cartan's structure equation 
(\ref{cartan2}) vanishes identically in $d=2$ leading to a formula for the
curvature
\begin{equation}
*(d \wedge \omega) = - \frac{R(e)}{2}, \label{omecurv}
\end{equation} 
where we have introduced the determinant $(e)$ defined by
\begin{equation}
(e) := e_0^-e_1^+ - e_1^-e_0^+ .
\end{equation}
The $*$-symbol denotes the \Ix{Hodge dual}:
\eqa{
&& p := p_{\left[ i_1 \dots i_p \right]} dx^{i_1} \wedge \dots \wedge 
dx^{i_p}, \\
&& *p := g^{i_1 j_{d-p+1}} \dots g^{i_p j_p} \tilde{\eps}_{j_1 \dots j_d} 
p_{\left[ i_1 \dots i_p \right]} dx^{j_1} \wedge \dots \wedge dx^{j_{d-p}},
}{hodge}
where $p$ is a p-form, $d$ the dimension (in our case d=2) and 
$\tilde{\eps}_{j_1 \dots j_d}$ the totally antisymmetric tensor density (the 
``volume form'').
 
Note that in (\ref{omecurv}) $\omega$ is the same as defined on the 
\ac{r.h.s.} of (\ref{omega2}) which explains the factor 2. 

In principle it is possible to eliminate the connection by virtue of the first
Cartan's structure equation (\ref{cartan1}) if the torsion 2-form is set
equal to zero, the advantage being the reduction of degrees of freedom in the
action. The price on has to pay is an ugly non-linearity arising in the
Lagrangian which is the reason why we abandon this idea.

\section[Hamiltonian formalism of constrained systems]{\Ix{Hamiltonian formalism} of constrained \linebreak[4] systems}

In this section we briefly summarize the main results of \Ix{Hamiltonian 
analysis} closely following \cite{het92}.

\subsection{Introduction and Notation}

Starting point of all common quantization schemes is an educated guess for the
Lagrangian, often motivated by symmetry considerations and desired properties
of the quantized theory (like e.g. the concept of renormalizability in 
ordinary \ac{QFT}).

The Lagrangian action is defined by
\begin{equation}
L_L = \int\limits_{t_0}^{t} L(q_i, \dot{q}_i, t')dt',
\end{equation}
and normally one assumes in ``fundamental theories'' that $\partial_t L = 0$ 
encoding the absence of ``friction''.

Iff the Hessian $\partial^2 L/\partial \dot{q}_i \dot{q}_j$
does not vanish it is possible to invert the equation for the \Ix{canonical 
momenta}
\begin{equation}
p^i = \frac{\partial L}{\partial \dot{q}_i},
\end{equation}
and express the velocities in terms of momenta and coordinates.

Because of
\begin{equation*}
\dim M = \text{rank } M + \text{kernel } M, 
\end{equation*}
with $M$ being a matrix, there must exist $d-k$ {\em constraints} 
$\Phi_j (p^i, q_i) = 0$ (where $d$ is the dimension of the Hessian and $k$ its 
rank) called {\em primary constraints} which can be added to the Lagrangian 
multiplied with {\em \Ix{Lagrange multipliers}} without changing the 
dynamics.

In field theories the Lagrangian is replaced by the {\em Lagrange density}
and the (countable) degrees of freedom labeled by the index $i$ become 
uncountable and will be denoted by some continuous parameter $x$ and some 
discrete parameter $i$. In order to simplify our notation we will not 
distinguish between continuous and discrete degrees of freedom -- thus the
expression $p^i \dot{q}_i$ has to be read as $\sum_i \int_x p^i(x) 
\dot{q}_i(x)$.

The Hamiltonian is obtained by a Legendre transformation of the 
``velocities\footnote{We assume that this transformation can be done which is
certainly true for our model.}''
\begin{equation}
H = p^i \dot{q}_i - L - \lambda^j \Phi_j, \hspace{1cm} i=1..k, \hspace{1cm} 
j=1..(d-k),
\end{equation}
where $\lambda^i$ are the Lagrange multipliers.

For consistency, the constraints must not change under the temporal evolution
of our system establishing the {\em \Ix{consistency equations}}
\begin{equation}
\dot {\Phi}_i = \left\{ \Phi_i, H \right\} \approx 0,
\label{consistency}
\end{equation}
introducing the {\em \Ix{Poisson bracket}} defined by
\begin{equation}
\left\{ A, B \right\} = \frac{\partial A}{\partial z_i} \Gamma_{ij} 
\frac{B}{z_j},
\end{equation}
where $\Gamma_{ij}$ is a symplectic 2-form and $z_i=(q_i, p^i)$ contains the
full set of (generalized) canonical variables.

We will only use canonical variables which have the trivial 
\Ix{fundamental Poisson bracket}\footnote{This is always possible via the 
{\em Darboux theorem} \cite{dar89}.}
\begin{equation}
\left\{ q_i, p^j \right\} = \delta_i^j, 
\label{fundabracket}
\end{equation}
which simplifies the 2-form $\Gamma_{ij}$ to the {\em symplectic normal 
form}.

Note that on the \ac{r.h.s.} of equation (\ref{fundabracket}) -- in accordance 
with our abbreviative index-notation -- the $\delta$-symbol contains 
$\delta$-functions over continuous and Kronecker-$\delta$'s over discrete 
degrees of freedom.

\subsection{Analysis of constraints}\index{constraint analysis}

For a profound discussion of constraint analysis we refer the reader to the
literature \cite{dir96}, \cite{frv77}, \cite{git90} and \cite{het92}. In brief,
one has to investigate the solutions of (\ref{consistency}). The three 
possibilities which may occur are a tautology, a contradiction or the 
existence of a new constraint. The first case is trivial because the 
Dirac-procedure stops. The second case is also trivial and shows the 
inconsistency of the Lagrangian\footnote{Not all thinkable functions are 
possible Lagrangians -- the easiest counterexample is the well-known 
pathological ``Lagrangian'' ${\cal{L}} (q, \dot{q}, t) = q$ leading to the \ac{EOM} 
1=0.}. The third possibility gives rise to secondary constraints which (from
the Hamiltonian point of view) are treated on an equal footing as the 
\Ix{primary constraints}. We will specialize to cases where at some step of the
algorithm either of the first two possibilities is reached (for physically 
relevant models it must be always the first!).
 
Constraints which ``commute'' with all other constraints\footnote{With 
``commute'' we mean the vanishing of the corresponding Poisson brackets.} are
called {\em \Ix{first class constraints}}. The rest is called {\em second
class constraints}. 

If the constraint algorithm never bifurcates in the sense of Henneaux and 
Teitelboim one can eliminate all second class constraints by a redefinition of
the Poisson bracket to the {\em Dirac bracket} which contains the second 
class constraints\footnote{Note: As long as we do not treat fermions in our 
model we will only get first class constraints.} \cite{het92}.

Hence we suppose that all second class constraints have been 
eliminated and proceed to the treatment of first class constraints.

For later convenience we define
\begin{eqnarray}
\left\{ \Phi_i, \Phi_j \right\} &=& C_{ij}{}^k \Phi_k, \label{phiphi} \\
\left\{ H_0, \Phi_i \right\} &=& V_i{}^j \Phi_j, \label{haphi}
\end{eqnarray}
where $H_0$ is the Hamiltonian without constraint terms.

\subsection{Gauge freedom}\index{gauge freedom}

The {\em Dirac conjecture}, which states that for every independent first 
class constraint there exists an independent gauge transformation, is not true
in general \cite{het92}. However, under rather general assumptions\footnote{
In the absence of second class constraints the most important assumption is
that the matrix $V_a^{\,b}$ (cf. (\ref{haphi})) has maximal rank where the 
index $a$ runs over all primary constraints and $b$ over all secondary 
constraints.} which apply to all relevant physical models known so far, one 
can prove it. 

Thus, the number of physical degrees of freedom in the phase space is equal to 
the total number of canonical variables minus the number of second class 
constraints minus {\em twice} the number of first class constraints (one for 
the constraint itself and one for the corresponding gauge degree of freedom).

In the configuration space we have only half the number of physical degrees of
freedom and therefore also half the number of gauge degrees of freedom as 
compared to the Hamiltonian formalism. Thus, in phase space the number of 
gauge degrees of freedom doubles as well as the number of physical degrees of
freedom. 

The first class constraints are the generators of 
{\em \Ix{gauge transformations}}
\begin{equation}
\delta_{\varepsilon} F(q_i, p^i) = \varepsilon^a \left\{ F, \Phi_a \right\}, 
\end{equation}
and it can be shown \cite{het92} that the action is invariant if we transform
simultaneously the Lagrange multipliers according to 
\begin{equation}
\delta_{\varepsilon} \lambda^a = \dot{\varepsilon}^a + \lambda^c \varepsilon^b
C_{bc}{}^a - \varepsilon^b V_b{}^a .
\end{equation}
Note that there always exist additionally {\em \Ix{trivial gauge 
transformations}} generated by the \ac{EOM}, which are disregarded in theories 
without non-trivial gauge symmetries or ``ordinary'' gauge theories like 
Yang-Mills theories. However, in order to make time parametrization invariance
manifest in the Hamiltonian formalism of a theory which possesses this 
symmetry (like all general covariant theories) one must use a certain 
combination of the \ac{EOM}  gauge symmetry generators with the non-trivial gauge 
symmetry generators \cite{het92}. 

A {\em gauge fixing condition} is an equation which reduces the number of
free parameters in the Hamiltonian (i.e. it fixes the Lagrange multipliers). A 
gauge is fixed completely if no arbitrary multipliers are left in the
Hamiltonian.

There are two classes of gauges, namely \Ix{canonical gauges} defined by
\begin{equation}
C_i (q_j, p^k) = 0 \label{cangauge},
\end{equation}
and \Ix{non-canonical gauges} which are defined by
\begin{equation}
N_i \left(q_j, p^k, \lambda^n, \dot{\lambda}^m, \ddot{\lambda}^l, ...\right) 
= 0, \label{noncangauge}
\end{equation} 
where $\lambda^n$ are the Lagrange multipliers.

Relevant for us is the (linear) \Ix{multiplier gauge}
\begin{equation}
\lambda^i + M^i (q_j, p^k) = 0,
\end{equation}
with the easiest choice being $M^i = 0$.

There are two restrictions on the canonical gauge conditions (\ref{cangauge}): 
They must fix the gauge completely\footnote{All Lagrange
multipliers and all gauge parameters (up to a (space-dependent) constant) must 
be fixed by the gauge conditions.} and the gauge must be accessible.

But for practical reasons there is an important further ``constraint'' on all 
gauge conditions: They must lead to an adequate action. That is
why finding a gauge is a technicality, but finding a good one is an art.

\subsection{\ac{BRST}-construction}

The \ac{BRST}-construction is a purely classical task and thus could have been 
invented long before quantum mechanics. However, only in the context of 
quantization of constrained systems (e.g. using non-canonical gauges) this 
construction is very useful, which is the reason why it has been developed 
there \cite{brs75, tyu75}.

Mathematically, the {\em \Ix{\ac{BRST}-differential}} $s$ is the sum of the 
(Koszul-Tate) differential $\de$, a differential $d$ modulo $\de$ and some 
``higher order'' derivations $s^{(i)}$ with higher resolution 
degrees\footnote{For the definitions used here cf. e.g. \cite{het92}.}. It has
two remarkable properties:
\blist
\item It is nilpotent\footnote{Since it is a differential, this statement seems
trivial. The non-trivial part is actually that it really {\em is} a 
differential, and not just a derivation.}: $s^2 = 0$
\item The $k$-th cohomology class of this differential $s$ is equivalent to
the $k$-th cohomology class of $d$ computed in $H_0 (\de)$, i.e. we have the
equations
\eq{
dx = \de y, \hspace{1cm} x \sim x + dz + \de z',
}{coho1} 
with
\eq{
r(x) = 0 = r(z), \hspace{1cm} r(y) = 1 =r(z'),
}{coho2}
where $r(...)$ is the resolution degree defining a $\mathbb{Z}$-grading of the
differential algebra.
\elist

The definition of $s$ is by no means unique: There is a huge freedom in its 
construction. One part of this freedom can be used to introduce a symplectic
structure in the extended space of the original (``physical'') variables and of
the new generators associated with $\de$ and $d$. In terms of this symplectic
structure the {\em \ac{BRST} transformation} is a canonical transformation, i.e.
\eq{
sx = \left\{ x, \Om \right\} .
}{brstcanon}
Once this is required, the {\em \Ix{\ac{BRST}-charge}} $\Om$ is unique up to 
canonical transformations in the ``minimal'' sector of physical variables. 
The nilpotency of $s$ together with the Jacobi-identity for the 
Poisson-(anti)bracket implies a cornerstone of \ac{BRST} theory
\eq{
\frac{1}{2}\left\{ \Om, \Om \right\} = \Om^2 = 0 .
}{brstcharge}
The differential $\de$ yields a Koszul-Tate resolution of the algebra 
$C^{\infty}\left(\Si\right)$ of smooth functions on the surface of constraints
denoted by $\Si$, while the derivation $d$ is s.t. $d^2$ vanishes identically 
only on $\Si$ and up to $\de$-exact terms in the total extended space.

Physically, the derivation $d$ is the longitudinal derivative on the surface of
constraints\footnote{When restricted to this surface it is upgraded to a 
differential.} mapping functions vanishing weakly to other functions of this 
type. Thus it can be interpreted as the generator of gauge orbits. The
differential $\de$ has also a physical interpretation: Quantities, that are
$\de$-closed but not $\de$-exact depend only on ``physical'' variables, while
$\de$-exact ones correspond to (linear combinations of) Lagrange multipliers.
$\de$-non-closed functions depend on ghosts. Functions, that coincide on  
$\Si$ are identified, since they have the same physical content.

The \ac{BRST}-charge can be constructed using homological perturbation theory
\eq{
\Om = c^i \Phi_i + \sum_{p \geq 1} \Om^{(p)}, \hspace{1cm} \text{antigh} 
\Om^{(p)} = p, 
}{brsthomper}
with $c^i$ being the {\em \Ix{ghosts}} possessing a pure ghost number of one. 
In the abelian case only the first term exists; in the Yang-Mills case only 
one further non-trivial term is present, leading to a \ac{BRST}-charge 
containing 
the structure constants of the constraint algebra\footnote{If there existed 
constraints with non-zero $\mathbb{Z}_2$-grading (``fermionic constraints'') 
then the second term in (\ref{brstYM}) would enter with a negative sign.}
\eq{
\Om = c^i \Phi_i + \frac{1}{2}c^jc^k C_{jk}{}^i p_i^c.
}{brstYM} 
The $p^c_i$ are the ghost-momenta, having an antighost number of one. Note
that the upper label $c$ is not a counting index but just a tag in order to
visualize its relation to the canonically conjugate $c^i$.

In general, also higher order structure functions are present in the 
\ac{BRST}-charge (the rank may even be infinite). However, in first order 
gravity it is always possible to construct the \ac{BRST}-charge s.t. it 
contains only the first order structure functions resembling the Yang-Mills 
case (see {\app appendix E}).

The total ghost number (or just ``the'' ghost number) is defined as the 
difference of the pure ghost number and the antighost number. The 
\ac{BRST}-charge has {\em per constructionem} always a ghost number of one.

Once the \ac{BRST}-charge has been constructed one can always find a 
\ac{BRST}-invariant extension of the Hamiltonian fulfilling
\eq{
\left\{ H_{\text{BRST}}, \Om \right\} = 0.
}{brsthamil}
Since $H_{\text{BRST}}$ is \ac{BRST}-closed, one can always add 
\ac{BRST}-exact terms without changing the dynamics of \ac{BRST}-invariant 
functions at the cohomological level. The redefinition
\eq{
H_{\text{BRST}} \to H_{\text{BRST}} + \left\{ \Psi, \Om \right\},
}{brstgf}
with some {\em \Ix{gauge fixing fermion}} $\Psi$ amounts to a different choice
of gauge. For consistency, $\Psi$ must be fermionic  and has a ghost
number of minus one.

\subsection{Quantization}\index{quantization}

For transparency, we undo the natural choice $\hbar = 1$ in this subsection.

We are going to use path integral quantization following the philosophy of
Henneaux and Teitelboim \cite{het92}: ``We will not try to give a rigorous
presentation of the path integral. Many formal manipulations will thus be
allowed without attempting to provide a mathematical justification. Experience
shows that when dealing with the path integral, it is best not to try to be
rigorous too early. This has, of course, its dangers.''

One of these dangers is the divergence of the path integral in the presence of
gauge degrees of freedom if one uses a naive definition of the path integral
as a ``sum over all paths''. There exist two ways of treating constrained
systems in this context: Either one imposes some canonical gauge working with
a reduced phase space path integral taking into account only physical degrees
of freedom, or one introduces a \ac{BRST}-extended gauge fixed action (which 
in the Yang-Mills case leads to the well-known Faddeev-Popov representation of 
the path integral due to the appearance of bilinear ghost terms in the action
\cite{fap67}) summing over {\em all} degrees of freedom consistent with the
boundary conditions. Since the first approach leads to a rather complicated
non-linear representation of first order gravity we prefer the latter. The 
second approach has also the advantage of allowing non-canonical gauges 
(e.g. multiplier gauges) provided that one treats the boundary conditions
properly.

The definition of an ordinary (unconstrained) \Ix{path integral} for $n$ 
physical degrees of freedom has been initiated by Dirac \cite{dir33} and 
developed by Feynman using Huyghens principle \cite{fey48, fey51}:
\eq{
Z\left(q_f,t_f;q_i,t_i\right) = \lim_{N\to\infty}\int\left(\prod_{t=1}^{\infty}
\frac{dq_tdp_t}{\left(2\pi\hbar\right)^n} \right)\frac{dp_i}{\left(2\pi\hbar
\right)^n} \exp \left(\frac{i}{\hbar}S^{\eps}\right),
}{pidef}
with
\eq{
S^{\eps} = \sum_{t=0}^{N} \left[ p_t\left(q_{t+1}-q_t\right) - \eps H 
\left(p_t, q_t\right) \right], \hspace{1cm} 
\eps = \frac{\left(t_f-t_i\right)}{N+1}, 
}{sdef}
and $p_0 = p_i, q_0 = q_i, q_{N+1} = q_f$.

Taking the limit in (\ref{pidef}) yields the path integral which is formally
written as
\eq{
Z \left( q_f, t_f; q_i, t_i \right) = \int \left({\cal D}q\right)\left({\cal D}
p\right) \exp \left(\frac{i}{\hbar}S\left[q(t), p(t)\right]\right),
}{piformdef}
with
\eq{
S\left[q(t), p(t)\right] = \int\limits_{t_i}^{t_f} \left(p\dot{q}-H\right)dt, 
\hspace{1cm} q(t_f) = q_f, q(t_i) = q_i.
}{sformdef}

Note that the definition (\ref{pidef}) yields a $p-q$ ordering of the 
corresponding operators. Alternatively, we could have used a definition leading
to a $q-p$ ordering or a Weyl ordering. Therefore the ``ordering problem'' of
canonical operator quantization sneaks into the path integral formalism via
the definition (\ref{pidef}), although no intrinsic operators are present.
For the rest of this work -- whenever it is important -- we will assume 
Weyl-ordering, because it is the most symmetric one.

As mentioned above, (\ref{pidef}) makes only sense if no constraints are 
present, which is why the reduced phase space quantization was historically
the first one in the context of path integrals. However, equipped with the
powerful tools of \ac{BRST}-quantization it is possible to circumvent this 
restriction and -- on the contrary -- use {\em extended} phase space 
quantization instead. The \ac{BRST} extended path integral in the Schr\"odinger 
representation reads
\eq{
Z \left(q, q'; t_2-t_1\right) = \int \left({\cal D}z\right) \left({\cal D}p_z
\right)\exp \left(\frac{i}{\hbar}S \left[z(t)\right]\right),
}{brstpi}
with 
\eq{
S \left[z(t)\right] = \int \left( p_z\dot{z} - H - \left\{\Psi, \Om\right\}
\right) dt,
}{brsts}
where the sum extends over all paths obeying the boundary conditions
\seq{3cm}{
&& q(t_2) = q', \\ 
&& q(t_1) = q,
}{3cm}{
&& \tilde{z} (t_2) = 0, \\
&& \tilde{z} (t_1) = 0,
}{brstboundary}
with $\tilde{z}$ containing the ghosts, the anti-ghosts and the 
Nakanishi-Lautrup fields\footnote{This is a synonym for ``canonical momentum 
of the Lagrange multipliers'' often used in \ac{QFT}.}, but not the 
ghost-momenta, the anti-ghost momenta or the Lagrange-multipliers.

In a multiplier gauge (which is {\em per definitionem} not a canonical gauge)
the boundary conditions (\ref{brstboundary}) have to be modified:
\seq{3cm}{
&& q(t_2) = q', \\
&& q(t_1) = q,
}{3cm}{
&& \tilde{z} (t_2)   = 0, \\
&& \tilde{p}_z (t_1) = 0,
}{multiboundary}
with $\tilde{z}$ defined as above and $\tilde{p}_z$ defined by the 
corresponding canonical conjugate quantities excluded above \cite{het92}. 

An important formula for our quantization procedure is given by the formal path
integral of a linear functional $L(\phi,\chi) = \int \phi M \chi$
\eq{
\int \left({\cal D}\phi\right)\left({\cal D}\chi\right)\exp i \int \phi M \chi 
= \int \left({\cal D}\chi\right) \de\left(M \chi\right) = \left( \det M 
\right)^{- 1}.
}{linearfunctional}
We have assumed that both fields are bosonic. In the fermionic case the 
well-known Faddeev-Popov result $\det M$ is obtained \cite{fap67}.

\clearplaindoublepage

\chapter{Spherical reduction}\index{spherical reduction}

\section{Introduction}

Starting with a $d$-dimensional Pseudo-Riemannian manifold $M$ with 
Lo\-rentz\-ian signature $(+, -, -, ..., -)$ and total spherical symmetry the 
coordinates  describing the manifold can be separated in a {\twod} 
Lorentzian part spanning the manifold $L$ and a $(d-2)$-dimensional Riemannian 
angular part constituting an $S^{(d-2)}$. Due to the spherical symmetry $L$ 
should 
{\em per definitionem} contain the whole physical content. The fact 
that $L$ has originally been embedded in $M$ leads to an extrinsic curvature 
term on $L$ that shall be computed here.

$M$ has a generalized spherically symmetric metric
\begin{equation}
ds^{2}=\sum_{a,b = 0,1}g_{ab}dx^{a}dx^{b}-\Phi^{2}\left(x_0,x_1\right) 
\sum_{i,j = 2}^{d-1} g_{ij}dx^{i}dx^{j}.
\end{equation}
For later convenience we have introduced the {\em \Ix{dilaton field}} $\Phi^2$ 
that reduces in Schwarz\-schild-like gauges to $\Phi = r$ in 4 
dimensions\footnote{Note that $\Phi^2$ corresponds to what we later will 
denote by $X$.}.
 
In the vielbein formalism we can write the line element
\begin{equation}
ds^{2}=\sum_{a,b = 0,1}\eta_{ab}e^{a}\otimes e^{b}-\sum_{i,j = 2}^{d-1}
\delta_{ij}e^{i}\otimes e^{j},
\end{equation}
and the metric $g=\eta_{uv}e^{u}\otimes e^{v}$. 

Objects with indices belonging to $L$\ or $S$ are marked by bars on top 
(e.g. $\bar{e}^{a}$ is a 1-form on $L$) or by an index $L, S$. All other 
quantities belong to $M$.

\section{Calculation}

Writing down the line-elements $ds^2_L = \eta_{ab}\bar{e}^a\otimes\bar{e}^b$, 
$ds^2_S = \de_{ij}\bar{e}^i\otimes\bar{e}^j$ with 
$\eta_{ab}=diag\left(1,-1\right)$ and comparing with $ds^2$ we find the 
transformations: 
\newline
\parbox{4cm}{\begin{eqnarray*}
&& e^{a} = \bar{e}^{a}, \\
&& e^{i} = \Phi \bar{e}^{i}, 
\end{eqnarray*}} \hfill
\parbox{4cm}{\begin{eqnarray*}
&& E_{a} = \bar{E}_{a}, \\
&& E_{i} = \Phi^{-1} \bar{E}_{i}.
\end{eqnarray*}} \hfill
\parbox{1cm}{\begin{equation} \label{vieltrafo} \end{equation}} \hfill
\newline
By demanding metric compatibility and torsionlessness\footnote{This is not 
justified {\em a priori}, but a nice shortcut and consistent {\em a 
posteriori}.} for the connection 1-forms on $M, L$ and $S$ we obtain
\begin{equation}
\omega^a{}_b = \bar{\omega}^a{}_b,\hspace{0.5cm} 
\omega^i{}_j = \bar{\omega}^i{}_j,\hspace{0.5cm} 
\omega^i{}_a = \left( \bar{E}_a \Phi \right) \bar{e}^i,\hspace{0.5cm}
\omega^a{}_i = \left( \bar{E}^a \Phi \right) \bar{e}_i,
\end{equation}
using the relations (\ref{vieltrafo}).

Putting our connection into Cartan's structure equation (\ref{cartan2}) yields 
the curvature 2-form on $M$:
\begin{eqnarray}
&& R^a{}_b = \bar{R}^a{}_b, \\
&& R^i{}_j = \bar{R}^i{}_j+\left( \bar{E}_{c}\Phi \right) 
\left( \bar{E}^{c}\Phi \right) \bar{e}^{i}\bar{e}_{j}, \\
&& R^a{}_i = \left( \bar{E}_{b}\bar{E}^{a}\Phi \right) \bar{e}^{b}
\bar{e}_{i}+\left(\bar{E}^{b}\Phi \right) \bar{\omega}_{\,\,\,b}^{a}
\bar{e}_{i}, \\
&& R^i{}_a = \left( \bar{E}_{b}\bar{E}_{a}\Phi \right) \bar{e}^{b}
\bar{e}^{i}-\left(\bar{E}_{b}\Phi \right) \bar{\omega}^b{}_a \bar{e}^{i},
\end{eqnarray}
where $\bar{R^a{}_b}$ and $\bar{R}^i{}_j$ are the curvature
two forms on $L$ and $S$, respectively.

Contracting the vector indices with the 2-form indices yields after a lengthy
but straightforward calculation the curvature scalar
\meq{
R = \bar{R}^{L}-\frac{\bar{R}^{S}}{\Phi ^{2}}-\frac{\left( d-2\right)
\left( d-3\right) }{\Phi ^{2}}\left( \bar{E}_{b}\Phi \right) \left( \bar{E}%
^{b}\Phi \right) \\
- 2\left( \frac{d-2}{\Phi }\right) \left[ \left( \bar{E}_{b}%
\bar{E}^{b}\Phi \right) +\left( \bar{E}^{b}\Phi \right) \bar{\omega}%
^a{}_b \left( \bar{E}_{a}\right) \right],
}{redcurva}
with $\bar{R}^{L}$ and $\bar{R}^{S}$ being the scalar curvatures on $L$ and
$S$, respectively.

Knowing the Riemann tensor of the sphere $\bar{R}^i{}_j=\bar{e}%
^{i}\wedge \bar{e}_{j}$ we can insert $\bar{R}^{S}=\left( d-2\right) \left(
d-3\right) \,$. Since $\Phi$ is a scalar field we can replace $\bar{E}_{b}$
acting directly on $\Phi$ by a covariant derivative $\bar{\nabla}_{b}$. The 
last term of (\ref{redcurva}) can consistently be interpreted as the
Laplacian acting on $\Phi$ -- so after the smoke clears we find\footnote{We 
omit from now on the bars as no confusion may arise anymore because no Cartan
variables will be present in the following formulae.}

\begin{equation}
R = R^L-\frac{\left( d-2\right) \left( d-3\right) }
{\Phi ^{2}} \left[ 1+\left( \nabla _{b}\Phi \right) \left( \nabla ^{b}\Phi 
\right) \right] -2\left( \frac{d-2}{\Phi }\right) \left( \bigtriangleup \Phi 
\right).
\label{redcurv}
\end{equation}
This equation is the starting point of spherically reduced gravity formulated
by a 2d effective action\footnote{Note that for the ``East coast convention''
the first part of the second term of (\ref{redcurv}) changes sign explicitly.
Other terms -- like $R^{L}$ -- change sign only implicitly due to 
contraction with a metric of opposite signature.}. For consistency one can 
check the vanishing of the curvature scalar for flat or Schwarzschild metric.

\section{Dilaton Lagrangian}\index{dilaton Lagrangian}

\subsection{General case}

In the previous section we have calculated the curvature scalar for the
$d$-dimensional original space $M$. The resulting terms can be assigned to the
submanifolds $L$ and $S$ except for the additional terms 
\begin{equation}
R^{add}=-\frac{\left( d-2\right) \left( d-3\right) }{\Phi ^{2}}\left( \nabla
_{b}\Phi \right) \left( \nabla ^{b}\Phi \right) -2\left( \frac{d-2}{\Phi }%
\right) \left( \bigtriangleup \Phi \right).
\end{equation}
which can be interpreted as the scalar curvature caused by the embedding of $L$
into $M$. The curvature in a gravitational 
field theory appears in the Lagrangian leading to the 
Einstein Hilbert action:
\begin{equation}
L_{EH}=\int d^{d}x\sqrt{-g^{(d)}}R.
\end{equation}

The determinant simplifies to
\begin{equation}
g^{(d)}=\det \left( g^{(d)}_{\mu \nu }\right) =\det \left( g_{\alpha \beta} 
\right) \cdot \Phi^{(2d-4)} \cdot f(angles) .
\end{equation} 
where $g_{\alpha \beta}$ is the 2d metric ``living'' on $L$.
The angular integration gives just the surface of $S^{(d-2)}$ denoted by
$S$ (in the important case $d=4$ this reduces to $S=4 \pi$):
\meq{
L_{\text{dil}} = S \int d^{2}x\sqrt{-g} \left[ \Phi^{(d-2)} R^{L}-(d-2)
(d-3)\Phi^{(d-4)} \right. \\
\left. \cdot \left( 1+\left( \nabla_b\Phi \right) \left(\nabla^b\Phi\right)
\right)-2(d-2)\Phi^{(d-3)} \left( \bigtriangleup \Phi \right)\right].
}{whatever1}

The last term can be converted by a partial integration dropping the boundary
term $-2 (d-2)(d-3) S \int d^2x \nabla_b \left[ \sqrt{-g} \Phi \nabla^b 
\Phi \right]$. Using also metric compatibility results in
\meq{
L_{\text{dil}} = S \int d^{2}x \sqrt{-g} \left[\Phi^{(d-2)} R^{L} + (d-2)(d-3)
\Phi^{(d-4)}\left(\nabla \Phi \right)^2 \right. \\ 
\left. - (d-2)(d-3) \Phi^{(d-4)} \right], 
}{whatever2}
in agreement with \cite{tih84, haj84, lau96}. 

It shall also be mentioned that it is common to introduce a new dilaton 
field\footnote{Note that our formulae make only sense for $\Phi \neq 0$. Thus
$\Phi$ must be strictly positive or negative in the whole manifold motivating
the redefinition.} $\tilde{\Phi}$ by
\begin{equation}
\Phi = \frac{(d-2)}{\lambda}e^{-\tilde{\Phi}},
\label{dilredef1} 
\end{equation}
where $\lambda$ is a constant of mass dimension 1. With this substitution it 
follows
\meq{
L_{\text{dil}} = S \frac{(d-2)^{(d-2)}}{\lambda^{(d-2)}} \int d^{2}x 
\sqrt{-g} e^{-(d-2)\tilde{\Phi}} {\Bigg [} R^{L} + (d-2)(d-3) \\
\left( \nabla_{b}\tilde{\Phi} \right) \left( \nabla ^{b}\tilde{\Phi} 
\right) -\frac{(d-3)}{(d-2)} \lambda^2 e^{2\tilde{\Phi}} {\Bigg ]}.
}{whatever3}
This is the standard form of dilaton gravity -- cf. e.g. \cite{kuv99}. It has 
the advantage that the original dilaton field automatically is strictly 
positive (apart from the limiting case where it vanishes) in the whole range 
of $\tilde{\Phi}$.

By further transformations one can alternatively introduce the field
\begin{equation}
X=\exp{\left(-(d-2)\tilde{\Phi}\right)}
\label{dilredef2}
\end{equation}
leading to the well-known dilaton action
\begin{equation}
L_{\text{dil}}= S \frac{2(d-2)^{(d-2)}}{\lambda^{(d-2)}} \int d^{2}x
\sqrt{-g} \left[ X \frac{R^{L}}{2} + V(X) - \frac{1}{2}(\nabla X)^2 U(X) 
\right]. \label{familiar}
\end{equation}
with
\begin{equation}
U(X) = - \frac{(d-3)}{(d-2)X} , \hspace{0.5cm} V(X) = - \frac{(d-3)}{2(d-2)} 
\lambda^2 X^{\frac{(d-4)}{(d-2)}}, 
\label{potentials}
\end{equation}
showing resemblance with (\ref{1action}). In our redefinitions 
(\ref{dilredef1}-\ref{dilredef2}) the dimension dependent constants have been
chosen s.t. the limit $\lim_{d \to \infty}$ promotes our effective action of 
\ac{SRG} to the \ac{CGHS} action \cite{cgh92}.

The ``physical'' gauge for the Dilaton field would be $\Phi = r$, i.e.
\eq{
X =  e^{-(d-2)\tilde{\Phi}} = \left(\frac{\la r}{d-2}\right)^{(d-2)},
}{dilgauge}
since $X$ is related to $\Phi$ via (\ref{dilredef1}) and (\ref{dilredef2}).

\subsection[Special case $d=4$]{Special case $\boldsymbol{d=4}$}

Anthropocentrically speaking, $d=4$ is the most important special case. 
Therefore we list for convenience the most important formulae of this section
for this choice of dimension, namely the Einstein Hilbert action with the
dilaton field expressed in terms of $\Phi$
\begin{equation}
L_{dil}= 4 \pi \int d^{2}x \sqrt{-g} \left[ \Phi^2R^{L} + 2\left(\nabla 
\Phi \right)^2 - 2 \right] , \label{4Phi}
\end{equation}
in terms of $\tilde{\Phi}$
\begin{equation}
L_{\text{dil}}= 16 \pi \lambda^{-2} \int d^{2}x \sqrt{-g} e^{-2\tilde{\Phi}} 
\left[ R^{L} + 2 \left( \nabla_{b}\tilde{\Phi} \right) \left( \nabla^b
\tilde{\Phi} \right) - \frac{\lambda^{2}}{2} e^{2\tilde{\Phi}} \right] , 
\label{4tildephi}
\end{equation}
and in terms of $X$
\begin{equation}
L_{\text{dil}}= -32 \pi \lambda^{-2} \int d^2x\sqrt{-g} \left[ X\left(-\frac
{R^L}{2}\right) - \frac{1}{2}\frac{(\nabla X)^2}{2X} + \frac{\lambda^2}{4} 
\right]. \label{4X}
\end{equation}

In Schwarzschild-like coordinates $\Phi^2$ equals to $r^2$ establishing
\begin{equation}
X = \frac{\lambda^2}{4} r^2 . \label{dilatongauge}
\end{equation}

\subsection{Matter part}\label{matter part}\index{matter part}

The prefactors of the action are important if we want to adjust the coupling 
of the matter Lagrangian properly: Suppose that in the original theory the 
matter Lagrangian respects spherical symmetry and is of the generic form
\eq{
L_m^{(d)} = \ka \int d^dx \sqrt{-g^{(d)}} {\cal L}^{(m)} .
}{matter}
Then the spherically reduced form is given by
\eq{
L_m = \ka S \frac{(d-2)^{(d-2)}}{\la^{(d-2)}} \int d^2x \sqrt{-g} X
{\cal L}^{(m)} .
}{matterreduced}
Comparison with (\ref{familiar}) yields a relative factor of
\eq{
\ka_{eff} := -\frac{\ka}{2},
}{kappaeff}
which has to be used as an effective coupling constant in \ac{SRG}. 

For the special case of the spherically symmetric \ac{EMKG} in $d=4$ we 
have\footnote{$\epsilon^{(4)}$ is the volume form in $d=4$.}
\eq{
L_{EMKG}^{(m)} = \int_{M_4} \left(dS \wedge * dS - \epsilon^{(4)} f(S) 
\right) .
}{EMKGmatter}
Therefore, our effective coupling constant to be used in the calculations 
(remember that $8\pi G_N = c = 1$) is given by
\eq{
\ka_{eff}^{(EMKG)} = -\frac{1}{2}.
}{kappaeff4}
Thus, the proper spherically reduced \ac{EMKG} Lagrangian\index{spherically 
reduced Lagrangian} in $d=2$ presented in first order form is given by
\meq{
L_{EMKG} = \int \left[ X^+ D \wedge e^- + X^- D \wedge e^+ + X d \wedge \omega 
- {\cal V} (X^+X^-, X) e^- \wedge e^+ \right. \\ 
\left.  + F(X) \left(dS \wedge * dS - e^- \wedge e^+ f(S) \right) \right] . 
}{EMKG}
The notation for the geometric part is explained in {\app appendix D}.
The coupling function must be chosen as
\eq{
F(X) = - \frac{X}{2}.
}{F(X)}

\clearplaindoublepage

\chapter{\Ix{Lagrangian formalism}}\label{Cartan formulation}

\section[Introduction and relation to Poisson-$\sigma$ models]{Introduction 
and relation to Poisson-$\boldsymbol{\sigma}$ models}\label{poisson sigma}
\index{Poisson-$\si$ models}

It has been shown that any Poisson structure on any finite dimensional manifold
naturally induces a two dimensional topological field theory \cite{sst94a}.
For example, pure gravity in two dimensions is a special case of a 
Poisson-$\sigma$ model since it contains only a Gauss-Bonnet term in the 
action. But also \ac{GDT} can be described with this formalism.

The action of such models is given by
\begin{equation}
L = \int_{M_3} F_{,i} \wedge DX^i,
\end{equation}
with
\begin{eqnarray}
&& D = d + A, \hspace{0.5cm} D \wedge D = F_{,i}P^{ij} \partial_j, \\
&& F = \left(A_{\nu , \mu}+\frac{1}{2}A_{\mu , i}A_{\nu , j} P^{ij} + 
C_{\mu \nu} \right) dx^{\mu} \wedge dx^{\nu},
\end{eqnarray}
where $X^i$ are the coordinates on the Poisson manifold, while $x^{\mu}$ are 
coordinates on $M_3$. Note that the term $C^{\mu \nu}dx^{\mu}dx^{\nu}$ must be 
set to 0 for gravity theory since it violates diffeomorphism invariance on 
the world sheet manifold $\partial M_3$.

As the integrand is exact this produces a {\twod} action
\begin{equation}
L = \int_{\partial M_3} \left( A_i \wedge dX^i + \frac{1}{2}P^{ij}A_i \wedge 
A_j \right),
\end{equation}
with rather simple \ac{EOM}
\eqa{
dX^i + P^{ij} A_j = 0 \label{PSM1}, \\ 
d\wedge A_i + \frac{1}{2} \frac{\partial P^{jk}}{\partial X^i} A_j \wedge A_k
= 0.
}{PSM2}

With the identifications
\seq{4cm}{
&& A_1 = \om, \\
&& A_2 = e^-, \\
&& A_3 = e^+,
}{4cm}{
&& X^1 = X, \\
&& X^2 = X^+, \\
&& X^3 = X^-, 
}{identifications}
and the following choice of the Poisson tensor
\eq{
P^{ij} = \left( \begin{array}{ccc} 
0 & -X^+ & X^- \\
X^+ & 0 & -{\cal V}(X^+X^-,X) \\
-X^-& {\cal V}(X^+X^-,X) & 0 
\end{array} \right),
}{poissontensor}
we finally obtain (after dropping a boundary term) the \Ix{first order action}
\begin{equation}
L_1 = \int_{M_2} \left[ X^+ D \wedge e^- + X^- D \wedge e^+ + X d \wedge 
\omega - {\cal V}(X^+X^-, X) e^- \wedge e^+ \right], \label{1st}
\end{equation}
with the gauge-covariant derivative
\begin{equation}
D \wedge e^a := d \wedge e^a + \varepsilon^a{}_b \omega \wedge e^b .
\end{equation}

A large class of 2d models of generalized gravity -- among them 
$d>3$-dimensional \ac{SRG}, the \ac{KV} model, the \ac{CGHS} model and other string inspired 
dilaton theories -- can be written in this form by properly adjusting 
${\cal V}(X^+X^-, X)$ \cite{kst96a, kst96b}. For convenience we list below 
some of these ``potentials'' using the special, but still general enough 
ansatz 
\eq{
{\cal V}(X^+X^-,X) = U(X)X^+X^- + V(X).
}{specialpot}

\begin{figure}
\setlength{\extrarowheight}{12pt}
\begin{center}
\begin{tabular}{|>{\large}l||>{\large $}c<{$}|>{\large $}c<{$}|} \hline
Model & V(X) & U(X) \\ \hline \hline
\acs{SRG} ($d>3$) & -\frac{(d-3)}{2(d-2)} \lambda^2 X^{\frac{(d-4)}{(d-2)}} & 
-\frac{(d-3)}{(d-2)X} \\ 
\acs{KV} & -\frac{\beta}{2}X^2-\Lambda & -\alpha \\ 
\acs{JT} & -\La X  & 0 \\ 
\acs{EBH} & - \frac{1}{\sqrt{X}} & 0 \\
\acs{CGHS} & -\frac{\lambda^2}{2}X & -\frac{1}{X} \\ 
$R^2$ gravity & -X^2+\La  & 0 \\ \hline
\end{tabular} \label{table1}
\end{center}
\setlength{\extrarowheight}{0pt}\index{twodimensional models}
\caption[Important dilaton theories]{A selection of important dilaton theories}
\end{figure}

Unfortunately, the inclusion of matter is problematic in this formalism: Since
Poisson-$\si$ models are tailor-made for topological theories, the inclusion 
of propagating modes is difficult. Thus, only chiral fermions 
\cite{kum92} or \mbox{(anti-)} selfdual scalars \cite{pes98} can be described
easily. This is the main reason, why I am not using this model very 
intensively in my thesis, despite of its elegance.

\section{``Equivalence'' of first and second order formulation}

First and second order formalism are strictly speaking not 
equivalent because they differ in the number of degrees of freedom. However,
it can be shown (and this is what we are going to perform in this section) that
the \Ix{first order action} (which contains additional fields as compared to 
the \Ix{second order action}) can be integrated out to a ``simpler'' action 
which is (up to surface terms) equivalent to the latter.

We start with the standard form of the {\em \Ix{first order Lagrangian}}
\eq{
{\cal{L}}_{FO} = X^+ D\wedge e^- + X^- D\wedge e^+ + 
X d\wedge \omega - {\cal V}\left(X^+X^-,X\right) e^- \wedge e^+,
}{1action}
where $X, X^{\pm}$ denote the \Ix{target space coordinates}, $\omega, e^{\pm}$ 
the \Ix{Cartan variables} and $d$ is the ``gauge''-covariant derivative, which 
in light cone components decomposes into
\begin{equation}
D \wedge e^{\pm} = d \wedge e^{\pm} \pm \omega \wedge e^{\pm} .
\end{equation}
From now on we specialize the form of the ``potential'' 
${\cal V}(X^+X^-, X) = V(X)+X^+X^-U(X)$.

Integration over the \Ix{spin connection} $\omega$ yields the Lagrangian
\begin{equation}
{\cal{L}} = X^+d \wedge e^- + X^- d \wedge e^+ - e^- \wedge e^+ 
\left(V(X)+X^+X^-U(X)\right), \label{lag1}
\end{equation}
and the constraint
\begin{equation}
dX-X^+e^-+X^-e^+ = 0,
\end{equation}
which can be solved for $X^{\pm}$:
\begin{equation}
X^{\pm} = \mp E_{\mp} X \label{Xpm},
\end{equation}
where $E_{\mp}$ is the inverse vielbein. This implies $(\nabla X)^2 = - 2 
X^+X^-$.

With Cartan's structure equations (written for our special case of a 
{\twod} model without torsion in light cone components) 
\begin{equation}
\tilde{\omega} \wedge e^{\pm} = \mp d \wedge e^{\pm}, \hspace{0.5cm} *(d 
\wedge \tilde{\omega}) = - \frac{R(e)}{2}, 
\end{equation}
we can present the first two expressions of the Lagrangian (\ref{lag1}) in 
terms of the curvature using the Levi-Civit\'a connection $\tilde{\om}$:
\begin{equation}
{\cal{L}}_{dil} = X\left(-\frac{R^S}{2}\right)e^- \wedge e^+ - e^- \wedge e^+ 
\left(V(X) - \frac{1}{2} \left(\nabla X\right)^2 U(X) \right), \label{lag2}
\end{equation}
where $\nabla$ denotes the ordinary covariant derivative. The last term of 
(\ref{lag2}) is a consequence of (\ref{Xpm}) and the property of the inverse 
vielbein $E_a f = \nabla_a f$. 

The action for this Lagrangian is given by
\begin{equation}
L_{dil} = c \int d^2x \sqrt{-g} \left[ X\left(-\frac{R^S}{2}\right) + 
\frac{1}{2} U(X) \left( \nabla X \right)^2 - V(X) \right], \label{2action}
\end{equation}
where $c$ is some arbitrary constant.

Note that (\ref{2action}) is already the second order Lagrangian being 
equivalent to equation (\ref{familiar}) if we substitute $X=\Phi^2$ and adjust
$U(X)$ and $V(X)$ properly for \ac{SRG}.

\section{Equations of motion}\label{equations of motion}\index{equations of 
motion}

The first order action (\ref{1action}) yields the following \ac{EOM} (in 
comparison with the simpler \ac{EOM} (\ref{PSM1},\ref{PSM2}) we have now 
additional matter 
contributions): 
\begin{eqnarray} 
\hspace{-1cm} \delta \omega &:& dX + X^-e^+ - X^+e^- = 0, \label{eom1} \\
\hspace{-1cm} \delta e^{\mp} &:& dX^{\pm} \pm \omega X^{\pm} \mp e^{\pm}
\left(V(X)+X^+X^-U(X)\right) + M^{\pm} = 0, \label{eom2} \\ 
\hspace{-1cm} \delta X &:& d \wedge \omega - e^- \wedge e^+ \left(V'(X)+X^+X^-
U'(X)\right) + \frac{\delta {\cal{L}}^{(m)}}{\delta X} = 0, \label{eom3} \\
\hspace{-1cm} \delta X^{\mp} &:& d \wedge e^{\pm} \pm \omega \wedge e^{\pm} - 
e^- \wedge e^+ X^{\pm} U(X) = 0, \label{eom4}
\end{eqnarray}
where we have supposed the presence of a matter Lagrangian ${\cal L}^{(m)}$ 
which does not couple to $X^{\pm}$ or $\om$ and used the 1-form abbreviation
\eq{
M^{\pm} = \frac{\delta {\cal{L}}^{(m)}}{\delta e^{\mp}} .
}{Mpm}
For this thesis I have assumed the following form of ${\cal L}^{(m)}$:
\eq{
{\cal L}^{(m)} = F(X) \left( dS\wedge *dS - e^-\wedge e^+f(S)\right),
}{Lm}
giving rise to a Klein-Gordon type equation: 
\eq{
\delta S : d \wedge \left( F(X) (2e^+S^- + 2e^-S^+) \right) - e^-
\wedge e^+ F(X) f'(S) = 0 .
}{eom5}
In the last equation the 0-form definition
\begin{equation}
S^{\pm} = * \left(dS \wedge e^{\pm}\right) = \frac{1}{(e)}
\varepsilon^{\alpha \beta} e^{\pm}_{\beta} 
\partial_{\alpha} S
\end{equation}
has been used which brings the matter action into the simple form
\eq{
L^{(m)}_{s} = - \int d^2x \sqrt{-g} F(X) \left( 2 S^+S^- + f(S) \right) .
}{mattersimple}

Eqs. (\ref{eom1}, \ref{eom2}) are equations of 1-forms
while the rest contains 2-forms (or -- if the hodge $*$ is used -- 0-forms).
Therefore, in components we have 10 \ac{EOM} (2+4+1+2+1).

\section{An absolute \Ix{conservation law}}\label{conservation law}

In the absence of matter the existence of the conservation law is a simple 
consequence of our choice of the Poisson tensor: It has only rank 2 and 
therefore exactly one conserved Casimir-function exists. The linear combination
\eq{
\eps_{ijk} P^{ij} (d X^k + P^{kl} A_l) = \eps_{ijk} P^{ij} d X^k = 0,
}{lincombPSM} 
multiplied by an integrating factor $I(X^+X^-, X)$ yields a closed quantity 
for the important special case (\ref{specialpot}).

Combining the \ac{EOM} for the Cartan variables in the same way, namely
\begin{equation}
X^+ \times [eq. (\ref{eom2})^-] + X^- \times [eq. (\ref{eom2})^+] + {\cal V} 
\left(X^+X^-,X\right) \times [eq. (\ref{eom1})] \label{lincombEOM}
\end{equation}
where eq. $(\ref{eom2})^{\pm}$ is the \ac{EOM} for $\de e^{\mp}$, we obtain 
with an integrating factor
\begin{equation}
I(X) = \exp\left[\int^XU(X')dX'\right], \label{intfactor}
\end{equation}
which simplifies to $I(X) = X^{-\frac{1}{2}}$ in the case of \ac{SRG}, the 
absolute conserved quantity
\begin{equation}
d\left( C^{(g)} + C^{(m)}\right) = 0, \label{conserv}
\end{equation} 
with
\begin{eqnarray}
&& C^{(g)} = I(X)X^+X^- + \int^X V(X') I(X') dX', \\
&& dC^{(m)} = I(X) \left( M^+X^- + M^-X^+ \right),
\end{eqnarray}
where $M^{\pm}$ is defined by (\ref{Mpm}) and reads explicitly
\eq{
M^{\pm} = -2 F(X) \left[S^{\pm}dS \pm e^{\pm}\left(S^+S^- + \frac{1}{2}f(S)
\right)\right].
}{Mpmexplicitly}

For the case of (in $d=4$) minimally coupled scalars we obtain
\begin{equation}
d {\cal{C}}^{(g)} + W^{(m)}= 0, \label{EMKGcons}
\end{equation}
where
\begin{eqnarray}
&& \hspace{-1.5cm} {\cal{C}}^{\left(g\right)} = \frac{X^+X^-}{\sqrt{X}} - 
\frac{\lambda^2}{2} \sqrt{X}, \\
&& \hspace{-1.5cm} W^{\left( m \right)} = \sqrt{X} \left[ dS \left(
S^-X^+ + S^+X^-\right) - dX \left( S^+S^- + \frac{1}{2} f(S) \right) \right].
\label{WEMKG} 
\end{eqnarray}
The factor $F(X) = -X/2$ is explained in {\em appendix C} on page 
\pageref{matter part}.

In the absence of matter (the 1-form $W^{\left( m \right)}$ vanishes) 
the 0-form ${\cal{C}}^{\left( g \right)} < 0$ is 
proportional to the mass of the \ac{BH}; in the presence of matter it 
becomes the so-called ``mass aspect function'' \cite{gru99, grk00}. In these
papers also the relevance of this conservation law for the initial conditions 
is emphasized and its relation to \ac{ADM}- and Bondi-mass is shown. The relation
to the \Ix{mass-aspect function} is given by
\eq{
m = -\frac{ 2 {\cal C}^{(g)}}{\la^3}.
}{massaspect}

Note that in the presence of matter the conservation law is related to a 
non-linear Noether current which up to recently was only known in its 
infinitesimal form \cite{kut99}.

\section{Conformal transformations}\label{conformal trafo}\index{conformal 
transformations}

For several reasons the behavior of (\ref{2action}) under conformal 
transformations 
\eq{
g_{\al\be} (x_{\ga}) \to \Om(x_{\ga})^{-2} g_{\al\be} (x_{\ga})
}{cotrafo}
is of interest: First of all, it is an interesting task simply because {\twod} 
models are conformally invariant, provided that no intrinsic scale exists. 
Second, for this reason they are often used in {\twod} field theories. Third, 
in \ac{GDT} it is always possible to eliminate the ``torsion term'' 
proportional to $U(X)$ but the global structure may be changed by this 
conformal transformation. We will show this property explicitly.

The curvature scalar transforms according to
\eq{
R \to R\Om^2 + 2 \frac{\nabla_{\ga} \nabla^{\ga} \Om}{\Om} - 2 
\frac{\nabla_{\ga} \Om \nabla^{\ga} \Om}{\Om^2} .
}{Rcotrafo}
Lets assume that the conformal factor depends on the dilaton field $X$. The 
choice
\eq{
\Om = \exp{\left[-\frac{1}{2} \int^X U(X')dX'\right]} = I(X)^{-\frac{1}{2}}
}{conformalfactor}
is found to cancel the torsion term. The transformed potentials read
\eq{
U(X)^{(CT)} = 0, \hspace{0.5cm} V(X)^{(CT)} = V(X) e^{Q(X)} \stackrel{CGHS}{=}
 const. 
}{trafopotential}
It is interesting to note the close relation of the conformal factor to the
integrating factor $I(X)$ of the conserved quantity. On this occasion we would 
like to note that conformal transformations on the
world-sheet metric can also be generated by {\em target space diffeomorphisms}
\cite{ert01}. Although this is an interesting result, in view of our remarks 
after \ref{table1} we will not explore this fact further. 

Now we specialize to \ac{SRG}, i.e. the \ac{SSBH}. The conformal factor has 
two singular points: $X = 0$ and $X = \infty$. 
Therefore, one cannot expect the singularity structure to be unchanged by such
a transformation. Indeed, inspection of the Killing-norm (\ref{C8}) for the
case $1/b \to 0$ yields
\eq{
K^2(X) \propto {\cal C}^{(g)} - \frac{\la^2}{2} \sqrt{X} .
}{confkilling}
The metric becomes singular at $X = \infty$ and regular at $X = 0$ -- thus 
regularity properties have swapped at the boundary due to the conformal 
transformation. The fact, that addition of a matter contribution to 
${\cal C}^{(g)}$ does not change the singularity structure (in sharp contrast
to the Schwarzschild case) is another important difference. This observation
has lead to the name \acf{EBH} for the conformally transformed 
\ac{SSBH} \cite{kumxx}. An investigation of geodesic (in-)completeness 
properties also can yield essential differences between conformally 
transformed theories \cite{kst96b}.

We conclude that conformal transformations with a conformal factor 
(\ref{conformalfactor}) can yield globally different theories. Since without
the inclusion of dynamical matter a {\twod} model is characterized by its 
global structure only, this means that the two conformally related theories are
{\em inequivalent}. Thus it is a dangerous route to use conformal 
transformations, afterwards quantize the ``simplified'' theory and finally 
claim, that one has quantized the original theory. Unfortunately, this is
frequently done in literature (cf. the discussion in \cite{ghk01,ghk00}).

\section{Static solutions for the \ac{EMKG}}\index{static solutions}
\label{static solutions}

One example where the \ac{EOM} -- a coupled system of \ac{PDE} -- yields a 
\ac{ODE} is given by the CSS solution of the \ac{EMKG} (cf. eqs. 
(\ref{C24}-\ref{C31})). Another example -- the static solutions -- have already 
been discussed in 1948 \cite{fis48} and have been rediscovered many times since 
then (cf. \cite{jnw68} for the most popular one of the older versions). In the 
first order formalism it is particularly easy to re-derive these results and 
clarify subtle issues in existing literature (e.g. an important error in the 
original work (eqs. 28-29 of \cite{fis48}) which is the result of an improper 
limit).

\subsection{The general solution}

With $x_0 = v$ and $x_1 = r$ staticity implies $\partial_0 f = 0$ and
$\partial_1 f = f'$ where $f$ is the scalar field or one of the geometric 
quantities. The ingoing \ac{SB} line element $(ds)^2 = e \left[hdv-2dr\right]
dv$ can be obtained via the gauge fixing \cite{grk00}
\eq{
e_1^+ = 0, \hspace{0.5cm} e_1^- = -1, \hspace{0.5cm} X = \frac{\la^2}{4}r^2.
}{BSgauge}
It is straightforward to solve the \ac{EOM} for the auxiliary fields and the 
connexion:
\seq{4cm}{
&& X^+ = -\frac{\la^2}{2} r, \\
&& X^- = -\frac{\la^2}{4} \frac{rh}{e},
}{4cm}{
&& \om_0 = \frac{h'}{2} + \frac{h}{2}\frac{e'}{e} + \frac{h}{r}, \\
&& \om_1 = -\frac{1}{r}-\frac{e'}{e}. 
}{xpxmom}
It is useful to introduce an auxiliary field defined by $Z = rh$, because the
KG equation and slicing condition simplify to
\eq{
S' = \frac{c_1}{rZ}, c_1 \in \mathbb{R} \hspace{0.5cm} e = Z' .
}{Zeom}
$c_1$ is an integration constant. Thus all geometric and matter quantities can
be expressed in terms of the function $Z(r)$.

The Hamiltonian constraint yields a nonlinear
second order \ac{ODE}
\eq{
Z^2Z'' = \frac{c_1^2Z'}{2r} .
}{Zeq}
This is equivalent to the equation (16) obtained in \cite{fis48}. The first 
integral of this equation yields the second integration constant which turns
out to be twice the \ac{ADM}-mass, as we will see later:
\eq{
rZ'-Z+\frac{c_1^2}{2Z} = c_2 = 2 m_{ADM} .
}{ZeqADM}
With the definitions
\seq{3cm}{
&& y = \frac{Z}{m_{ADM}}, \\
&& y_0 = p^{-1} - 1, \\
&& y_1 = p^{-1} + 1, 
}{5cm}{
&& \eta = \frac{c_1^2}{2m_{ADM}^2} \in \mathbb{R}_0^+ \label{Zeta}, \\
&& p = \frac{1}{\sqrt{1 + \eta}} \in [0,1],
}{Zdefs}
the general solution which yields asymptotically Minkowski space has a 
\linebreak[4] unique third integration constant and reads
\eq{
\left(y+y_1\right)^{1+p}\left(y-y_0\right)^{1-p} = \frac{r^2}{m_{ADM}^2}.
}{Zsolution}
For a given \ac{ADM}-mass this is a one-parameter family of solutions with 
family parameter $p$. We would like to emphasize that even in the limit 
$y \to \infty$, i.e. for large radii, and for small $p$ it is not possible to 
neglect the term $(y+y_1)^p(y-y_0)^{-p}$ and simply solve the 
hyperbola equation $(y+y_1)(y-y_0) = r^2/m_{ADM}^2$ if one is 
interested in the next to leading order in a $m_{ADM}/r$-expansion, as 
it has been done in \cite{fis48}. Below, we will treat this asymptotic region 
in more detail.

The general solution for the metric
\eq{
e = \frac{m_{ADM}}{r} \frac{(y-y_0)(y+y_1)}{y}, \hspace{0.5cm} h = 
\frac{m_{ADM}}{r} y
}{Zmetric}
has already a remarkable feature: Killing horizons ($h=0$) are singular. 
However, for $p \in [0,1)$ it is straightforward to check that no Killing 
horizons exist, implying that the singularity at the origin is a naked one.
The case $p=1$ corresponds to the Schwarzschild solution. Note that the limit
$\lim_{\eps \to 0} p = 1 - \eps$ differs from the Schwarzschild solution. Thus,
the point $p=1$ is very special in this one-parameter family. This property 
has even led to speculations that not the Schwarzschild solution but this 
limiting solution is the ``physical one'' \cite{jnw68}.

The mass aspect function is given by
\eq{
m = \frac{r}{2} \left(1-\frac{m^2_{ADM}}{r^2}y^2\left(\frac{y+y_1}{y-y_0}
\right)^p\right).
}{Zmass}

The scalar field
\eq{
S = - \frac{c_1}{2m_{ADM}} p \ln{\left(\frac{y+y_1}{y-y_0}\right)}
}{ZS}
contains another constant of integration which has been fixed such that 
$\lim_{r \to \infty} S(r) = 0$. For small radii the singular behavior 
$S \propto \ln{r}$ has improved as compared to the non-relativistic solution
$S \propto 1/r$.

\subsection[Large $r/m_{ADM}$]{Large $\boldsymbol {r/m_{ADM}}$-expansion} 

After some trivial calculations we obtain
\eqa{
&& y = \frac{r}{m_{ADM}} - 2 + \frac{\eta}{2} \frac{m_{ADM}}{r} + 
{\cal O}\left(\frac{m_{ADM}}{r}\right)^2, \\
&& e = 1 - \frac{\eta}{2}\frac{m_{ADM}^2}{r^2} + 
{\cal O}\left(\frac{m_{ADM}}{r}\right)^3, \\
&& \al^2  = 1-\frac{2m_{ADM}}{r} + {\cal O}\left(\frac{m_{ADM}}{r}\right)^2, \\
&& a^{-2} = 1-\frac{2m_{ADM}}{r} + {\cal O}\left(\frac{m_{ADM}}{r}\right)^2, \\
&& m = m_{ADM} \left(1-\frac{m_{ADM}}{r} \frac{\eta}{2}\right) + 
{\cal O}\left(\frac{m_{ADM}}{r}\right)^2 \le m_{ADM}, \\
&& S = -\frac{c_1}{r} + {\cal O}\left(\frac{m_{ADM}}{r}\right)^2.
}{Zlarger}
We have included the result for the diagonal line element 
$(ds)^2 = \al^2 (dt)^2 - a^2 (dr)^2$ in order to pinpoint that indeed the next
to leading order in this expansion gives the Schwarzschild solution, in 
contrast to what is claimed in \cite{fis48}.

\subsection[Small $r/m_{ADM}$ and non-existence of Killing horizons]
{Small $\boldsymbol{r/m_{ADM}}$-expansion and non-existence of Killing 
horizons}

Since at the origin all quantities become singular this expansion is of little
use. Therefore, we present just the result for $y$ which implies all other 
quantities:
\eq{
y = y_0 + \left(\frac{p}{2}\right)^{\frac{1+p}{1-p}}\left(\frac{r}{m_{ADM}}
\right)^{\frac{2}{1-p}} + \dots
}{Zsmallr}
Note that we have to choose $y = y_0$ rather than $y = -y_1$ because $y$
cannot be negative for the following reason: For large $r$ it is positive by
choice (this choice was triggered by the requirement of asymptotic flatness).
For reasons of continuity it it ever becomes negative somewhere, there must 
also exist a point $r \in (0,\infty)$ where it vanishes. Thus, the equation
\eq{
(y_1)^{1+p}(-y_0)^{1-p} = \frac{r^2}{m_{ADM}^2}
}{Znokill}
must have a real and positive solution for $r$. But since $y_1 \ge 2$, 
$y_0 \ge 0$ and $p \in [0,1]$ such a solution can only exist for $y_0 = 0$, 
corresponding to the Schwarzschild case $p=1$. Thus, only the trivial family 
endpoint $p=1$ allows the existence of a Killing horizon (note that in that 
case there is a subtle limit of the form $0^0$).

\subsection[Special values of $p$]{Special values of $\boldsymbol{p}$}

\subsubsection{$\boldsymbol{p = 0}$}

The unique solution
\eq{
y = \sqrt{\eta + 1 + \frac{r^2}{m_{ADM}^2}} - 1 
}{Zp0}
is useless because $\eta$ diverges like $p^{-1/2}$. For finite $r$ one can 
expand the square root and obtain some more or less sensible results for the 
geometric quantities and the scalar field, which coincide precisely with the 
small $r$-expansion results for $p=0$. But for large $r$ this expansion breaks 
down. 

\subsubsection{Small $\boldsymbol{p = \eps}$}

We get perturbatively
\eq{
y = \frac{1}{\eps} - 1 + {\cal O}\left(\eps\frac{r^2}{m_{ADM}^2}\right) .
}{Zpsmall}
This serves as another way of explaining the error in \cite{fis48}: No matter,
how small $\eps$, there exists a radius beyond which this perturbation breaks
down. I.e. for large $r$ one cannot use the small $p$ expansion any more.

\subsubsection{$\boldsymbol{p = \frac{1}{2}}$}

As an algebraically simply soluble example we treat the ``heart'' of the 
family $p = \frac{1}{2}$. Note that apart from the endpoints $p=0$ and $p=1$
this is the only point in the parameter space which allows an exact 
treatment\footnote{Only for $p \in \mathbb{Q}$ we obtain an algebraic 
equation. In this case, we have a representation $p = a/b$ with 
$a,b \in \mathbb{N}$. The highest power of $y$ appearing in this equation 
turns out to be $2b$. Only equations up to quartic order can be solved 
exactly. Therefore, we must have either $b=1$ or $b=2$. The first case leads to
the two endpoint solutions and the last one to $p=1/2$.}.

The resulting quartic equation
\eq{
y^4 + 8y^3 + 18y^2 - 27 - \frac{r^4}{m^4_{ADM}} = 0,
}{Zquartic}
together with the branch condition $y > 0$ and the reality condition 
$y \in \mathbb{R}$ establishes uniquely
\eqa{
\hspace{-1cm} y = - 2 + d_1 + d_2, \hspace{0.5cm} 
&& d_1 = \sqrt{1 + \frac{r^2}{2\sqrt{3}m_{ADM}^2}(w-w^{-1})}, \\
&& d_2 = \sqrt{2+\frac{r^2}{2\sqrt{3} m_{ADM}^2}(w^{-1}-w)+8d_1^{-1}}, 
\nonumber
}{Zp12}
with $w = \left(\frac{m_{ADM}^2}{r^2}\left(\sqrt{\frac{r^4}{m_{ADM}^4}+27}-
3\sqrt{3}\right)\right)^{1/3}$. For large radii we obtain
\eq{
y = \frac{r}{m_{ADM}} - 2 + \frac{3}{2}\frac{m_{ADM}}{r} +
{\cal O}\left(\frac{m_{ADM}}{r}\right)^2,
}{ylarger}
in accordance with our results for the large $r$ expansion (\ref{Zlarger})
 --  note that for this specific value of $p$ we have $\eta = 3$. For small 
radii we get
\eq{
y = 1 + \frac{1}{64}\frac{r^4}{m_{ADM}^4} + 
{\cal O} \left(\frac{r}{m_{ADM}}\right)^{16/3}, 
}{ysmallr}
which is consistent with (\ref{Zsmallr}). 

\subsubsection{$\boldsymbol{p = 1}$}

The exact result
\eq{
y = \frac{r}{m_{ADM}} - 2
}{Zp1}
implies the Schwarzschild solution for geometry and matter ($S=0$).

\subsubsection{``Large'' $\boldsymbol{p = 1 - \eps}$}

Since this expansion is again problematic for large radii and gives little new 
insight we do not bother to treat this case separately.

\subsection{Artificial horizons}

We have seen that (non-trivial) static solutions allow no Killing horizons.
However, there is a ``loophole'' in the derivation of this fact: We have 
assumed positivity of $\eta$, which is justified for physical reasons: The
\ac{ADM} mass is real and also $c_1$ has been assumed to be real. Further, 
implicitly the sign of the gravitational coupling constant entered the 
definition (\ref{Zeta}). Mathematically, however, there is no reason for this 
restriction. For curiosity, we investigate in this section this (unphysical)
region for various values of $\eta$ (assuming still a positive \ac{ADM} mass).

A new interesting feature will occur: The solution breaks down at the 
``horizon''\footnote{We put horizon under quotation marks because the 
determinant $(e)$ diverges on it s.t. $(e)h$ is constant. Thus, strictly 
speaking, the point $h=0$ is not a usual Killing-horizon.}. There is no
possibility to extend the solution beyond this point. This can be interpreted
as a breakdown of the assumption of staticity on (and beyond) the ``horizon''.
Note that although $(e)$ diverges and $h$ goes to $0$ it is possible to extract
finite quantities like $(e)h$ or $S'/(e)$.

\subsubsection{$\boldsymbol{\eta = - \eps^2}$}  

The position of the Killing horizon is located at
\eq{
r_h = 2m_{ADM}\left( 1 - \frac{\eps^2}{2} \ln{\eps} + {\cal O}(\eps)^2 
\right) > 2m_{ADM} .
}{artificialkilling1}
The limit $\lim_{\eps \to 0} \eps^2 \ln{\eps} = 0$ implies the Schwarzschild 
metric as a limiting case. Thus, from this side of the parameter space it is
possible to reach the Schwarzschild solution continuously. Note that the mass
aspect function near $i_0$ is always greater than $m_{ADM}$ -- thus there is
more energy stored in geometry than in the whole system, which shows that the
matter part of the energy must be negative, in accordance with our previous 
discussion.

\subsubsection{$\boldsymbol{\eta = - 1 + \eps^2}$}

After a delicate limit the equation
\eq{
(y+1)\exp{\frac{1}{y+1}} = \frac{r}{m_{ADM}}\left(1 + {\cal O}(\eps)^2\right)
}{delicatey}
establishes a Killing horizon at $r = em_{ADM}$. In the limit $\eps \to 0$ $p$ 
goes to $\infty$.

\subsubsection{$\boldsymbol{\eta = -1 - \eps^2}$}

Apart from the fact, that in this case $p$ goes to $i\infty$ the same 
considerations as before are valid. The location of the Killing horizon is
continuous at the singular point $\eta = -1$.

\subsubsection{The general case for $\boldsymbol{\eta < - 1}$}

If we set $r_h = \ga m_{ADM}$ with $\ga \in \mathbb{R}_0^+$ and let $p=ix$
with $x \in \mathbb{R}$ the general solution for $\ga$ reads
\eq{
\ga^2 = \frac{x^2+1}{x^2} \exp\left[x(\pi-2\arctan{x})\right] .
}{generalga1} 
It is interesting to look for extrema, but the equation 
$\arctan{x}+x^{-1} = \pi/2$ has no real solution for $x$. Only the
limiting case $x \to \infty$ can fulfill this equation. It corresponds to the
case $\eta = - 1 -\eps^2$ for $\eps \to 0$. But for this case also all further
derivatives vanish. 

\subsubsection{The general case for $\boldsymbol{-1 < \eta < 0}$}

With the same definitions as before and the modification $p = x \in \mathbb{R}$
we obtain
\eq{
\ga^2 = \frac{(x+1)^{x+1}}{x^2(x-1)^{x-1}} .
}{generalga2}
Again, there are no extrema, apart from the singular endpoint $x = 0$ which 
corresponds to the Schwarzschild solution. The limiting case 
$\eta = -1 + \eps^2$ for $\eps \to 0$ yields again vanishing derivatives for 
$\ga$. Hence, $\ga(\eta)$ has a critical point at $\eta = -1$: From both 
sides, all derivatives vanish. This implies, that small variations of the 
Killing horizon lead to huge changes in $\eta$ and thus in eventual non-static
fluctuations. For this reason and the numerical coincidence 
$\ga_{crit} \approx \ga_{Choptuik}^{-1}$ it may be tempting to seek a relation 
to \Ix{critical collapse}. Unfortunately, all our solutions are unphysical 
(remember that they imply a negative matter energy density) and hence a direct 
relation does not seem to exist. 

\clearplaindoublepage

\chapter{\Ix{Hamiltonian formalism}}\index{Hamiltonian analysis}

\section{Introduction}

As opposed to the previous appendices, this one contains new results and 
therefore some of the calculations are performed more explicitly than in the
reviewing sections before. Nevertheless I have decided to put this technical
part in the appendix, because I believe that it increases the readability of
the main part.

\section{Analysis of constraints}\index{analysis of constraints}

Starting with the first order Lagrangian (\ref{1action}) together with a 
matter part (\ref{Lm}) and defining
\seq{4cm}{
&& (e) = \left( e^-_0 e^+_1 - e^+_0 e^-_1 \right), \\
&& \bar{q}_i = \left( \omega_0, e^-_0, e^+_0 \right),  \\
&& q_i = \left( \omega_1, e^-_1, e^+_1 \right), 
}{4cm}{
&& p^i = \left( X, X^+, X^- \right), \\
&& P = \frac{\partial \cal{L}}{\partial \partial_0 S} = 
\frac{\partial {\cal{L}}^{\left( m \right)}}{\partial \partial_0 S},
}{h6}
we obtain in components
\begin{eqnarray}
&& P = \frac{F(X)}{(e)}\left[\left(\partial_1 S\right)\left(e^+_0e^-_1 + e^+_1 
e^-_0 \right) - 2 \left(\partial_0 S\right) e^+_1 e^-_1 \right], \label{h11} \\
&& \left(\partial_0 S \right) = \frac{1}{2 e^+_1 e^-_1} \left[\left(e^+_0e^-_1+
e^+_1e^-_0 \right) \left( \partial_1 S \right) - \frac{(e)}{F(X)} P \right],
\label{h12} \\
&& {\cal{L}}^{\left( m \right)} = F(X)(e)\left[\frac{1}{4e^+_1e^-_1} \left[ 
\left( \partial_1 S \right)^2 - \frac{P^2}{F^2(X)} \right] - f(S) \right].
\label{SP}
\end{eqnarray}
The interaction-function $f(S)$ will not be specified in this chapter thus 
allowing for a rather general treatment of mass-terms and/or arbitrary 
(non-derivative) self-interactions of the scalar field.

The Hamilton density is given by
\begin{equation}
H = p^i \dot{q}_i + P \dot{S} - \cal{L}, \label{h13}
\end{equation}
and can be expressed in terms of a sum over constraints\footnote{In theories
which are invariant with respect to time reparametrization this is believed to
be a general feature of such theories. It is not quite true -- a remarkably
simple counterexample can be found in chapter 4 of \cite{het92}, but it is
true if $q$ and $p$ transform as scalars under time reparametrization.}
\begin{equation}
H = \alpha^i \tilde{G}_i.
\end{equation}
We will show in the next subsection that
\begin{equation}
\alpha^i = - \bar{q}_i, \hspace{0.5cm}\tilde{G}_i = G_i,
\end{equation}
with $G_i$ being the secondary constraints. Therefore -- by strict analogy 
with \ac{QED} -- the $\bar{q}_i$ can be interpreted as Lagrange multipliers 
and hence (in our notation) will get an upper index. Thus, from now on we will 
use upper indices for $\bar{q}^i$ and lower indices for $\bar{p}_i$. 

The fundamental Poisson brackets are
\begin{eqnarray}
&& \{ q_i, {p^j}' \} = \delta_i^j \delta(x-x'), \label{h14} \\
&& \{ \bar{q}^i, \bar{p}_j' \} = \delta^i_j \delta(x-x'), 
\label{h15} \\
&& \{ S, P' \} = \delta (x-x'), \label{h16}
\end{eqnarray}
where primed functions denote functions of $x'$ and unprimed functions depend 
on $x$.

Because the $\bar{q}$-fields have no ``time\footnote{We refer to the zero 
component as ``time'', although it need not have the physical meaning of it.
It is just the parameter with respect to which our dynamical systems 
evolves.}'' derivative we get the \Ix{primary constraints}
\begin{equation}
\bar{p}_i = 0. \label{h17}
\end{equation}

\subsection{Poisson brackets with primary constraints}
Using the Dirac procedure we check first the Poisson brackets of the primary 
constraints with themselves
\begin{equation}
\left\{ \bar{p}_i, \bar{p}_j' \right\} = 0, \label{h18}
\end{equation}
and obtain that they are trivial. Next we compute the brackets of the 
primary constraints with the Hamiltonian\footnote{Of course, the Hamiltonian is
an integral over the Hamilton density but we will omit the integrals because 
they are always ``killed'' by $\delta$-functions.} (\ref{h13})
\begin{equation}
\left\{ \bar{p}_i, H' \right\} = \left\{ \bar{p}_i, \left( \partial_0 S' 
\right) \right\} P - \left\{ \bar{p}_i, {\cal{L'}} \right\} =: G_i. \label{h19}
\end{equation}

We find for the \Ix{secondary constraints}
\begin{eqnarray}
&& G_1 = G_1^g, \label{notnew} \\
&& G_2 = G_2^g + \frac{F(X)}{4e^-_1} \left[ \left( \partial_1 S 
\right) - \frac{P}{F(X)} \right]^2 - F(X)e^+_1f(S), \label{new1} \\
&& G_3 = G_3^g - \frac{F(X)}{4e^+_1} \left[ \left( \partial_1 S 
\right) + \frac{P}{F(X)} \right]^2 + F(X)e^-_1f(S), \label{new2}
\end{eqnarray}
where the geometric part of the constraints is determined by
\eqa{
&& G_1^g = \partial_1 X + X^-e_1^+ - X^+e_1^-, \label{geom1} \\
&& G_2^g = \partial_1 X^+ + \om_1 X^+ - e_1^+ {\cal V}, \label{geom2} \\
&& G_3^g = \partial_1 X^- - \om_1 X^- + e_1^- {\cal V}, 
}{geom3}
with the common definition
\eq{
{\cal V} = V(X) + X^+X^- U(X) .
}{calV}

Now we see indeed that (\ref{h13}) can be expressed in terms of 
the secondary constraints
\begin{equation}
H [q_i, p^i, S, P, \bar{q}^i] = - \bar{q}^i G_i. \label{hamilcon}
\end{equation}

\subsection{Poisson brackets with secondary constraints}

\subsubsection{Brackets with primary constraints}

Because of $\{ \bar{p}_i, {G_j^{(g)}}' \} = 0$, $\forall (i, j)$
we need only the brackets with the ``new'' part of the secondary 
constraints. Because of the $\bar{q}^i$-independence of eqs. (\ref{new1}) and 
(\ref{new2}) that contribution vanishes also. Thus we conclude
\begin{equation}
\left\{ \bar{p}_i, G_j' \right\} = 0, \hspace{0.5cm}\forall i, j,
\end{equation} 
establishing the first class property of the primary constraints $\bar{p}_i$.

\subsubsection{Brackets with the Hamiltonian}

The bracket with the Hamiltonian splits into a sum of four terms:
\begin{equation}
\{ G_j, H' \} = \{ G_j, {\dot{q}_i'} {p^i}' \} + \{ 
G_j, \left( \partial_{0'} S' \right) P' \} - \{ G_j, 
{{\cal{L}}^{\left( g \right)}}' \} - \{ G_j, {{\cal{L}}^{\left( m 
\right)}}' \}.
\end{equation}

However, with (\ref{hamilcon}) it is 
sufficient (and easier!) to work with the Hamiltonian expressed in terms of 
secondary constraints multiplied with $\bar{q}^i$. Because of
\begin{equation}
\left\{ \bar{q}^iG_i,  G_j' \right\} = \bar{q}^i \left\{ G_i, G_j' \right\}, 
\hspace{0.5cm}\forall ( i, j ),
\end{equation}
it suffices to work with the Poisson algebra of secondary constraints with
themselves.

\subsection{\Ix{Poisson algebra of secondary constraints}}

For convenience we split the bracket between two secondary constraints into 
four terms
\begin{equation}
\left\{ G_i, G'_j \right\} = \left\{ G_i^g, {G^g_j}' \right\} + 
\left\{ G_i^g, {G^m_j}' \right\} + \left\{ G_i^m, {G^g_j}' \right\} + \left\{ 
G_i^m, {G^m_j}' \right\},
\end{equation}
the first of which has been calculated in \cite{klv97}.

The brackets with $G_1$ give
\begin{equation}
\left\{ G_1^g, {G^m_2}' \right\} = - G_2^m, \hspace{0.5cm}\left\{ G_1^g, 
{G_3^m}' \right\} = G_3^m,
\end{equation}
showing the consistency of the generalization of the algebra in \cite{klv97} 
to the non-minimally coupled case even in the presence of nontrivial 
interaction terms.

Finally we need the brackets between $G_2$ and $G_3$:
\meq{
\left\{ G_2^g, {G_3^m}' \right\} + \left\{ G_2^m, {G_3^g}' \right\} = 
- X^+ U(X) G_3^m - X^- U(X) G_2^m \\ 
- \frac{X^+}{4e^+_1} F'(X) \left(\left(\partial_1 S\right)^2 - \frac{P^2}
{F^2(X)}\right) + \frac{X^-}{4e^-_1} F'(X) \left(\left(\partial_1 S\right)^2 - 
\frac{P^2} {F^2(X)}\right) \\
- f(S) F'(X) \left( e_1^-X^+-e_1^+X^--\partial_1 X \right),
}{g2g3new1}
\eq{
\left\{ G_2^m, {G_3^m}' \right\} = \frac{F'(X)\partial_1X}
{4e^-_1e^+_1} \left[ \left(\partial_1 S\right)^2 - \frac{P^2}{F^2(X)} \right].
}{g2g3new2}

Our algebra has therefore the following fundamental Poisson brackets between
its generators:
\begin{eqnarray}
&& \left\{ G_1, G'_2 \right\} = - G_2 \de, \label{algebra1} \\
&& \left\{ G_1, G'_3 \right\} = G_3 \de, \label{algebra2} \\
&& \left\{ G_2, G'_3 \right\} = - \frac{d {\cal V}}{d p^i} G_i \de
+\frac {F'(X)}{(e)F(X)}{\cal{L}}^{(m)} G_1 \de. \label{algebra3}
\end{eqnarray}
We have used the abbreviation $\de = \de (x-x')$.
For completeness we need the Poisson brackets
\begin{equation}
\left\{ G_1, G'_1 \right\} = 0, \hspace{0.5cm} \left\{ G_2, G'_2 \right\} 
= 0, \hspace{0.5cm} \left\{ G_3, G'_3 \right\} = 0. \label{g3g3}
\end{equation}
Note that despite of their simplicity eqs. (\ref{g3g3}) are not 
trivial since first derivatives of the $\de$-function could be present on the 
\ac{r.h.s.}, as it is the case for the Virasoro constraints 
(\ref{virasoroalgebra}).

The nonvanishing structure functions of $rank=1$ (cf. e.g. \cite{het92} for 
the definition of rank) are therefore given by
\newline \parbox{3cm}{\begin{eqnarray*}
&& C_{12}{}^2 = -1, \\
&& C_{13}{}^3 = 1,
\end{eqnarray*}} \hfill
\parbox{7cm}{\begin{eqnarray*}
&& C_{23}{}^1 = -\frac{\partial{\cal{V}}}{\partial X}+\frac{F'(X)}
{(e)F(X)}{\cal{L}}^{(m)}, \\
&& C_{23}{}^2 = -\frac{\partial {\cal{V}}}{\partial X^+}, \\
&& C_{23}{}^3 = -\frac{\partial {\cal{V}}}{\partial X^-}.
\end{eqnarray*}} \hfill
\parbox{1cm}{\begin{equation} \label{structure} \end{equation}} \hfill
\newline
This is a non-trivial generalization of the matterless algebra \cite{sst94b,
klv97a} and the algebra with minimally coupled matter \cite{klv97}. Both of
them are finite W-algebras (for a definition of W-algebras cf. e.g. 
\cite{bht96}) with generators $p_i$ and $G_i$. In our case, however, we do not 
have such a W-algebra anymore because the matter Lagrangian contains also 
$q_i$, $S$ and $P$.

\subsection{Relation to Poisson-$\si$ models}\index{Poisson-$\si$ models}

The matterless algebra can be obtained most simply in the Poisson-$\si$ 
formulation (cf. {\app appendix D} ): The secondary constraints are
\eq{
G^i = \partial_1 p^i + P^{ij}q_j,
}{PSMsecondary} 
and the structure functions are therefore given by 
\eq{
C^{ij}{}_k = \frac{\partial P^{ij}}{\partial p^k},
}{PSMstructure} 
with the Poisson tensor $P^{ij}$ given by (\ref{poissontensor}). By comparison 
with (\ref{structure}) one may be tempted to replace ${\cal V} \to
{\cal V} - {\cal L}^{(m)}/(e)$ in the Poisson tensor and add a 
Poisson coordinate $S$ corresponding to the scalar field to the Poisson 
manifold. This works well for static scalars (i.e. a scalar field with a
Lagrangian containing no derivatives of it), which is the deeper reason why 
Mann was able to generalize the conservation law to static matter 
\cite{man93}. It also works for topological matter, e.g. chiral fermions 
\cite{kum92} or (anti-)selfdual scalars \cite{pes98}. But there seems to be no 
way to treat propagating matter in the framework of Poisson-$\si$ models, since
they can cope with topological degrees of freedom only.

\section{Different sets of constraints and canonical variables}

\subsection{The relation to the \Ix{Virasoro algebra}}

By a certain linear combination of our constraints, namely
\begin{eqnarray}
&& G = G_1, \\
&& H_0 = q_1 G_1 - q_2 G_2 + q_3 G_3, \\
&& H_1 = q_i G_i,
\end{eqnarray}
we obtain a slightly different algebra \newline
\parbox{5cm}{\begin{eqnarray*}
&& \left\{ G, G' \right\} = 0, \\
&& \left\{ G, H_0' \right\} = -G \delta', \\
&& \left\{ G, H_1' \right\} = -G \delta',
\end{eqnarray*}} \hfill
\parbox{5cm}{\begin{eqnarray*}
&& \left\{ H_0, H_0' \right\} = \left( H_1 + H_1' \right) \delta', \\
&& \left\{ H_0, H_1' \right\} = \left( H_0 + H_0' \right) \delta', \\
&& \left\{ H_1, H_1' \right\} = \left( H_1 + H_1' \right) \delta',
\end{eqnarray*}} \hfill
\parbox{1cm}{\begin{equation} \label{virasoroalgebra} \end{equation}} \hfill
\newline which is a semidirect product of the \Ix{Virasoro algebra} (or 
conformal algebra) generated by $H_i$ and an invariant abelian subalgebra 
generated by $G$ \cite{kat93}. We have used the abbreviation 
$\de' = \partial \de (x-x')/\partial x'$.

Note that the natural algebra closes with the $\de$-function and the Virasoro
algebra with its first derivative. The main advantage of this algebra is its
simplicity in the case of (non-minimally coupled) matter: The structure 
functions are always constant!

In string theory usually the Fourier-transformed version of 
(\ref{virasoroalgebra}) is called (classical) ``Virasoro-algebra'' 
\cite{gsw86}:
\eq{
\left\{ L_n, L_m \right\} = i(n-m) L_{n+m}.
}{fouriervirasoro}
The (inessential but unusual) factor $i$ appears, because we have used 
classical (Poisson) brackets instead of commutator relations. This is also the
reason why no central charge appears.

\subsection{Abelianization}\index{abelianization}

It is well-known that the algebra of constraints can always be abelianized 
(at least locally) -- cf. e.g. \cite{het92} -- although for practical 
calculations this feature is of no great use. We will show that our 
algebra (\ref{algebra1}-\ref{algebra3}) -- which contains field-dependent 
structure functions -- can be abelianized in a patch containing no zeros of
the Killing-norm by a proper redefinition of the  constraints in the 
matterless case. Since this set of constraints proves useful also in the case
with matter, we discuss them together in the next section.

\subsection{The ``\Ix{energetic constraints}''}

In {\twod} models of first order gravity there exists always a conservation law
(cf. {\app appendix D} on p. \pageref{conservation law}).  
The $\partial_1$-derivative of the conserved quantity is a linear
combination of the constraints $G_i$
\eq{
\partial_1 {\cal C} = I(p_1)\left({\cal V}G_1+p_3G_2+p_2G_3\right),
}{en10}
with $I(p_1)$ being an integrating factor the explicit form of which is 
irrelevant for our purposes. Lets define a new set of constraints
\eqa{
&& G_1^{(e)} = G_1, \label{en11} \\
&& G_2^{(e)} = \frac{1}{2p_2p_3I(p_1)} \left(p_3G_2-p_2G_3\right), 
\label{en12} \\
&& G_3^{(e)} = I(p_1)\left({\cal V}G_1+p_3G_2+p_2G_3\right),
}{en13}
which we would like to call ``energetic'' since the third constraint is 
equal to $\partial_1 {\cal C}$ which has a very close relation to
the \ac{ADM}-mass \cite{gru99,grk00}.

The factors in the new constraints are chosen such that
\eq{
\det \left|\frac{\partial(G_1^{(e)}, G_2^{(e)}, G_3^{(e)})}
{\partial(G_1, G_2, G_3)} \right| = 1, 
}{en14} 
showing their linear independence.

Note that the redefinitions (\ref{en11}-\ref{en13}) behave well in a patch 
where $p_2 \neq 0 \neq p_3$. Particularly, they break down on a 
Killing-horizon. But actually, this is not a problem at the classical 
level, because it is possible to transform back canonically after a convenient 
gauge has been chosen.

Since the geometric parts of the energetic constraints forms an abelian algebra
(see below) one could try to use them as canonical coordinates and find their 
corresponding momenta. This is possible for $G_1^{(e)}$ and $G_2^{(e)}$ 
straightforwardly:
\seq{3cm}{
&& q_1^c := G_1^{(e)} I(p_1)^{-1}, \\
&& q_2^c := G_2^{(e)},
}
{3cm}{
&& p_1^c := - \frac{1}{2}\ln{p_2/p_3}, \\
&& p_2^c := p_1.
}{en100}
In the framework of \ac{PSM} this amounts to finding the 
Casimir-Darboux\footnote{Since the Poisson tensor is degenerate simple Darboux
coordinates are not accessible because the manifold is not symplectic. But it
contains symplectic leaves ``counted'' by the value of the Casimir function. So
each point in the Poisson manifold is uniquely defined by the Darboux 
coordinates on a given symplectic leave labeled by a certain value of the 
Casimir function. This justifies the name ``Casimir-Darboux'' coordinates.} 
coordinates on the Poisson manifold \cite{sst94b}.

Unfortunately, $\partial_1 {\cal C}^{(g)}$ contains unwieldy 
$\partial_1$-derivatives. Therefore, it is convenient to identify the third 
coordinate with ${\cal C}^{(g)}$ (up to the integrating factor) rather than 
with its derivative:
\eq{
q_3^c := {\cal C}^{(g)}/I(p_1).
}{en101}
The canonical momentum reduces to the simple expression
\eq{
p_3^c := - \frac{q_2p_2 + q_3p_3}{2p_2p_3} .
}{en102}

We have obtained a nice interpretation for these variables in the absence of 
matter:
\blist
\item $p_1^c$ is canonically conjugate to the Lorentz-constraint, which is
the expected result, since it parametrizes the Lorentz-angle.
\item $p_2^c$ is the canonical momentum for the second energetic constraint, 
for which we have had no physical interpretation up to now: This constraint is 
the canonical conjugate of the dilaton field -- so if the latter is to be 
interpreted as a radius, the former plays the r{\^o}le of radial translations.  
\item $p_3^c$ is essentially the canonical conjugate for the geometric part of 
the  conserved quantity -- thus, a natural choice of gauge would be 
$p_3^c = x_0$ or $q = x_1$, depending whether $x_0$ or $x_1$ are timelike 
coordinates, since qualitatively the canonical conjugate of energy is time. 

\elist
Possible anomalies of the constraints correspond thus to 1. quantum violation 
of frame-rotation symmetry, 2. quantum violation of translation invariance and 
3. quantum violation of energy conservation, respectively.

Of course, matter modifies these relations according to 
(\ref{enm1}-\ref{enm3}).
Since we do not like to have integrals in the constraint $G_3^{(e)}$ we will 
again return to the original definition (\ref{en13}) for the case with matter.
Before we investigate the properties of the set (\ref{en11}-\ref{en13}) in the 
presence of matter, it is very convenient to introduce a new set of canonical 
variables.

\subsubsection{Energetic coordinates}\index{energetic coordinates}

We perform a canonical transformation\footnote{This is a special case of a
transformation introduced in \cite{kat94}. Similar coordinates (in the 
matterless case) have already been used in \cite{kul97}.}
\eq{
(q_2, q_3, p_2, p_3) \to (q, q_{\bot}, p, p_{\bot}),
}{en20}
defined by the relations\footnote{It is straightforward to prove that all 
fundamental Poisson brackets are conserved by this transformation.}
\seq{3cm}{
&& q := - p_3^c, \label{21} \\
&& \qb := q_2p_2 - q_3p_3, \label{22} \\
}{3cm}{
&& p := p_2p_3, \label{23} \\
&& \pb := -p_1^c.
}{en24}
It is noteworthy, that $q, \qb, p$ are Lorentz-covariant variables while $\pb$
is proportional to the Lorentz-angle. 

The determinant
\eq{
\det \left|\frac{\partial(q_2, q_3, p_2, p_3)}{\partial (q, q_{\bot}, p, 
p_{\bot})} \right| = 1 
}{en25} 
again equals unity and the canonical transformation is regular in the same 
range where the new set of constraints (\ref{en11}-\ref{en13}) is defined. 
Consistently, the inverse transformation
\seq{3cm}{
&& p_2 = \sqrt{p} e^{\pb}, \label{en26} \\
&& p_3 = \sqrt{p} e^{-\pb}, \label{en27} \\
}{4cm}{
&& q_2 = \frac{1}{2\sqrt{p}} e^{\pb} \left(2qp-\qb\right), \label{en28} \\
&& q_3 = \frac{1}{2\sqrt{p}} e^{-\pb} \left(2qp+\qb\right),
}{en29}
is regular as long as $p \neq 0$. For later convenience we introduce the 
quantity
\eq{
E := 1/(4q_2q_3) = 1/(4e_1^-e_1^+) = p/\left(4q^2p^2-\qb^2\right).
}{ene1e1}

The geometric part of the new constraints (\ref{en11}-\ref{en13}) has now the 
very simple form
\eqa{
&& G_1^{(e,g)} = \partial_1 p_1 - \qb, \label{eng1} \\
&& G_2^{(e,g)} = I(p_1)^{-1} \left( \partial_1 \pb + q_1 - {\cal V}q \right),
\label{eng2} \\
&& G_3^{(e,g)} = I(p_1) \left( \partial_1 p + {\cal V}\partial_1 p_1 \right), 
}{eng3}
while the matter part is given by
\eqa{
&& G_1^{(e,m)} = 0, \label{enm1} \\
&& G_2^{(e,m)} = I(p_1)^{-1}\left(-qF(p_1)f(S) + qE T^+ + \frac{\qb E}{2p} T^* 
\right), \label{enm2} \\
&& G_3^{(e,m)} = I(p_1) \left(\qb F(p_1) f(S) - \qb E T^+ - 2qpE T^* \right), 
}{enm3}
resembling in structure the matter part of the natural constraints.
We have used the definitions
\eq{
T^{\pm} := F(p_1) \left( \left(\partial_1 S\right)^2 \pm \frac{P^2}{F^2(p_1)}
\right), \hspace{0.5cm} T^* := 2 \left(\partial_1 S\right) P,
}{Tdefs}
which yield the Virasoro-like algebra (it closes with $\de'$)
\eqa{
&& \hspace{-1cm} f\left\{ T^{\pm}, {T^{\pm}}' \right\} g' = \pm 2 T^* 
\de(x-x') f \overleftrightarrow{\partial} g, \\ 
&& \hspace{-1cm} f\left\{ T^*, {T^*}' \right\} g' = 2 T^* \de(x-x') f 
\overleftrightarrow{\partial} g, \\ 
&& \hspace{-1cm} f\left\{ {T^{\pm}}, {T^*}' \right\} g' = 2T^{\pm}\de(x-x') f 
\overleftrightarrow{\partial} g - 
2 T^{\mp} \de(x-x') fg \frac{F'}{F}, \\
&& \hspace{-1cm} f\left\{ T^{\pm}, {T^{\mp}}' \right\} g' = 2T^*\de(x-x') fg 
\left( 2 \frac{F'}{F} - \frac{P'}{P} + \frac{S''}{S'} \right),
}{Talgebra}
with $a\overleftrightarrow{\partial}b := a (\partial b) - b (\partial a)$.

All structure function vanish apart from 
$C_{23}{}^i = - C_{32}{}^i$ which has non-vanishing matter 
contributions. For simplicity we assume $f(S) = 0$ and obtain:
\eqa{
&& \hspace{-1cm} C_{23}{}^1 = \frac{1}{(e)}\frac{F'(p_1)}{F(p_1)}
{\cal L}^{(m)} = {\cal O}\left(S^2\right), \label{enc231} \\
&& \hspace{-1cm} C_{23}{}^2 = 0, \label{enc232} \\
&& \hspace{-1cm} C_{23}{}^3 = -\frac{E}{p}\left(T^+(8Epq^2-1)+4ET^*q\qb\right) 
I(p_1)^{-1} = {\cal O}\left(S^2\right).
}{enc233}
This implies in particular the abelianization of the algebra in the matterless 
limit.

\subsection{A \Ix{canonical transformation}}

In the case of non-minimal coupling ($F(X) \neq const$) the scalar field $S$ 
and its momentum $P$ appear unsymmetrically in the Lagrangian. Even more
important than this esthetic remark is the problem with the coupling function
in the path-integral: It leads to a set of coupled \ac{PDE} 
(\ref{Q101}-\ref{Q102}) as opposed to the minimally coupled case, where these 
\ac{PDE} decouple. This is the motivation for a canonical transformation 
leading to a symmetrized set of fields. 

Let $\tilde{S} = \left[ F(X) \right]^a S$ with $a \in \mathbb{R}$. Because of
$\left\{ S, P \right\} = \{ \tilde{S}, \tilde{P} \}$ with
$\tilde{P} = \partial {\cal{L}}/\partial \partial_0 \tilde{S} = 
P F^{-a}$ this seems to be a canonical transformation at first glance -- and 
a good one, too, because the asymmetry between $S$ and $P$ due to the factor 
$F(X)$ vanishes for $a=\frac{1}{2}$.

However, because of
\begin{equation}
\left\{ \tilde{S}, \omega_1 \right\} = F^{-a}\left\{ S, \omega_1 \right\} - aS
F^{-a-1} F' \left\{ X, \omega_1 \right\} \neq 0
\end{equation}
the transformation
\begin{equation}
S \to \tilde{S} = S \left[ F(X) \right]^a, \hspace{0.5cm}P \to \tilde{P} = 
P \left[ F(X) \right]^{-a}
\end{equation}
is {\em not} canonical.

A transformation of canonical variables is called ``canonical'' if all the 
Poisson  brackets are invariant under this transformation or if all brackets 
are multiplied by the same constant \cite{arn78}. We define the transformation 
$T$:
\begin{equation}
T:(q_i,p^i,\bar{q^i},\bar{p_i},S,P) \to (\tilde{q_i},\tilde{p^i},\tilde{\bar
{q^i}},\tilde{\bar{p_i}},\tilde{S},\tilde{P}), \label{trafo}
\end{equation}
with

\parbox{3cm}{\begin{eqnarray*} 
&& \tilde{q}_1 = q_1 + a\frac{F'(X)}{F(X)}SP, \\
&& \tilde{q}_2 = q_2, \\
&& \tilde{q}_3 = q_3, \\
&& \tilde{p}^i = p^i, \end{eqnarray*}} \hfill
\parbox{3cm}{\begin{eqnarray*} 
&& \tilde{\bar{q}}^i = \bar{q}^i, \\
&& \tilde{\bar{p}}_i = \bar{p}_i, \\
&& \tilde{S} = S\left[F(X)\right]^a, \\
&& \tilde{P} = P\left[F(X)\right]^{-a}, \end{eqnarray*}} \hfill
\parbox{1cm}{\begin{equation} \label{trafo1} \end{equation}} \newline
and prove that it is canonical:
We show that all Poisson brackets are invariant under $T$. Obviously, this is 
true for all brackets of unchanged variables with other unchanged variables. 
Thus it is sufficient to show that all brackets of changed variables ($S$, $P$
and $q_1 = \omega_1$) with all other variables are invariant under $T$.
The brackets of $S$ and $P$ are obviously the same as the brackets of 
$\tilde{S}$ and $\tilde{P}$ with all variables except $\omega_1$.
The brackets of $\omega_1$ are obviously the same as the brackets of 
$\tilde{\omega}_1$ with all variables except $S$, $P$ and itself.
Therefore, it is sufficient to check only the brackets between 
$\tilde{\omega_1}$ and the three changed variables explicitly:
\begin{eqnarray}
&& \{ \tilde{S}, \tilde{\omega}_1' \} = F^{-a}\left\{ S, \tilde
{\omega_1}' \right\} - aSF^{-a-1}F'\left\{ X, \tilde{\omega}_1'\right\} = 0, \\
&& \{ \tilde{P}, \tilde{\omega}_1' \} = F^{a}\left\{ P, \tilde
{\omega_1}' \right\} + aSF^{a-1}F'\left\{ X, \tilde{\omega}_1' \right\} = 0, \\
&& \left\{ \tilde{\omega}_1, \tilde{\omega}_1' \right\} = 0,
\end{eqnarray}
with the last equation being a consequence of the absence of space derivatives 
in $\tilde{\omega}_1$. 

Thus we get a new Hamiltonian
\begin{equation}
H(q,p,...) \to T \circ H(q,p,...) = \tilde{H}(\tilde{q},\tilde{p},...),
\end{equation}
where
\begin{equation}
\tilde{H} = -\bar{\tilde{q}^i}\tilde{G}_i(\tilde{q},\tilde{p},...) .
\end{equation}

Of course we also get a new effective Hamiltonian
\begin{equation}
H_{eff} \to T \circ H_{eff}(q,p,...) = \tilde{H}(\tilde{q},\tilde{p},...)
+ \tilde{\lambda}^i\bar{\tilde{p}_i} + \tilde{\tau}^i\tilde{G}_i(\tilde{q},
\tilde{p},...) .
\end{equation}
Unfortunately the action of space-derivatives on the coupling function in the
secondary constraints spoils the simplicity of this Hamiltonian. Thus, the law 
of conservation of misery prevents us from simplifying the path
integral by this canonical transformation.

\section{Gauge invariance and \Ix{gauge fixing}}\label{gauge invariance}

There exists an infinite number of possible gauge fixings -- but even
by restricting oneself to ``meaningful gauges'' -- in the sense that they 
result in a computable, non-trivial and physically interpretable quantized 
theory -- there is a variety of choices. We will just pick a few of them and 
discuss their advantages and drawbacks.

To start with, we have to count our gauge degrees of freedom: We have got 6 
independent first class constraints and therefore need 6 gauge conditions if
the {\em Dirac conjecture} holds for our model. If there are no second
class constraints involved in the Dirac algorithm and no bifurcations arise,
then it is neccessary and sufficient that the matrix $V_a^b$, defined by
\begin{equation}
\left\{ H, \Phi_a \right\} = V_a{}^b \Phi_b,
\end{equation}
is of maximal rank, where the index $a$ runs over all primary constraints
and $b$ over all secondary constraints \cite{het92}. Indeed, for our model 
we get
\begin{equation}
V_a{}^b = - diag \left( 1, 1, 1 \right),
\end{equation}
which proofs the validity of the Dirac conjecture in our case.

Bearing in mind the number of gauge conditions in the Lagrangian formulation
(2 from diffeomorphisms and 1 from local Lorentz transformations) we see that 
the number of necessary gauge conditions has doubled, us usual in the 
Hamiltonian formalism.

One rather simple choice is the multiplier gauge
\begin{equation}
\mu^i = 0,\hspace{0.5cm}\forall i, \label{multigauge}
\end{equation}
which leads after elimination of $\lambda^i$ and $\bar{p}_i$ to a Hamiltonian 
action where only the unconstrained momenta appear
\begin{equation}
L_H = \int d^2x \left[ p^i\dot{q}_i+P\dot{S}+\bar{q}^iG_i \right].
\label{hamiaction}
\end{equation}
Because of (\ref{hamiaction}) it makes sense to call $\bar{q}^i$ a 
``Lagrange multiplier'' -- by strict analogy with \ac{QED} where
$A_0$ plays this r{\^o}le \cite{het92}. 
Of course, one still has three gauge degrees of freedom left the fixing of 
which will be the issue of the next subsections.

But before we would like to investigate the action of a general gauge 
transformation 
\begin{equation}
\int dx \left( \varepsilon^i_1 \bar{p}_i + \varepsilon^i_2 G_i \right)
\end{equation}
on the canonical variables and the Lagrange multipliers: \newline
\parbox{5cm}
{\begin{eqnarray*}
&& \delta q_i = \varepsilon^j_2 \left\{ q_i, G_j \right\}, \\
&& \delta \bar{q}^i = \varepsilon^i_1, \\
&& \delta \lambda^i = \dot{\varepsilon}^i_1,
\end{eqnarray*}} \hfill
\parbox{5cm}
{\begin{eqnarray*}
&& \delta p^i = \varepsilon^j_2 \left\{ p^i, G_j \right\}, \\
&& \delta \bar{p}_i = 0,
\end{eqnarray*}}\hfill
\parbox{1cm}{\begin{equation}\end{equation}}
\newline and
\begin{equation}
\delta \mu^i = \dot{\varepsilon}^i_2 + \left( \mu^k - \bar{q}^k \right)
\varepsilon^j_2 C_{jk}{}^i - \varepsilon^i_1.
\end{equation}

Since we have chosen the multiplier-gauge $\mu = 0$ we get as consistency 
condition
\begin{equation}
\varepsilon^i_1 = \dot{\varepsilon}^i_2 + \left( \mu^k - \bar{q}^k \right)
\varepsilon^j_2 C_{jk}{}^i.
\label{multitrafo}
\end{equation}
Therefore, we obtain
\begin{eqnarray}
&& \delta \bar{q}^i = \dot{\varepsilon}^i - \bar{q}^k \varepsilon^j 
C_{jk}{}^i, \label{qbartrafo} \\
&& \delta \lambda^i = \ddot{\varepsilon}^i - \bar{q}^k \dot{\varepsilon}^j 
C_{jk}{}^i - \dot{\bar{q}}^k \varepsilon^j C_{jk}{}^i - 
\bar{q}^k \varepsilon^j \dot{C}_{jk}{}^i, \label{lambdatrafo}
\end{eqnarray}
having defined $\varepsilon = \varepsilon_2$. Eq. (\ref{qbartrafo}) shows that 
$\bar{q}$ now really transforms like a Lagrange multiplier.

\subsection{Extended phase space approach}\label{BRST}\index{extended phase 
space approach}

We simplify our action from the beginning starting with a trivial shift in 
$\bar{q}^i$ and hence a canonical transformation where we have joined 
$\bar{q}^i$ and $\mu^i$ to a single variable denoted again by $\bar{q}^i$.

Apart from some minor differences in the notations, we follow closely 
\cite{het92}. Starting with
\begin{equation}
L_H = \int \left[ \dot{Q}_i P^i - \lambda^i \bar{p}_i + \bar{q}^i G_i \right],
\label{easyaction}
\end{equation}
where $Q_i = (q_i, \bar{q}^i, S)$ and $P^i$ are the corresponding 
momenta. Note that we treat now the Lagrange multipliers $\bar{q}^i$ as 
canonical variables thus enlarging our phase space. We obtain two quadruples 
of constraint/canonical coordinate/ghost/ghost momentum\footnote{Note that the
ghost momenta have ghost number $-1$ and therefore sometimes are called 
``antighosts'' (especially if they are part of a Lagrange multiplier 
quadruple); the conjugate momenta to the multipliers are sometimes (especially
in \ac{QFT}) called ``St\"uckelberg fields'' or ``Nakanishi-Lautrup'' fields.}:
\begin{equation}
\left( \bar{p}_i, \bar{q}^i, b^i, p^b_i \right), \hspace{0.5cm}\left( G_i, -, 
c^i, p^c_i \right). \label{qua2a} 
\end{equation}
Note that in (\ref{qua2a}) we do not have any ``coordinate'' conjugate to the
secondary constraints $G_i$, although one could try to construct some 
quantities which fulfill canonical Poisson bracket relations with them. But 
since we do not need these quantities for the \ac{BRST} procedure we will skip this.

The total extended gauge fixed Hamiltonian is given by
\begin{equation}
H_{gf} = H_{BRST} + \left\{ \Psi, \Omega \right\},
\end{equation}
with $H_{BRST}$ being a \Ix{\ac{BRST}-invariant} extension of the ``physical 
Hamiltonian'' (in our case it vanishes as the Hamiltonian equals to zero on the
surface of constraints), $\Psi$ being the so-called ``\Ix{gauge fixing 
fermion}'' and $\Omega$ is the \Ix{\ac{BRST}-charge} \cite{brs75, tyu75}.

A useful class of \Ix{gauge fixing fermion}s is given by \cite{het92}
\begin{equation}
\Psi = p^{b}_i \chi^i + p^c_i \bar{q}^i, \label{fixfer}
\end{equation}
where $\chi^i$ are some functions of $(Q^i,P_i)$ and are called the
``\Ix{gauge fixing functions}''.  

The \ac{BRST} charge is given by
\begin{eqnarray}
\Omega &=& \Omega_{min} + \Omega_{nonmin}, \label{brst} \\
\Omega_{min} &=& c^i G_i + \frac{1}{2}c^i c^j C_{ij}{}^k p^c_k, 
\label{brstmin} \\
\Omega_{nonmin} &=& b^i \bar{p}_i. \label{brstnonmin}
\end{eqnarray}
Note that the separation between the ``minimal'' and the ``nonminimal'' sector
of the theory is by no means unique: If we would have treated $\bar{q}^i$ not
as a Lagrange multiplier but as an ordinary canonical variable, then we 
would have put (\ref{brstnonmin}) into (\ref{brstmin}).

Since the structure functions (\ref{structure}) are field dependent it is 
non-trivial that the homological perturbation series stops at $rank=1$. In
general one would expect the presence of higher order ghost terms (``ghost 
self interactions''). However, we will prove that already $\Omega$ as
defined in (\ref{brst}) is nilpotent. 

For the matterless case this is a simple consequence of the Poisson structure:
The Jacobi identity for the Poisson tensor (\ref{poissontensor}) implies that 
the homological perturbation series stops already at the Yang-Mills level
\cite{sst94b}. Here I will show that the inclusion of (dynamical) scalars does
not change this feature.

An important point to note is that only the $C_{23}{}^i$ components 
may lead to non-trivial terms. Thus, after a straightforward calculation which 
is essentially equivalent to the matterless case we obtain:
\begin{equation}
\Omega^2 = \frac{1}{2}c^nc^ic^jp^c_k\Delta_{[nij]}^{\quad\,\,k} + 
\frac{1}{4}c^ic^j\Delta_{n[ij]}^{\quad\,\,n} + A^{ijnm}_{\quad\,\,\,\,kl} 
\left\{ C_{ij}{}^k, C_{nm}{}^l \right\}, \label{omegasqr}
\end{equation}
with 
$\Delta_{nij}{}^k = \left\{ G_n, C_{ij}{}^k \right\},$  
and
$A^{ijnm}_{\quad\,\,\,\,kl} = \frac{1}{4}c^ic^jc^nc^mp^c_kp^c_l.$
Note that the terms which contain Poisson brackets of $C_{ij}{}^k$ 
with themselves vanish either trivially since the corresponding components are 
constant or they vanish due to the appearance of at least on $c_i^2$ term.

The first expression of (\ref{omegasqr}) vanishes identically and after some 
small calculation it is found that also the second term vanishes, although the
individual Poisson brackets do not. The essential point is that in the 
nontrivial structure functions the matter part yields zero.
Therefore, $\Omega^2=0$ is true for (\ref{brst}) and no higher order structure 
functions appear even for the case of nonminimally coupled matter.
Thus we get for the gauge fixed Hamiltonian:
\begin{equation}
H_{gf} = p^b_i \left\{ \chi^i, \Omega \right\} - \chi^i \bar{p}_i 
- \bar{q}^i G_i + \bar{q}^n \left\{ p^c_n, \frac{1}{2}c^i c^j 
C_{ij}{}^k p^c_k \right\} + p^c_i b^i. \label{hgf1}
\end{equation}

We are free to choose the gauge fixing fermion as we wish without changing the 
``physical observables'' in the sense of Fradkin and Vilkovisky\footnote{This 
statement is true only for \ac{BRST}-invariant quantities, but not when 
source terms are included (i.e. on the quantum level). Since we are not trying 
to prove the gauge independence of the $S$-matrix \cite{kum01}, the 
Fradkin-Vilkovisky theorem is sufficient for our purposes.} because the 
resulting path integral is invariant \cite{frv75, bav77, frf78}.

\subsection{Temporal gauges}\label{temporal gauges}\index{temporal gauge}

By ``temporal gauges'' we mean the following class of gauge fixings:
\begin{equation}
\bar{q}^i = a^i,
\end{equation}
where $a^i$ is some constant triple leading to a non-trivial 
metric\footnote{E.g. the choice $a^i = (a, 0, 0), a \in \mathbb{R}$ leads to a 
singular metric and thus has to be rejected.}.
For sake of definiteness and simplicity we will choose $a^i = (0, 1, 0)$ in
accordance with \cite{klv97a,klv97,klv99}.

We specify the gauge fixing functions as follows:
\begin{equation}
\chi^i = \frac{1}{\varepsilon} \left( \bar{q}^i - a^i \right), \hspace{0.5cm}
a^i = (0, 1, 0), 
\label{gauge}
\end{equation}
with $\varepsilon$ being a (not necessarily small) positive constant.
With this choice equation (\ref{hgf1}) reduces to
\begin{equation}
H_{gf} = \frac{1}{\varepsilon} p^b_i b^i - \frac{1}{\varepsilon} 
\left( \bar{q}^i - a^i \right) \bar{p}_i - \bar{q}^i G_i - \bar{q}^i c^j 
C_{ij}{}^k p^c_k + p^c_ib_i. \label{hgf2}
\end{equation}

The \ac{EOM} for $\bar{q}^i$
\begin{equation}
\dot{\bar{q}}^i = - \frac{1}{\varepsilon} \left( \bar{q}^i - a^i \right),
\end{equation} 
yield some homogeneous contribution for all three
components and (in our gauge) only one inhomogeneous component:\newline
\parbox{3.3cm}{\begin{equation*}\bar{q}^1 = e^{- \frac{1}{\varepsilon} 
\left( x_0-y_0\right)}, \end{equation*}} \hfill
\parbox{3.3cm}{\begin{equation*}\bar{q}^2 = 1 + e^{- \frac{1}{\varepsilon} 
\left( x_0-y_0\right)}, \end{equation*}} \hfill
\parbox{3.3cm}{\begin{equation*}\bar{q}^3 = e^{- \frac{1}{\varepsilon} 
\left( x_0-y_0\right)}, \end{equation*}} \hfill
\parbox{1cm}{\begin{equation} \label{fixings} \end{equation}}
\newline
Clearly, in any of the following limits

\parbox{3.3cm}{\begin{equation*}\lim_{\varepsilon \to 0},\end{equation*}}\hfill
\parbox{3.3cm}{\begin{equation*}\lim_{x_0 \to \infty},\end{equation*}}\hfill
\parbox{3.3cm}{\begin{equation*}\lim_{y_0 \to - \infty},\end{equation*}}\hfill
\parbox{1cm}{\begin{equation} \end{equation}}
\newline we get vanishing homogeneous parts as long as $x_0 > y_0$ and 
$\varepsilon$ remains positive. 

It should be mentioned that equation (\ref{hgf2}) is (apart from different 
signs due to a different convention in the gauge fixing fermion) identical to 
equation (2.28) of \cite{hak94} and equation (24) of \cite{klv97}.

Because we wish to take the limit $\lim_{\varepsilon \to 0}$ later, we must get
rid of the $1/\varepsilon$-terms. This goal can be achieved by a 
redefinition of the canonical momenta
\begin{equation}
\bar{p_i} \to \hat{\bar{p}}_i = \varepsilon \bar{p}_i, \hspace{0.5cm} p_i^b \to
\hat{p}_i^b = \varepsilon p_i^b, 
\end{equation}
which has a unit super-Jacobian in the path integral. The \ac{EOM} for 
$\bar{q}^i$ do not change under this (generalized canonical) transformation.

The \Ix{path integral} is given by
\begin{eqnarray}
W &=& \int \left({\cal D}q_i\right)\left({\cal D}p^i\right)\left({\cal D}
\bar{q}^i\right)\left({\cal D}\hat{\bar{p}}_i\right)\left({\cal D}S\right)
\left({\cal D}P\right)\left({\cal D}c^i\right)\left({\cal D}p^c_i\right)
\left({\cal D}b^i\right)\left({\cal D}\hat{p}^b_i\right) \nonumber \\
&&\times \exp \left[ i \int \left( {\cal L}_{\text{eff}} + J_i p^i + j_i q_i 
+ Q S \right) d^2x \right], \label{pi1}
\end{eqnarray}
with
\begin{equation}
{\cal L}_{\text{eff}} = p^i \dot{q}_i + \varepsilon \hat{\bar{p}}_i 
\dot{\bar{q}}^i + P \dot{S} + \varepsilon \hat{p}^b_i \dot{b}^i + p^c_i 
\dot{c}^i - H_{gf} . 
\end{equation}
It turns out to be very useful to introduce sources not only for $q_i$ and $S$,
but also for the momenta $p_i$, which we have denoted by $J_i$. The reason of 
this unusual treatment is that we are going to use also an unusual order or 
path integrations: First we will integrate out the geometric coordinates and 
then the geometric momenta. In a certain sense we are ``forgetting'' that the 
target space coordinates $X, X^{\pm}$ are the canonical momenta of the 
Cartan-variables and introduce therefore sources for both quantities. This will
turn out to be very helpful in the detection of an ambiguity (see below).

Integrating over $\hat{\bar{p}}_i$ and $\bar{q}^i$ yields the gauge fixed
Hamiltonian where $\bar{q}^i$ is replaced by the fixed quantities 
(\ref{fixings}).
Next, we perform the integration over $b^i$ and $\hat{p}_i^b$ and get a
simpler expression for the \Ix{path integral}:
\begin{eqnarray}
W &=& \int \left({\cal D}q_i\right)\left({\cal D}p^i\right)\left({\cal D}S
\right)\left({\cal D}P\right)\left({\cal D}c^i\right)\left({\cal D}p^c_i
\right) \nonumber \\
&&\times \exp \left[ i \int \left( {\cal L}_{\text{eff}} + J_i p^i + j_i q_i 
+ Q S \right) d^2x \right], \label{pi2}
\end{eqnarray}
where
\begin{equation}
{\cal L}_{\text{eff}} = p^i \dot{q}_i + P\dot{S} + p^c_i \dot{c}^i + 
\left. \bar{q}^i \right|_{gf} G_i - \left. \bar{q}^i \right|_{gf} c^j 
C_{ij}{}^k \left( p^c_k + \varepsilon \dot{p}^c_k \right). 
\end{equation}
We define the ``Faddeev-Popov-matrix''
\begin{equation}
M_i^j = \left( \delta_i^j \partial_0 + \left. \bar{q}^k \right|_{gf} 
C_{ik}{}^j \right) \left( 1 + \varepsilon \partial_0 \right),
\end{equation}
and by partial integration rewrite the Lagrangian as
\begin{equation}
{\cal L}_{\text{eff}} = p^i \dot{q}_i + P\dot{S} + \left. \bar{q}^i 
\right|_{gf} G_i - c^i M_i^j p^c_j .
\end{equation}

Finally we integrate over the last ghost-momentum pair and obtain 
\begin{eqnarray}
W &=& \int \left({\cal D}q^i\right)\left({\cal D}p_i\right)\left({\cal D}S
\right)\left({\cal D}P\right) \det M_i^j \nonumber \\
&&\times \exp \left[ i \int \left( {\cal L}_{\text{eff}} + J_i p^i + j_i q_i 
+ Q S \right) d^2x \right], \label{pi3}
\end{eqnarray}
with
\begin{equation}
{\cal L}_{\text{eff}} = p^i\dot{q}_i+P\dot{S}+\left.\bar{q}^i\right|_{gf} G_i. 
\end{equation}
The factor $(1+\varepsilon \partial_0)$ yields a field 
independent determinant and thus it makes no difference for this term whether
we take the limit $\lim_{\varepsilon \to 0}$ or not. The rest of the matrix
takes the form
\begin{equation}
M = \left( \begin{array}{ccc} 
\partial_0 & -\bar{q}^2 & \bar{q}^3 \\
-K\bar{q}^3 & \partial_0 + \bar{q}^1 - X^-U(X)\bar{q}^3 & -X^+U(X) \bar{q}^3 \\
K \bar{q}^2 & X^-U(X) \bar{q}^2 & \partial_0 - \bar{q}^1 + X^+U(X) \bar{q}^2 
\end{array} \right), \label{matrix}
\end{equation}
with
\begin{equation}
K = V'(X) + X^+X^- U'(X) - \frac{F'(X)}{F(X)} \frac{{\cal L}^{(m)}}{(e)},
\end{equation}
and the $\bar{q}^i$ are understood to be the gauge fixed quantities 
(\ref{fixings}).
Letting $\varepsilon \to 0$ from the beginning simplifies (\ref{matrix}) 
considerably:
\begin{equation}
\lim_{\varepsilon \to 0} M = \left( \begin{array}{ccc} 
\partial_0 & -1 & 0 \\
0 & \partial_0 & 0 \\
K & X^-U(X) & \partial_0 + X^+U(X) \bar{q}^2 
\end{array} \right), \label{matrix0}
\end{equation}
leading to the \Ix{path integral}
\begin{eqnarray}
W_{\varepsilon=0} &=& \int \left({\cal D}q_i\right)\left({\cal D}p^i\right)
\left({\cal D}S\right)\left({\cal D}P\right) \left( \det \partial_0 
\right)^2 \det \left( \partial_0 + X^+U(X) \right) \nonumber \\
&&\times \exp \left[ i \int \left( {\cal L}_{\text{eff}}^{\varepsilon = 0} + 
J_i p^i + j_i q_i + Q S \right) d^2x \right], \label{pi0}
\end{eqnarray}
with
\begin{equation}
{\cal L}_{\text{eff}}^{\varepsilon = 0} = p^i \dot{q}_i + P\dot{S} + G_2. 
\label{eff0}
\end{equation}

\subsubsection{A shortcut}

Treating temporal gauge directly as a multiplier gauge simplifies considerably
the amount of calculational effort. With the \Ix{gauge fixing fermion}
\eq{
\Psi = p^c_2
}{shortfixfer}
we obtain immediately and without some $\eps$-trick
\eq{
H_{gf} = - G_2 - c^j C_{2j}{}^k p_k^c.
}{shorthgf}
Integrating out $c^i$ and $p^c_i$ leads to the same Faddeev-Popov determinant
as above giving rise to the same effective Lagrangian as before.

\subsubsection{The ``lost'' equations}\label{lost equations}\index{lost 
equations}

If one is used to canonical gauges one may wonder how it is possible to impose
temporal gauge in the Hamiltonian formalism and why the secondary constraints
now ``disappeared''. In order to investigate this issue it is convenient to
regard some ``toy-model'' -- e.g. \ac{QED}.

Lets perform the same steps as we have done for first order gravity (fixing
$A_0=0$) and we arrive (after integrating out the ghosts and ghost momenta)
at the analogue of the Hamiltonian \Ix{path integral} (\ref{pi3}):  
\begin{equation}
W=\int \left({\cal D}A_i \right) \left({\cal D} \Pi^i \right) \det \partial_0 
\times \exp{\left[i\int{\cal L}_{\text{eff}}+\text{sources}\right]}, 
\end{equation}
with
\begin{equation}
{\cal L}_{\text{eff}} = \Pi^i \dot{A}_i - H_{phys}, \label{LQED}
\end{equation}
where the only essential difference to first order gravity is the appearance of
a weakly nonvanishing (``physical'') Hamiltonian
\begin{equation}
\label{Hphys}
H_{phys} = \frac{1}{2}\Pi^i\Pi_i + \frac{1}{4}F^{ij}F_{ij}
\end{equation}
in the action. The standard notation for \ac{QED} has been chosen in (\ref{LQED}) 
and (\ref{Hphys}) -- cf. e.g. \cite{het92}.

It seems at first glance that the Gauss-law is absent. We will show that this
is not the case -- it is present, although in a rather unusual form: It is a
consequence of a Ward-like identity. To prove this we perform a trivial shift 
in the coordinates $A_i$
\begin{equation}
A_i \to A_i + \partial_i \Lambda, \label{ward}
\end{equation}
which does not change the measure of the path integral. The specific form of
this shift shows that it is a gauge transformation which generally moves away
from the gauge fixing constraint $A_0 = 0$ (whenever $\partial_0 \Lambda \neq 
0$). Note that the physical Hamiltonian (\ref{Hphys}) is invariant under this 
shift, but the expression $\Pi^i\dot{A}_i$ transforms into 
\begin{equation}
\Pi^i\dot{A}_i \to \Pi^i\dot{A}_i + \Pi^i\partial_0\partial_i \Lambda.
\end{equation}

Due to the invariance of the path integral under variable substitution we 
obtain therefore (after a partial integration dropping the surface term) the 
condition
\begin{equation}
\partial_i \Pi^i = 0, \label{gauss}
\end{equation}
which is the ``lost'' constraint!

Note that dropping the surface term is possible because we impose some boundary
conditions on $A_i$ (e.g. natural boundary conditions) and want them left 
unchanged by (\ref{ward}). Thus, $\Lambda$ has to be zero at the boundary 
leading to a vanishing surface term in (\ref{gauss}).

It is noteworthy that we must choose a time dependent function $\Lambda$ in 
order to obtain (\ref{gauss}) -- this is due to the fact that a shift with a
time independent function $\Lambda$ is just a residual gauge transformation
within the class of temporal gauges.

Analogously, for first order gravity one must perform ``shifts'' in the
coordinates (and momenta) the specific form of which is given by the gauge 
group $SO(1,1) \times \text{Diff}_2$ and derive such Ward-like identities 
which restore the ``lost equations''. Also, certain boundary conditions on the
canonical variables have to be imposed. For sake of simplicity we fix them by
assuming asymptotic flatness of the metric as the most natural choice 
\cite{gkv00}. We will investigate this more explicitly in the next paragraph. 

\subsubsection{Residual gauge freedom}\label{residual gauge freedom}
\index{residual gauge freedom}

For reasons discussed above, we are now interested in the residual gauge 
freedom, i.e. a gauge transformation which leaves the set of gauge- and 
consistency conditions
\begin{equation}
\delta \bar{q}^i = 0, \hspace{0.5cm} \delta \mu^i = 0, \hspace{0.5cm} \delta 
\lambda^i = 0
\end{equation}
invariant\footnote{Note that we have used $\delta \bar{q}^i = 0$ which is 
sufficient, but not neccessary for our residual gauge transformation. In fact, 
also transformations which change $\bar{q}^i$ according to
$\bar{q}^i \to \tilde{\bar{q}}^i = a^i + \left( \bar{q}^i - a^i \right) 
e^{\frac{1}{c} \left( x_0 - y_0 \right)}$
are consistent with our gauge fixings as long as we do not freeze the small 
parameter $\varepsilon$ in (\ref{gauge}). In this sense it actually does make a
difference whether we fix $\varepsilon$ or not. In the following we will assume
this $\eps$ to be constant.}. Because of 
(\ref{multitrafo}-\ref{lambdatrafo}) the relations
\begin{equation}
\delta \bar{q}^i = \dot{\varepsilon}^i - \bar{q}^k \varepsilon^j 
C_{jk}{}^i = 0
\end{equation} 
are sufficient and necessary conditions for a residual gauge transformation.
In components this reads
\begin{eqnarray}
&& \delta \bar{q}^1 = \dot{\varepsilon}^1 - \varepsilon^3 C_{32}{}^1, \\
&& \delta \bar{q}^2 = \dot{\varepsilon}^2 - \varepsilon^3 C_{32}{}^2 - 
\varepsilon^1 C_{12}{}^2, \\
&& \delta \bar{q}^3 = \dot{\varepsilon}^3 - \varepsilon^3 C_{32}{}^3.
\end{eqnarray}
This is a linear system of first order \ac{PDE} which can be solved easily:
\begin{eqnarray}
&& \varepsilon^1 = f^1 (x_1) - \int^t dt' \varepsilon^3 (t') \left( \frac{d 
{\cal V}}{d X} + \frac{F'(X)}{(e)F(X)}{\cal L}^{(m)} \right), \label{eps1} \\
&& \varepsilon^2 = f^2 (x_1) - \int^t dt' \left( \varepsilon^3 (t') X^-U(X) 
\right) + \int^t dt' \left( \varepsilon^1 (t') \right), \label{eps2} \\
&& \varepsilon^3 = f^3 (x_1) \cdot \exp \int^t dt' \left( X^+U(X) \right). 
\label{eps3}
\end{eqnarray}

The main difference to Yang-Mills lies in the non-vanishing and field
dependent structure functions of the secondary constraints' algebra and in the 
fact, that for physical reasons we could not choose $\bar{q}_i = 0$, 
$\forall i$. Therefore, we have obtained a nontrivial contribution beside the
integration constants. Fixing these three constants freezes the residual 
gauge freedom.

It is noteworthy, that the classical \ac{EOM} for $q_i$ (expressed in temporal 
gauge) are completely equivalent to (\ref{eps1}-\ref{eps3}), with the 
identification $\eps^i = q_i$. This is the reason, why the ambiguous term (see 
below) plays a dominant r{\^o}le in the quantum counterpart of the \ac{EOM} 
for $q_i$ (\ref{qEOM}).

Once we have fixed $\eps^3$ we also have determined the essential part of 
$\eps^1$ and $\eps^2$. The remaining (space-dependent) constants in these two 
parameters are just additive. However, the structure of $\eps^3$ is 
qualitatively very different. Indeed, $\eps^3$ is equivalent to the first part 
of the ``ambiguous term'' (\ref{Q112}). Thus, it seems plausible that there 
exists a relation between this ambiguity and the residual gauge freedom, and 
that this ambiguity must be fixed by the integration constants, which is 
possible via proper boundary conditions.

In terms of Lorentz boosts and diffeomorphisms the residual gauge 
transformations  have already been investigated (cf. eqs. (71-75) in 
\cite{klv99}).

\subsubsection{The \Ix{ambiguous terms}}\label{ambigous terms}

We let the residual gauge transformation (\ref{timecan}) with gauge 
parameters given by (\ref{eps1}-\ref{eps3}) act on the gauge fixed action
(\ref{eff0}):  
\eq{
\de_{\eps} \left( p^i \dot{q}_i + P\dot{S} + G_2 \right) = \frac{dM}{dt} +
\dot{\eps}^i G_i + \eps^i C_{2i}{}^k G_k = \frac{dM}{dt}.
}{g2trafo}
The first term is a surface term, the second arises due to the action of the
explicit time derivative on the generator of the time dependent canonical
transformation (\ref{timecan}) and the last one simply follows from the 
transformation of $G_2$. The sum over all terms cancels precisely if we plug in
the solutions (\ref{eps1}-\ref{eps3}) and only the (inessential) boundary term 
remains. Thus, the only non-trivial contribution may arise due to source 
terms, which we have neglected so far. But we have achieved an important
intermediate cross-check: The gauge fixed action (\ref{eff0}) is invariant 
under residual gauge transformations (\ref{timecan}) with (\ref{eps1}-
\ref{eps3}).

Now what about the source terms? In the following, we give a heuristic 
argument, why the terms $J_i p^i$ are potentially interesting: We do not 
care about the matter source at the moment, because the ambiguity arises 
before matter is integrated out perturbatively and also in the limit of 
vanishing matter. Gauge variation of a linear function of the coordinates 
$q_i$ yields again a linear function of them, as inspection of 
(\ref{residual}) reveals (modulo matter contributions). But gauge variation of 
the momenta $p_i$ results in something non-linear (both, the gauge parameter 
$\eps^i$ and the transformed momenta (\ref{residual}) show these 
non-linearities). Thus, the source terms $J_i p^i$ transform into some 
non-linear combination of the momenta and can yield something non-trivial in 
the action after such a residual gauge transformation. Of course, this is only
hand-waving, but at least we are led to investigate whether we find something
interesting in the source terms $J_i p^i$ or not.
  
Indeed, additional terms arise after all momenta and all geometric 
coordinates have been integrated out exactly due to the following ambiguity:
\eq{
\int_x \int_x' J_{ix} \nabla_{0xx'}^{-1} f_{x'} = - \int_x \int_x' 
\tilde{\nabla}_{0xx'}^{-1} J_{ix}  f_{x'}.
}{ambiguity}
We have decorated the ``inverse derivative'' in the term on the \ac{r.h.s.} with a
tilde, because the regularized version of this operator will differ from the
regularized version of $\nabla_{0xx'}^{-1}$, in general \cite{klv99}.
Terms of that form are encountered in the solution for the momenta 
(\ref{Q104}-\ref{Q106}).

This ambiguity implies in particular, that homogeneous solutions
\eq{
\tilde{\nabla}_{0xx'}^{-1} J_{ix} \to \tilde{\nabla}_{0xx'}^{-1} J_{ix} + h_i, 
\hspace{0.5cm}\tilde{\nabla}_0 h_i = 0
}{Jhom}
may be added to the sources. These homogeneous solutions (together with 
$f_{x'}$) constitute the ambiguous terms which have been introduced in
(\ref{Q110}-\ref{Q112}). 

After this ``sleight of mind'' one may wonder what would have happened, if we
had not bothered about this ambiguity at all: First of all, in the limit of
vanishing sources we would have obtained a trivial action for the scalar 
field. Also, in the matterless case, the quantum \ac{EOM} for the coordinates $q_i$
\eq{
\left< q_i \right> = \left. \frac{\de}{i\de j_i} \int {\cal L}^{(amb)} 
\right|_{j_i = 0, J_i = 0}
}{qEOM}
would have been trivial if the ambiguous terms were absent. Further, a 
traditional approach (i.e. $J_i = 0$ from the beginning and with the 
traditional order of path integration) for the \ac{KV} model produces an 
action which is precisely of the form of the ambiguous term \cite{hak94}, thus 
it would have been puzzling to lose it using a different order of 
integrations. Actually, the last fact triggered the search for the ambiguity 
\cite{kumxx}.

Finally we would like to clarify why in ordinary Yang-Mills theory this 
phenomenon of ambiguity is apparently absent. A technical argument is simply, 
that in Yang-Mills theory there is no reason why in temporal gauge one of the
gauge fixed vector potential components should be $\neq 0$. In fact, one of the
main advantages of temporal gauge is the vanishing of certain terms due to
$\bar{q}_i = 0$, $\forall i$. But we have already seen that gravity does not 
allow such a choice, because the metric would be singular. A more substantial
reason is the constance of the structure functions in the Yang-Mills 
case. Therefore, the residual gauge transformation parameters 
(\ref{eps1}-\ref{eps3}) do not involve non-linearities in the momenta. Finally,
we would like to emphasize that the trivial part of the ambiguous term actually
{\em does} occur in Yang-Mills theory. It is usually (and quite often tacitly)
fixed by natural boundary conditions \cite{het92}. This is again in contrast
to gravity, because for physical reasons it is impossible to achieve 
$q_i \to 0$ for $r \to \infty$ simply because the metric would be singular at 
$i_0$ in that case.

Thus, this ambiguous term (or better: the fact that is plays an important r{\^o}le 
in the action) is really a novel feature of gravity theories, which explains 
why most persons (including myself) feel uneasy when they encounter it for the 
first time.

\subsection{Energetic gauges}\index{energetic gauges}

Due to its nice properties, it is tempting to use the energetic set of 
constraints as a starting point for quantization. Since it is not very 
convenient to reformulate the canonical Hamiltonian (which vanishes weakly 
anyway) in terms of these constraints, we will apply reduced phase space
quantization, closely following \cite{kat01}.

\subsubsection{Reduced phase space quantization}\index{reduced phase space 
quantization}

The total Hamiltonian generating functional in a canonical gauge
\eq{
\bar{\chi}_i = 0 , \hspace{0.5cm} \chi_i = 0
}{41}
is given by
\meq{
W = \int ({\cal D}{\cal Q})({\cal D}{\cal P}) \de^{(3)}(\bar{\chi}_i)
\de^{(3)}(\chi_i) \de^{(3)}(\bar{p}_i) \de^{(3)}(G_i^{(e)})  \det 
\left\{\bar{\chi}_i, \bar{p}_j'\right\} \\
\times \det \left\{\chi_i, G_j^{(e)}{}'\right\} \exp{i\int 
\left({\cal L}^{(eff)} + \text{sources} + \text{boundary}\right)},
}{42}
with
\eq{
{\cal L}^{(eff)} = {\cal P}^i \dot{{\cal Q}}_i - {\cal H}_{ext}, 
\hspace{0.5cm} {\cal H}_{ext} = \la^i \bar{p}_i + \mu^i G_i^{(e)} \approx 0,
}{43}
and
\eq{
{\cal Q} = (q_1, q, \qb, \bar{q}^i, S) , \hspace{0.5cm} {\cal P} = (p_1, p, 
\pb, \bar{p}_i, P),
}{44}
having restricted ourselves to ``block''-gauges of the form
\eq{
\bar{\chi}_i = \bar{\chi}_i ({\cal Q}, {\cal P}) , \hspace{0.5cm} \chi_i = 
\chi_i (q_1, p_1, q, p, \qb, \pb, S, P) .
}{45}
For the time being we will neglect source and boundary contributions. 

\subsubsection{Useful gauges}

For the gauge fixing of the primary constraints we choose
\eqa{
&& \bar{\chi}_1 = \bar{q}_1 - f_1(q_1, p_1, q, p, \qb, \pb, \bar{q}_3), 
\label{primcongf1} \\  
&& \bar{\chi}_2 = \bar{q}_2 - \bar{q}_3 - f_2(q_1, p_1, q, p, \qb, \pb), 
\label{primcongf2} \\  
&& \bar{\chi}_3 = \bar{q}_3 - f_3(q_1, p_1, q, p, \qb, \pb, S, P), 
}{primcongf3}
with some (yet unspecified) functions $f_i$. $f_1$ ($f_2$) we will fix s.t. 
it is equivalent with the \ac{EOM} (\ref{eom4}) resp. (\ref{eom1}). $f_3$ will 
be fixed later explicitly by requiring consistency between the matter \ac{EOM} 
following from the gauge fixed action (\ref{effactgf}) and the original 
\ac{EOM} (\ref{eom5}) in the same gauge.

By $\chi_1$ we will fix $p_1$, because $p_1$ enters the constraints in a 
nonpolynomial way and thus it is very convenient to fix it. There are two
important special cases: $p_1 = p_1(x_0)$ and $p_1 = p_1 (x_1)$. The latter 
can lead to a trivial effective bulk action and hence the ``dynamics'' 
will be hidden in a complicated boundary structure (corresponding to a gauge 
with the ``time not flowing''). Therefore we restrict 
ourselves to the former, which will lead to a nonvanishing gauge fixed 
Hamiltonian. $\chi_1$ fixes part of the diffeomorphism invariance,
\eq{
\chi_1 = p_1 - p_1 (x_0).
}{chi1}

With $\chi_2$ we would like to fix Lorentz invariance and thus it must be a 
function of $q_1$ and/or $\pb$. Since the canonical partner of $\pb$ enters the
constraints only linearly -- as opposed to the partner of $q_1$ -- it is more
convenient to choose 
\eq{
\chi_2 = \pb - \pb(x_0,x_1).
}{chi2}

The last gauge degree of freedom will be fixed by
\eq{
\chi_3 = q - q(x_0,x_1).
}{chi3}
In this case $G_3^{(e)}$ provides an integral solution for $p$ with an 
undetermined integration function $p_0(x_0)$. It amounts to a residual gauge
degree of freedom:
\eq{
p = \int\limits_{-\infty}^{x_1} \frac{4}{q}(\partial_1 S)P dx_1 + p_0(x_0) =: 
p_0(x_0) \left(1-\eta\right).
}{p} 
Solving the constraint $G_1^{(e)}$ establishes $\qb = 0$. $G_2^{(e)}$ 
determines $q_1$ as a function of $p$.

The gauge fixed action (up to boundary and source terms) is given by
\eq{
{\cal L}_{g.f.} = P\dot{S} - \dot{p}_1 q_1 + p\dot{q}.
}{effactgf}
The Faddeev-Popov determinant is canceled completely by the corresponding 
terms coming from the $\de$-functions.

This class of gauges is what we call ``useful gauges'' -- not implying that all
other choices are useless, but still indicating that this specific class
has nice properties and nontrivial bulk dynamics.

\subsubsection{A special class of useful gauges}

An interesting three parameter family of gauge fixing conditions is given by
\seq{4cm}{
&& p_1 = \frac{\la^2}{4} x_0^2, \\
&& \pb = 0,
}{4cm}{
&& q = \frac{q_0}{x_0}, \\
&& p_0 = \hat{p}_0 x_0^2, 
}{3gauges}
where we have fixed also the residual gauge transformations. The exponents of 
the monomials $q$ and $p_0$ have been chosen s.t. the leading order of 
the effective Hamiltonian in a weak matter expansion is proportional to the 
Klein-Gordon Hamiltonian on a flat (auxiliary) background. Note that we could
add arbitrary NLO terms to $q$ and/or $p_0$ without changing this result. While
this change in $q$ just amounts to a different gauge choice additional terms in
$p_0$ correspond to different (physically motivated) boundary conditions. In 
our particular case it means that for $x_1 \to -\infty$ the manifold is flat
and contains a freely propagating KG field, while for $x_1 \to \infty$ there is
a nontrivial NLO contribution from the \ac{ADM} mass (we will see below that $\eta$
is proportional to the mass-aspect function divided by $x_0$).

We list some consequences of our gauge choice:
\seq{3cm}
{
&& e_1^- = e_1^+, \\
&& e_0^- = e_0^+ + \frac{2q_0}{e_1^+},
}{5cm}{
&& X^- = X^+ = \sqrt{p} = \frac{x_0}{q_0} e_1^+, \\
&& \sqrt{-g} = \frac{\la^2}{2} |q_0| = const. 
}{consequ}
Especially the property $\sqrt{-g} = const.$ is very convenient and allows a
nice geometrical interpretation. Moreover, it automatically yields the proper
diffeomorphism invariant measure in the path integral since 
${\cal D} S \propto {\cal D} (-g)^{1/4} S$ in that gauge.

For \ac{SRG} the gauge fixed Hamiltonian shows a complete decoupling from the 
``potential'' $V$ (which is just a constant in our case) and resembles the 
free KG Hamiltonian. The cancellation of the $U(p_1)$ terms is coincidence and 
valid for \ac{SRG} only\footnote{For other choices of $U(p_1)$ the gauge 
$p_1 \propto x_0^2$ has no simple physical interpretation and hence one is 
free to choose a different gauge leading again to such a cancellation.}.

It is possible to fix the 2D volume element to 1 by choosing 
$q_0 = \pm \frac{2}{\la^2}$. This reduces the original set to a two parameter 
family of gauge fixings (modulo the sign, which is fixed by a positivity
requirement below). The second condition arises because we assume 
$\lim_{x_0 \to \infty} g_{11} = 1$ in order to obtain asymptotically Minkowski 
space. This implies as second relation $\hat{p}_0 = 1/2q_0^2$. The parameter
$\la^2$ will not be fixed. 

This implies a (trivial) one-parameter family of (nontrivial) Hamiltonians
\eq{
H_{gf} = -F(p_1) \left[ \frac{1}{2} \left((\partial_1 S)^2 + 
\frac{P^2}{F^2(p_1)}\right) \frac{1}{1-\eta} - f(S) \right],
}{hgf}
with $F(p_1) = -\frac{\ka}{2}\frac{\la^2}{4}x_0^2$ and the (only) nonlocal and 
nonpolynomial quantity
\eq{
\eta := \frac{1}{\la x_0} \int\limits_{-\infty}^{x_1} (\partial_1 S)P dx_1' .
}{eta}
Observe that (\ref{hgf}) is 
well-defined for small $\eta$ {\em and} for large $\eta$, although a 
``perturbation'' series in $\eta$ is not well-defined in the latter 
case\footnote{Note, however, that for $\eta \ge 1$ one has to bother about 
properly defining $e_1^{\pm}$ and $X^{\pm}$}. This is consistent with 
perturbative non-renormalizability and possible non-perturbative finiteness of 
quantum gravity. The value $\eta = 1$ corresponds to a coordinate singularity 
at a Killing horizon\footnote{The determinant $\sqrt{-g}$ equals still 1 by 
construction, but $e_1^{\pm}$ both vanish and hence $e_0^{\pm}$ are divergent. 
Moreover, $X^{\pm}$ vanish in that limit, clearly indicating a horizon and the 
breakdown of the energetic set of constraints (cf. (\ref{en12}) and the 
comment below (\ref{en14})).}.

If it is possible to obtain an exact solution for the matter fields (which are 
now the only continuous degrees of freedom we have got) we could use them 
for a reconstruction of the complete geometry. However, this approach is not 
very practical, because exact solutions cannot be obtained straightforwardly,
apart from certain special cases.

In the weak matter limit $\eta \to 0$ we obtain as leading order 
the spherically symmetric KG equation (with possible self-interactions 
$f(S)$) on a 4D Minkowski background. Nonlocalities vanish in that limit. In 
the strong matter limit $\eta \to \infty$ we obtain an almost static matter 
distribution with a very large \ac{ADM} mass. 

The Cartan variables and auxiliary fields are given by
\seq{4cm}{
&& e_1^{\pm} = \frac{1}{\sqrt{2}}\sqrt{1-\eta},\\
&& \om_1 = \frac{2}{\la^2 x_0}\left(H_{gf}-\hat{p}_0q_0(1-\eta)\right), \\
&& X = \frac{\la^2}{4} x_0^2,
}{4cm}{
&& e_0^- = e_0^+ + \frac{\la^2 q_0}{2e_1^+} , \\
&& \om_0 = \text{EOM (\ref{eom4})}, \\
&& X^{\pm} = \sqrt{\hat{p}_0} x_0 \sqrt{1 - \eta}.
}{cartanlist}
The geometric part of the conserved quantity is
\eq{
{\cal C}^{(g)} = \frac{p}{\sqrt{p_1}}-\frac{\la^2}{2}\sqrt{p_1} = - 
\frac{2\hat{p}_0 x_0}{\la} \eta.
}{geomconsquant}

The effective \ac{EOM} following from the gauge fixed action are
\eqa{
&& \dot{S} = -\frac{P}{F(p_1)}\frac{1}{1-\eta} + (\partial_1 S) \La, \\
&& \dot{P} = - F(p_1) \partial_1 \left(\partial_1 S\frac{1}{(1-\eta)}\right) +
\partial_1 \left( P \La\right), \\
&& \La := \frac{1}{2\la x_0} \int\limits_{x_1}^{\infty} F(p_1)
\frac{(\partial_1 S)^2 + \frac{P^2}{F^2(p_1)}}{(1-\eta)^2}dx_1',
}{effeom}
while the ordinary \ac{EOM} following from the original action in the same 
gauge 
read
\eqa{
&& \dot{S} = -\frac{P}{2F(p_1)(e_1^+)^2} + (\partial_1 S) \hat{\La}, \\
&& \dot{P} = -F(p_1) \partial_1 \left( \frac{\partial_1 S}{2(e_1^+)^2}\right) +
\partial_1 \left( P \hat{\La}\right), \\
&& \hat{\La} := \frac{2e_0^+ - \frac{1}{e_1^+}}{2e_1^+}.
}{realeom}
Comparing this with (\ref{cartanlist}) we see that the equations are 
consistent, iff
\eq{
\La = \tilde{\La} .
}{consitent}
This constitutes the last gauge fixing condition (e.g. $\bar{\chi}_3$) and 
again the crucial variable (in this case $e_0^+$) enters only linearly.
Note that this last gauge condition contains a nonlocal matter 
part\footnote{It should be possible to overcome this nonlocality in the 
following way: Instead of using (\ref{eom5}) we could use one of the zero 
component equations of (\ref{eom2}) as the sixth {\em local} and algebraic 
gauge fixing condition. The disadvantage of this approach is the difficulty to
verify explicitly the equivalence of (\ref{effeom}) with (\ref{realeom}).}.

Since asymptotically $\La$ vanishes this yields for the zweibein zero 
components
\eqa{
e_0^{\pm} = \mp \frac{1}{\sqrt{2}} + {\cal O}(1/x_0) .
}{e0as}
Thus, the consistency condition is equivalent to the requirement that 
asymptotically free Klein-Gordon modes propagate on a flat background. 
The asymptotic line element reads
\eqa{
(ds)^2 &=& \left(1 + {\cal O}\left(r^{-1}\right)\right) (dt)^2 - \left(1 
+ {\cal O}\left(r^{-1}\right)\right) (dr)^2 \nonumber \\
&& - r^2 (d\Om^2) + 2{\cal O}\left(r^{-1}\right)(dr)(dt),
}{ds}
if we identify $r := x_0$, $t := x_1$.
Hence our Hamiltonian evolution is with respect to the physical radius as
``time'' coordinate rather than to ``real'' time. ``Initial'' conditions 
correspond to data on a timelike slice.

\subsection{``Relativistic'' gauges}\index{relativistic gauges}

The important class of gauge fixings
\eq{
p_1 = p_1 (x_{\al}), \hspace{0.5cm} \chi_i(q_j, \bar{q}_k) = 0, \,\,i = 1..4; 
\,\,j,k = 1..2,
}{relgauge}
where the $\chi_i$ are chosen s.t. they fix the four quantities $e^{\pm}_{\mu}$
in terms of two arbitrary functions $\al (x_{\al})$ and $a (x_{\al})$, leads
(up to residual gauge transformations) to a complete fixing of the 
fourdimensional metric
\eq{
g_{\mu\nu} = \left( \begin{array}{cc}
g_{\al\be} (\al (x_{\al}, a (x_{\al}))) & 0 \\
0 & - \frac{4}{\la^4} p_1 (x_{\al})
\end{array} \right) .
}{g4} 
Thus a general relativistic interpretation of such gauges is straightforward,
which is their main advantage. All gauges frequently
(or not so frequently) used in general relativity can be reached by 
(\ref{relgauge}). We re-emphasize, that our first
order approach allows a more general treatment, since also the spin-connection
$\om$ or the other target space coordinates $X^{\pm}$ can be part of the
gauge fixing conditions. 

Local Lorentz transformations have to be fixed separately by a proper 
condition, since neither the metric, nor $p_1 = X$ depend on the Lorentz 
angle.

\subsubsection{Example 1: Diagonal gauges}\index{diagonal gauges}

The gauge fixings \newline
\parbox{3cm}{\begin{equation*}
X = \frac{\lambda^2}{4}r^2,
\end{equation*}} \hfill
\parbox{3cm}{\begin{eqnarray*}
e_1^- = -e^{- \Lambda} \frac{a}{\sqrt{2}}, \\
e_0^- = e^{- \Lambda} \frac{\alpha}{\sqrt{2}},
\end{eqnarray*}} \hfill
\parbox{3cm}{\begin{eqnarray*}
e_1^+ = e^{\Lambda} \frac{a}{\sqrt{2}}, \\
e_0^+ = e^{\Lambda} \frac{\alpha}{\sqrt{2}},
\end{eqnarray*}} \hfill
\parbox{1cm}{\begin{equation} \label{diagonal} \end{equation}} \hfill
\newline where $\Lambda$ is some fixed and $a$, $\alpha$ are some free 
functions of $(x_0, x_1)$ lead to a diagonal metric of the form
\begin{equation}
g_{\alpha \beta} = \left( \begin{array}{cc}
\alpha^2 & 0 \\
0 & -a^2
\end{array} \right).
\end{equation}
$\Lambda$ fixes the local Lorentz transformations and the other equations fix
the {\twod} diffeomorphisms up to time-reparametrization which remains as a 
residual gauge transformation.

This gauge (with the simple choice $\Lambda = 0$) has been used e.g. in 
\cite{gru99, grk00} and has the advantage of showing resemblance with 
Schwarzschild coordinates thus simplifying the physical interpretation of 
quantities expressed in these coordinates. In numerical relativity parlance 
this is the polar-areal or polar-radial system.

\subsubsection{Example 2: \Ix{\ac{SB} gauge}}

The choice
\begin{equation}
X = \frac{\la^2}{4} r^2, \hspace{0.5cm} e^+_1 = 0, \hspace{1cm} e^-_1 = -1,
\label{BS}
\end{equation}
leads to an ingoing \ac{SB} gauge for the {\twod} metric
\begin{equation}
g_{\alpha \beta} = \left( \begin{array}{cc}
K(v,r) & -e(v,r) \\
-e(v,r) & 0
\end{array} \right),
\end{equation}
with the determinant $\sqrt{-g} = e = e^+_0$ and the Killing-norm 
$K = 2e_0^-e_0^+$. An analogous choice leads to an outgoing \ac{SB} gauge.

Its main advantage as opposed to the Schwarzschild metric is its regular 
behavior on a Killing-horizon. For details (e.g. the \ac{EOM} in this gauge) we 
refer to \cite{grk00}.

By exchanging $e_0^{\pm} \to \pm e_1^{\pm}$ we obtain the temporal gauge 
(\ref{gauge}) which we have chosen for the quantization procedure 
corresponding to an outgoing \ac{SB} gauge (cf. eq. (\ref{Q123})).

\subsubsection{Example 3: \Ix{Painlev{\'e}-Gullstrand gauge}}

The interesting feature of this coordinate system is on the one hand its 
regular behavior at a Killing-horizon (an advantage it shares with the 
\ac{SB} gauge), on the other hand hypersurfaces of $T = const$ are flat.

The choice
\eq{
X = \frac{\la^2}{4} r^2, \hspace{0.5cm} e^+_1 = \frac{1}{\sqrt{2}}, 
\hspace{1cm} e^-_1 = -\frac{1}{\sqrt{2}},
}{paingull}
(which also fixes local Lorentz transformations) leads to the metric
\begin{equation}
g_{\alpha \beta} = \left( \begin{array}{cc}
K(T,r) & -a(T,r) \\
-a(T,r) & -1 
\end{array} \right),
\end{equation}
with $e_0^{\pm} = (\sqrt{a^2+f^2}\pm a)/\sqrt{2}$. It has been discussed for
the first time in \cite{pai21,gul22} and shows
resemblance with the \ac{SB} gauge. Indeed the Killing-norm is again
proportional to $K(T,r)$. The technical disadvantage as compared to the 
\ac{SB} gauge is the complication arising due to the nonvanishing diagonal 
element: Terms that vanished in the latter (e.g. in the gauge fixed action) do 
not disappear anymore.

\subsection{Action of \Ix{residual gauge transformations}}
\label{residual gauge transformations}

In the gauge (\ref{multigauge}) the action of a residual gauge transformation 
on a function of canonical variables is given by
\begin{equation}
\delta_{\varepsilon} F\left(q_i, p^i, \bar{q}^i, \bar{p}_i, S, P \right) = 
\varepsilon^j \left\{ F\left(q_i, p^i, \bar{q}^i, \bar{p}_i, S, P \right), 
G_j' \right\} \label{timecan}.
\end{equation}

In order to investigate the action of such a residual gauge transformation on 
an arbitrary function of the canonical variables, it suffices to calculate all 
brackets of them with the secondary constraints\footnote{All $\de$-functions
have been omitted -- only when its derivative enters an equation it has been
written down explicitly.}: \newline
\parbox{6.5cm}{\begin{eqnarray*} 
&& \hspace{-0.5cm} \left\{ \bar{q}^i , G_j' \right\} = 0, \hspace{0.5cm} 
\forall i,j, \\
&& \hspace{-0.5cm} \left\{ q_1 , G_1' \right\} = \delta', \\
&& \hspace{-0.5cm} \left\{ q_1 , G_2' \right\} = -e^+_1 {\cal \hat{V}}', \\
&& \hspace{-0.5cm} \left\{ q_1 , G_3' \right\} = e^-_1 {\cal \hat{V}}', \\
&& \hspace{-0.5cm} \left\{ q_2 , G_1' \right\} = e^+_1, \\
&& \hspace{-0.5cm} \left\{ q_2 , G_2' \right\} = -e^+_1 X^+ U(X), \\
&& \hspace{-0.5cm} \left\{ q_2 , G_3' \right\} = \delta' - \omega_1 + 
e^-_1X^+U(X), \\
&& \hspace{-0.5cm} \left\{ q_3 , G_1' \right\} = -e^-_1, \\
&& \hspace{-0.5cm} \left\{ q_3 , G_2' \right\} = \delta' + \omega_1 - 
e^+_1X^-U(X), \\
&& \hspace{-0.5cm} \left\{ q_3 , G_3' \right\} = e^-_1X^-U(X), \\
&& \hspace{-0.5cm} \left\{ P   , G_1' \right\} = 0, \\
&& \hspace{-0.5cm} \left\{ P   , G_2' \right\} = \frac{F(X)}{2e_1^-}M_+\de'+
F(X)e_1^+f'(S), \\
&& \hspace{-0.5cm} \left\{ P   , G_2' \right\} = \frac{F(X)}{2e_1^+}M_-\de'-
F(X)e_1^-f'(S),
\end{eqnarray*}} \hfill
\parbox{3.5cm}{\begin{eqnarray*}
&& \hspace{-0.5cm} \left\{ \bar{p}_i , G_j' \right\} = 0, \hspace{0.5cm} 
\forall i,j, \\
&& \hspace{-0.5cm} \left\{ p^1 , G_1' \right\} = 0, \\
&& \hspace{-0.5cm} \left\{ p^1 , G_2' \right\} = -X^+, \\
&& \hspace{-0.5cm} \left\{ p^1 , G_3' \right\} = X^-, \\
&& \hspace{-0.5cm} \left\{ p^2 , G_1' \right\} = X^+, \\
&& \hspace{-0.5cm} \left\{ p^2 , G_2' \right\} = K_+, \\
&& \hspace{-0.5cm} \left\{ p^2 , G_3' \right\} = -{\cal \tilde{V}}, \\
&& \hspace{-0.5cm} \left\{ p^3 , G_1' \right\} = -X^-, \\
&& \hspace{-0.5cm} \left\{ p^3 , G_2' \right\} = {\cal \tilde{V}}, \\
&& \hspace{-0.5cm} \left\{ p^3 , G_3' \right\} = K_-, \\
&& \hspace{-0.5cm} \left\{ S   , G_1' \right\} = 0, \\
&& \hspace{-0.5cm} \left\{ S   , G_2' \right\} = \frac{1}{2e_1^-}M_+, \\
&& \hspace{-0.5cm} \left\{ S   , G_3' \right\} = \frac{1}{2e_1^+}M_-,
\end{eqnarray*}} \hfill
\parbox{1cm}{\begin{equation} \label{residual} 
\end{equation}} \hfill
\newline with the definitions
\begin{eqnarray}
&& \hspace{-1cm} \delta' = \frac{\partial}{\partial x_1'} \delta(x-x'), \\
&& \hspace{-1cm} {\cal \tilde{V}} = V(X) + X^+X^-U(X) + F(X) f(S), \\
&& \hspace{-1cm} {\cal \hat{V}}' = V'(X) + X^+X^-U'(X) + \frac{F'(X)}{F(X)}
\frac{{\cal L}^{(m)}}{(e)}, \\
&& \hspace{-1cm} K_{\pm} = \pm \frac{F(X)}{4(e^{\mp}_1)^2} \left[\left( 
\partial_1 S \right) \mp \frac{P}{F(X)} \right]^2, \\
&& \hspace{-1cm} M_{\pm} = \pm \left(\frac{P}{F(X)} \mp \partial_1 S\right).
\end{eqnarray}

\subsection{Covariant approach}\index{covariant approach}

Often it is very useful to investigate gauge transformations covariantly -- 
especially in general relativity. We will only consider infinitesimal gauge 
transformations\footnote{One has to bear in mind that large gauge 
transformations may play an important r{\^o}le, but for our purposes we can 
neglect these subtleties.}. The (infinitesimal) gauge parameters will be denoted
by $\de \xi^{\mu}$ for the Lie-variation and by $\de \ga$ for the 
Lorentz-rotation. The total variation yields
\eq{
\de e^{\pm}_{\mu} = \pm \de \ga e^{\pm}_{\mu}  + \de \xi^{\nu} \partial_{\nu}
e^{\pm}_{\mu}  + \partial_{\mu}\left(\de \xi^{\nu}\right)e^{\pm}_{\nu},
}{variationvielbein}
\eq{
\de \om_{\mu} = -\partial_{\mu}\de\ga + \de \xi^{\nu}\partial_{\nu}\om_{\mu} 
+ \partial_{\mu}\left(\de \xi^{\nu}\right)\om_{\nu},
}{variationconnexion}
\eq{
\de X = \de \xi^{\nu}\partial_{\nu} X. 
}{variationdilaton}

\subsubsection{\ac{SB} gauge}

As a special case we discuss once again the \ac{SB} gauge, and look at possible
residual gauge transformations which leave the gauge conditions (\ref{BS}) 
invariant. We obtain:
\eqa{
&& \de e^+_1 = \partial_1 \left(\de\xi^0\right) e^+_0= 0, \label{resgtBS1}, \\
&& \de e^-_1 = -\de\ga + \partial_1 \de \xi^1 + \partial_1 \left(\de \xi^0
\right) e^-_0= 0, \label{resgtBS2}, \\
&& \de X = \de \xi^1 \frac{\la^2}{2} r  = 0.
}{resgtBS3}
The last condition implies $\de \xi^1 = 0$. The first condition leads to 
$\de \xi^0 = \de \xi^0(t)$ and the second one establishes $\de \ga = 0$. Thus 
we see, that (\ref{BS}) allows only residual time reparametrizations.

\subsubsection{Conformal gauge}\index{conformal gauge}

It is also instructive to consider the conformal gauge:
\eq{
e_0^+ = e_0^- = e^{\phi}, \hspace{0.5cm}e_1^+ = -e_1^- = e^{\phi}.
}{confgauge}
The residual gauge transformations compatible with (\ref{confgauge}) are given 
by:
\eqa{
\de \ln{e^+_0} = \de\ga + \de\xi^{\nu}\partial_{\nu}\phi + \partial_0 \left(
\de \xi^0 + \de \xi^1\right)=0, \\
\de \ln{e^-_0} =  -\de\ga + \de\xi^{\nu}\partial_{\nu}\phi + \partial_0 \left(
\de \xi^0 - \de \xi^1\right)=0, \\
\de \ln{e^+_1} = \de\ga + \de\xi^\nu\partial_{\nu}\phi + \partial_1 \left(
\de \xi^0 + \de \xi^1\right)=0, \\
\de \ln{e^-_1} = -\de\ga + \de\xi^\nu\partial_{\nu}\phi - \partial_1 \left(
\de \xi^0 - \de \xi^1\right)=0.
}{confresgt}
One can immediately extract the relations $(\partial_0\mp\partial_1)
(\de\xi^0\pm\de\xi^1)$, $\partial_0\de\xi^0=\partial_1\de\xi^1$ and 
$\partial_1\de\xi^0=\partial_0\de\xi^1$. The condition 
$\de\ga+\partial_0\de\xi^1=0$ shows that $\de\ga$ does not vanish necessarily.

More convenient is the light-cone representation of conformal gauge:
\eq{
e^+_0 = e_1^- = e^\phi, \hspace{0.5cm}e_0^- = e_1^+ = 0.
}{lcconfgauge} 
Now the residual gauge transformations are given by:
\eqa{
&& \de \ln{e^+_0} = \de\ga + \de\xi^\nu\partial_\nu\phi + \partial_0\de\xi^0
= 0, \\
&& \de e^-_0 = \partial_0 \left(\de\xi^1\right) e^\phi = 0, \\
&& \de e^+_1 = \partial_1 \left(\de\xi^0\right) e^\phi = 0, \\
&& \de \ln{e^-_1} = -\de\ga + \de\xi^\nu\partial_\nu\phi + \partial_1\de\xi^1 
= 0.
}{lcconfresgt}
This implies\footnote{Using complex coordinates $\xi^0$ must be holonomic and
$\xi^1$ antiholonomic (or vice versa).} $\xi^\mu=\xi^\mu(x_\mu)$, 
$2\de\xi^\nu\partial_\nu\phi+\partial_0\de\xi^0+\partial_1\de\xi^1=0$ and 
$2\de\ga+\partial_0\de\xi^0-\partial_1\de\xi^1=0$. Again, one obtains a 
combination of residual diffeomorphisms and residual Lorentz transformations.
However, if $\phi$ depends arbitrarily\footnote{I.e. it cannot be split into 
the sum of a holonomic and an antiholonomic function.} on $x_0$ and $x_1$ then 
no residual gauge transformations are possible, i.e. 
$\de\xi^0=\de\xi^1=\de\ga=0$.

\subsubsection{Temporal gauge}

Finally, we apply this scheme to the temporal gauge (\ref{gauge}):
\eqa{
&& \de e^+_0 = \partial_0 \left(\de \xi^1\right) e^+_1 = 0, \label{resgttg1} \\
&& \de e^-_0 = -\de\ga + \partial_0 \de \xi^0 + \partial_0 \left(\de \xi^1
\right) e^-_1 = 0, \label{resgttg2} \\
&& \de \om_0 = -\partial_0 \de \ga + \partial_0 \left(\de\xi^1\right) \om_1 
= 0.
}{resgttg3}
The first line provides $\de\xi^1 = \de\xi^1(x_1)$. The second one leads to 
$\de \ga = \partial_0 \de \xi^0$ and the last one produces 
$\de\ga = \de\ga(x_1)$. This promotes the missing gauge parameter to 
$\de\xi^0 = x_0\de\ga(x_1)+\de\bar{\xi}^0(x_1)$. We see clearly, that the gauge
has not yet been fixed completely. These residual gauge degrees of
freedom must be fixed by asymptotic conditions. One of these conditions has 
been $\left. X \right|_{x_0 \to \infty} = x_0$:
\eq{
\left. \de X \right|_{x_0 \to \infty} = \left. \de \xi^0 \right|_{x_0 \to 
\infty} = 0.
}{asympgftg} 
This implies automatically $\left. \de \ga \right|_{x_0 \to \infty} = 0$. 
Moreover, it leads to the conclusion that $\de\ga$ and $\de\xi^0$ must vanish
everywhere. Thus, the only residual gauge degree of freedom is 
$x_1$-reparametrization invariance, completely analogous to the \ac{SB} case.

\subsection{Conclusion}

We conclude this appendix with the remark that the best strategy seems to be
analogous to \ac{QED}: Fixing only the Lagrange multipliers married with the 
secondary constraints establishes a simpler action (\ref{easyaction}) upon 
which various ``standard'' gauge fixings can be applied, such as 
``relativistic'' gauges or temporal gauges. We focused especially on the 
latter for two reasons: It is a regular gauge at Killing-horizons (as
opposed to conformal gauge or Schwarzschild gauge) and it simplifies the
calculations drastically (as opposed to other regular gauges of the 
Painlev{\'e}-Gullstrand type or relativistic \ac{SB} gauges). 

\clearplaindoublepage

\chapter{The scattering amplitude}

\section{Introduction}

This appendix contains one of the principal calculations of this thesis, the 
lowest order tree graph scattering amplitude (sections F.2-F.4). But as 
mentioned in the introduction to {\app appendix E}, I wanted to put as few 
formulae as possible in the main text.

In addition to this appendix, there exists a documented
{\em Mathematica} notepad available at the URL\\
{\URL{http://stop.itp.tuwien.ac.at/$\sim$grumil/projects/myself/thesis/s4.nb}} 
\\
containing trivial, but rather lengthy summations, differentiations and 
simplifications. 

Peter Fischer's diploma thesis is also an important reference in 
this context, providing supplementary material \cite{fis01}. We did most of 
these calculations together and shared as well depression (``The symmetric 
amplitude diverges...'') as enthusiasm (``...but the sum of both is 
finite!'').

\section{Integrals}

\subsection[Fourier transformation of $\Th(x)x^{\la}$]
{\Ix{Fourier transformation} of $\boldsymbol{\Th(x)x^{\la}}$}

In the derivation of the scattering amplitude we are using heavily the 
integral formula \cite{ges64}
\eq{
\int\limits_0^{\infty} x^{\la} e^{i(\si+i\eps)x} dx = i e^{\frac{i\la\pi}{2}} 
\Gamma(\la+1) (\si+i\eps)^{-\la-1},
}{F1}
with a regulator $\eps$. Of special interest are the (singular) values 
$\la \in \mathbb{Z}^-$. They can be approached through a limit 
$\lim_{\de \to 0} (-n+\de)$ with $n \in \mathbb{N}$. Useful limits in this 
context are
\eq{
\lim_{\de \to 0} (\si+i\eps)^{-\de} = 1 - \de \left(\ln{\left|\si\right|}+i\pi
\Th(-\si)+{\cal O}(\eps)\right) + {\cal O}(\de^2),
}{F2}
and
\eq{
\lim_{\de \to 0} \Gamma(\de) = \frac{1}{\de} - \ga + {\cal O}(\de),
}{F3}
where $\ga = 0.577\dots$ is the Euler-Mascheroni constant. Note the 
appearance of a singular $1/\de$-factor, which is the reason why we have to
expand all other quantities up to linear order in $\de$. The divergent part
will always cancel exactly in the calculation of scattering amplitudes because
they are very specific linear combinations of such integrals. We will now
present the two most important special cases.

\subsubsection{Example $\boldsymbol{\la = -1+\de}$}

The result of the integration and expansion in $\de$ is
\eq{
\left(\frac{1}{\de}+\frac{i\pi}{2}-\ga\right) - \ln{(\si + i\eps)} + 
{\cal O}(\de).
}{F4}

\subsubsection{Example $\boldsymbol{\la = -2+\de}$} 

The result of the integration and expansion in $\de$ is
\eq{
i \si \left[\left(\frac{1}{\de}+1+\frac{i\pi}{2}-\ga\right) - 
\ln{(\si + i\eps)} + {\cal O}(\de)\right] .
}{F5}

\subsection[The \Ix{dilogarithm $Li_2(z)$}]
{The dilogarithm $\boldsymbol{Li_2(z)}$}\label{dilogarithm}

In the final integration of the vertices the dilogarithm \cite{lew81}
\eq{
Li_2 (z) := \int\limits_z^0 dt \frac{\ln{(1-t)}}{t}
}{F100}
appears. We will need an expansion for large $z$ with small imaginary 
part and the derivative of the dilogarithm. 

\subsubsection{The derivative of $\boldsymbol{Li_2(z)}$}

From the definition (\ref{F100}) it follows trivially
\eq{
\frac{d}{dz} Li_2 (z) = - \frac{\ln{(1-z)}}{z}.
}{F103}

\subsubsection{Expansion for large $\boldsymbol{z}$}

Lets suppose that $z \in \mathbb{C}$ with small imaginary part. Then we have
\eq{
\lim_{z \to \infty \pm i\tilde{\eps}} Li_2 (z) = -\frac{1}{2} \ln^2{|z|} \mp 
i\pi \ln{|z|} + {\cal O}(1),
}{F101} 
and
\eq{
\lim_{z \to -\infty \pm i\tilde{\eps}} Li_2 (z) = -\frac{1}{2} \ln^2{|z|} 
+ {\cal O}(1).
}{F102}

\section{Symmetric part of the scattering amplitude}\label{symmetric part}
\index{symmetric part of the scattering amplitude}

Using the result of the vertex calculation (\ref{Q121}) together with the 
asymptotic mode expansion (\ref{Q127}) as well as the redefinitions 
(\ref{Q132}) and the integral representation of the step function
\eq{
\theta(x) = \lim_{\eps \to 0} \frac{1}{2\pi i} \int\limits_{-\infty}^{\infty} 
\frac{d\tau}{\tau-i\eps} e^{i\tau x},
}{step}
the trivial $u$- and $u'$-integrations yield 
\meq{
T_a(q,q';k,k') = \frac{1}{4} \int\limits_0^{\infty} dr \int\limits_0^{\infty} 
dr' \frac{1}{2\pi i} \lim_{\eps \to 0} \int\limits_{-\infty}^{\infty} \left(
\frac{d\tau}{\tau-i\eps}+\frac{d\tau}{\tau+i\eps}\right)
e^{i\tau\left(r'-r\right)} \\ 
\frac{2\pi \de\left(k+k'-q-q'\right)}{\sqrt{16kk'qq'}} \frac{r'-r}{4}
\left(3r^2+3{r'}^2+2rr'\right) \sum_{legs} I(k,k';q,q'),
}{F6}
with
\meq{
I(k,k';q,q') = \left(R_{k0}ike^{ikr}+R_{k1} (-k) e^{ikr}\right)\left(R_{k'0}
ik'e^{ik'r}+R_{k'1} (-k') e^{ik'r}\right) \\
\left(R_{q0}(-iq)e^{-iqr'}-R_{q1} q e^{-iqr'}\right)\left(R_{q'0}(-iq')
e^{-iq'r'}-R_{q'1} q' e^{-iq'r'}\right) .
}{F7}
The spherical Bessel-function $R_{k1}$ reads explicitly
\eq{
R_{k1} = 2k \frac{\sin(kr)-kr\cos(kr)}{(kr)^2} = -\frac{k}{r}\partial_k\left(
\frac{R_{k0}}{k}\right).
}{F8}
From now on we treat the quantities $k, k', q, q'$, $E_x=0$, $E_y=0$,
$\tilde{k} = k, \tilde{k}'=k', \tilde{q} = q, \tilde{q}' = q'$ as independent
variables, which allows us to differentiate with respect to them. This trick
shortens the calculations considerably. Using the definitions
\eq{
c = \frac{\de\left(k+k'-q-q'\right)}{64\sqrt{kk'qq'}},
}{F9}
and 
\eq{
J(k,k',\si,\la) := 2 \lim_{\eps \to 0} \int\limits_{0}^{\infty} dx 
e^{i(\si+i\tilde{\eps})x} x^{\la} \frac{\cos((k-k')x)-\cos((k+k')x)}{x^2},  
}{F11}
establishes a deceivingly simple expression for $T_a$:
\meq{
T_a = c \lim_{\eps \to 0} \int\limits_{-\infty}^{\infty} \left(\frac{d\tau}
{\tau-i\eps} + \frac{d\tau} {\tau+i\eps}\right) \frac{\partial}{\partial \tau} 
\left(3\frac{\partial^2}{\partial E_x^2}+3\frac{\partial^2}{\partial E_y^2}- 
2\frac{\partial}{\partial E_x}\frac{\partial}{\partial E_y}\right) \\
\sum_{legs} \frac{\partial}{\partial k}\frac{\partial}{\partial k'}
\frac{\partial}{\partial q}\frac{\partial}{\partial q'} \left(
\frac{\tilde{k}^2\tilde{k'}^2\tilde{q}^2\tilde{q'}^2}{kk'qq'} J(k,k',\si_k,-2)J(-q,-q',\si_q,-2)\right).
}{F10} 
with $\si_k = k+k'+E_x-\tau$ and $\si_q = -q-q'-E_y+\tau$. Applying several 
times the formula (\ref{F1}) for the integral $J$ yields\footnote{The small
quantity $\tilde{\eps}$ appearing in (\ref{F12}) has nothing to do with the 
$\eps$ introduced in the integral representation of the $\theta$-function in the 
formulae above.}
\meq{
J(k,k',\si,-2) = \frac{i}{6} \sum_{\pm k\pm k'} (-)^{kk'-1} (\si\pm k\pm k')^3
\ln{(\si\pm k\pm k'+i\tilde{\eps})} \\
+4i\si kk'\left(\frac{1}{\tilde{\eps}}+\frac{i\pi}{2}-\ga+\frac{11}{6}\right) .
}{F12}
where we have introduced the suggestive shorthand notation
\meq{
\sum_{\pm k \pm k'} (-)^{kk'-1} f(x \pm k \pm k') := f(x+k-k') + f(x-k+k') \\
- f(x+k+k') - f(x-k-k') .
}{F19}
This sum has the following properties, which will be used later on:
\eqa{
&& \sum_{\pm k \pm k'} (-)^{kk'-1} (\pm k \pm k')^{2n-1} = 0, \hspace{0.5cm} 
\forall n \in \mathbb{Z}, \label{F40} \\
&& \sum_{\pm k \pm k'} (-)^{kk'-1} (x) = 0, \label{F32} \\
&& \sum_{\pm k \pm k'} (-)^{kk'-1} (x \pm k \pm k') = 0, \label{F23} \\
&& \sum_{\pm k \pm k'} (-)^{kk'-1} (x \pm k \pm k')^2 = -8kk', \label{F24} \\
&& \sum_{\pm k \pm k'} (-)^{kk'-1} (x \pm k \pm k')^3 = -24xkk'.
}{F25} 
Therefore, mixed polynomials in $(x \pm k \pm k')^n$ and $(y \pm q \pm q')^m$
yield either 0 or something which is either proportional to $kk'$ or to $qq'$
without further dependence on $x$ or $y$, respectively, provided that 
$n+m \leq 5$ and $n,m \in \mathbb{N}$. This insight helps to get control over
pseudo-divergent terms appearing in the $\tau$-integration later on.  

It is crucial, that the divergent part of (\ref{F12}) -- together with the
fancy constants -- plays no r{\^o}le due to the cancellation of $kk'$ with
the $1/(kk')$ in (\ref{F10}). Therefore, differentiation with respect to 
these variables yields 0 for these terms. The deeper reason for this 
``miracle'' is the regularity of $R_{k0}$ and $R_{k1}$ at $r=0$. Thus, no
UV divergencies are present.

\subsection[The result up to $\tau$-integration]
{The result up to $\boldsymbol{\tau}$-integration}

Now we pull out all constants, rearrange some terms and obtain
\meq{
T_a = -\frac{c}{36} \left(3 \frac{\partial^2}{\partial E_x^2} + 
3 \frac{\partial^2}{\partial E_y^2} - 2 \frac{\partial}{\partial E_x}
\frac{\partial}{\partial E_y}\right) \sum_{legs} D_{kk'} D_{qq'} 
\lim_{\eps \to 0} \int\limits_{-\infty}^{\infty}  d\tau \\
\left(\frac{1}{\tau-i\eps}+\frac{1}{\tau+i\eps}\right) \frac{\partial}
{\partial \tau} \left(F_{kk'}(k+k'+E_x-\tau) F_{qq'}(-q-q'-E_y+\tau) \right),
}{F13}
where we have introduced again new abbreviations:
\eq{
D_{xy} := xy \frac{\partial^2}{\partial x \partial y} - y\frac{\partial}
{\partial y} - x \frac{\partial}{\partial x} + 1 ,
}{F14}
and
\eq{
F_{xy}(z) := \sum_{\pm x \pm y} (-)^{xy-1}(z \pm x \pm y)^3\ln{(z \pm x \pm y' 
+ i\tilde{\eps})}.
}{F15}
The differential operator $D_{xy}$ has the properties
\eqa{
\hspace{-2cm} && D_{xy} \left(abx^ny^m\right) = ab(m-1)(n-1)x^ny^m, \hspace{0.5cm}
\forall m,n,a,b \in \mathbb{R}, \label{F28} \\
\hspace{-2cm} && D_{xy} \left(af(x,y)+bg(x,y)\right) = a D_{xy} \left(f(x,y)\right) + 
b D_{xy} \left(g(x,y)\right),  \label{F29} \\
\hspace{-2cm} && D_{xy} \left(f(x,y)g(x,y)\right) = f D_{xy} \left(g(x,y)\right) + g D_{xy} 
\left(f(x,y)\right) + xy \nonumber \\
\hspace{-2cm} && \quad \left(\frac{\partial f(x,y)}{\partial x} \frac{\partial g(x,y)}
{\partial y} + \frac{\partial f(x,y)}{\partial y} \frac{\partial g(x,y)}
{\partial x} \right) - f(x,y)g(x,y).
}{F30}

\subsection{Remarks regarding the final integration}

Lets focus on the integral
\meq{
I = \lim_{\eps \to 0} \int\limits_{-\infty}^{\infty} \left(\frac{d\tau}
{\tau-i\eps}+\frac{d\tau}{\tau+i\eps}\right)\frac{\partial}{\partial \tau} \\
\left(F_{kk'}(k+k'+E_x-\tau) F_{qq'}(-q-q'-E_y+\tau)\right).
}{F16}
Branch cuts of logarithms put aside, there are no singularities in
the function $F_{xy}(z)$ in the complex plane, apart from a possible divergence
at infinity. Therefore, first of all we investigate its behavior at infinity:
\eq{
\lim_{\left|z\right| \to \infty}F_{xy}(z) = -4xy\left|z\right|
\left(5+6\ln{\left|z\right|}\right) - \frac{4xy(x^2+y^2)}{\left|z\right|} + 
{\cal O} \left(\frac{1}{\left|z\right|^2}\right) .
}{F17}
At first glance, this seems a bit catastrophic: $F_{xy}(z)$ diverges stronger 
than linearly, hence the derivative of its square diverges also stronger than
linearly. Thus the integral -- in its present form -- is also divergent.
However, looking more closely, we see that the integrand vanishes like
$\left|\tau\right|^{-4}$ apart from terms proportional to $kk'$ or 
$qq'$. By virtue of (\ref{F28}) they vanish identically. Thus, an auxiliary 
path at $\left|\tau\right|=\infty$ can be regarded as harmless (apart from 
possible problems with the branch cuts), because we can isolate the 
non-harmless part and kill it with our differential operators. Still, we have 
to introduce an intermediate regulator (we will use a cutoff $R$) in order to
evaluate the integrals.

\begin{figure}
\centering
\epsfig{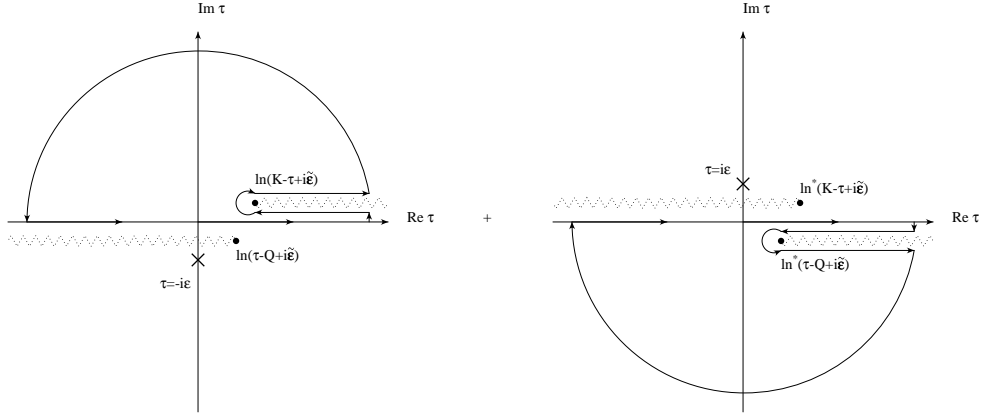}
\caption[Integration contour]
{The contour encircles no pole and all auxiliary paths are harmless, 
unless $Q=0$ or $K=0$; thus the integral over the real axis equals in the 
non-singular case the \Ix{branch cut contribution}.}
\label{fig:contour}
\end{figure}

The remaining subtlety is the appearance of two logarithms in the complex plane
with branch cuts located s.t. it is impossible to draw a closed half-circle
contour at infinity without crossing one of the cuts. We can get rid of one of
the logarithms using the contour depicted in figure \ref{fig:contour}. Before 
actually performing the integration we would like to discuss the action of the
differential operators since they cancel most of the terms.

\subsection{Harmless terms}\index{harmless terms}

Some of the terms in the following list vanish trivially, in some cases
the proof of harmlessness is more involved ($K$ and $Q$ are defined in 
(\ref{F1n})):
\blist
\item All terms of ${\cal O} \left(R^{-1}\right)$, where $R$ is the cutoff, 
clearly vanish in the limit $R \to \infty$.
\item All terms of the order $\eps$ and $\tilde{\eps}$ are harmless. 
\item All terms of the form $K^n f(Q)$ and $Q^n g(K)$ with $n = 0,1,2,3$ and
$f,g$ being arbitrary (smooth) functions. This is 
a straightforward consequence of the sign-sum (\ref{F32}-\ref{F25}) together 
with the property (\ref{F28}). As an important consequence all polynomials in 
$K$ and $Q$ up to seventh order and all polynomials up to fifth
order multiplied by an arbitrary function of $K$ or $Q$ alone are harmless. 
\elist
Especially the last property is very useful, because our polynomial prefactors
are of fifth order or smaller. Hence only non-polynomial functions depending
on $K$ {\em and} $Q$ survive all differentiations.

\subsection{The final integral up to harmless terms}

Applying $\partial/\partial \tau$ dropping all single logarithmic 
terms (they yield only harmless contributions) and using 
\eq{
K := k+k'\pm k\pm k'+E_x , Q := q+q'\pm q\pm q'+E_y,
}{F1n} 
we obtain
\meq{
I = -3 \lim_{\eps \to 0} \int\limits_{-\infty}^{\infty} \left(\frac{d\tau}
{\tau-i\eps}+\frac{d\tau}{\tau+i\eps}\right) \ln{\left(K-\tau+i\tilde{\eps}
\right)} \ln{\left(\tau-Q+i\tilde{\eps}\right)} \\ 
\left(K-\tau\right)^2 \left(Q-\tau\right)^2 \left(K+Q-2\tau\right),
}{F2n}
modulo harmless terms.
There is potentially a problem at $\tau=0$ whenever the signs are s.t. $K=0$ 
and/or $Q=0$ after differentiation. We will discuss this point below. 
The integration over the cut yields
\meq{
I = 6i\pi \lim_{R \to \infty} \left[\int\limits_K^R\ln{\left(\tau-Q+i
\tilde{\eps}\right)}-\int\limits_Q^R\ln^*{\left(K-\tau+i
\tilde{\eps}\right)}\right] \\
\left(K-\tau\right)^2\left(Q-\tau\right)^2\left(K+Q-2\tau\right)
\frac{d\tau}{\tau},
}{F3n}
having introduced an intermediate cutoff $R$ in order to make sense of the 
integrals (but note that all terms proportional to some positive power of $R$
are harmless and all negative powers of $R$ vanish in the limit). The $*$ in 
the second logarithm indicates that for this logarithm we rotate the branch cut
by $\pi$: $\lim_{\tilde{\eps}\to 0}\ln^*{(x \pm i\tilde{\eps})} := 
\ln{|x|} \mp i\pi \theta{(x)}$. We obtain two contributions: An integral of 
the form $\ln{(x+a)}\cdot\text{polynomial}(x)$ and some dilogarithmic 
contribution.
Using the asymptotic expansion (\ref{F101},\ref{F102}) we see that 
indeed all divergencies in the cutoff $R$ cancel modulo harmless terms. Also,
the difference between using $\ln^*$ and $\ln$ is a harmless contribution.
With  these considerations the integration yields
\meq{
I = 6i\pi K^2Q^2(K+Q)\left[Li_2\left(\frac{K}{Q},0\right)-Li_2\left(\frac{Q}{K}
,0\right)\right] - \frac{i\pi}{10} \\
\cdot \left(K^4-14K^3Q-94K^2Q^2-14KQ^3+Q^4\right)(K-Q)\ln{|K-Q|},
}{F7n}
where $Li_2(x,0)$ is the dilogarithm for a complex argument in the limit of 
vanishing imaginary part. Harmlessness implies, that all logarithms appearing 
after differentiation contain absolute values of their arguments, in 
particular:
\eq{
\frac{d}{dx}\left[Li_2\left(x,0\right)-Li_2\left(x^{-1},0\right)\right] = 
\frac{1}{x} \left[\ln{|1-x|}+\ln{\left|1-x^{-1}\right|}\right].
}{F11n}

\subsection{The \Ix{divergent part}}

First of all, after differentiating the $(K-Q) \ln{|K-Q|}$-term of (\ref{F3n}),
divergencies could appear. However, after a simple calculation it turns out 
that no such divergencies exist (these calculations have been performed with 
{\em Mathematica} and were checked explicitly in \cite{fis01}). As an example
one might want to look at (\ref{F8n}), where already two differentiations
have been performed. The polynomial prefactor contains again $(K-Q)$ and thus
the logarithmic singularity is compensated by a polynomial zero.

Further potential divergencies exist for $K=E_x$ and/or $Q=E_y$. This can 
already be seen from the expression (\ref{F12}), if the sign are chosen s.t.
the logarithm is singular. In the case of (\ref{F12}), there is a compensating
polynomial zero, but if we differentiate the whole expression at least thrice,
a (logarithmic) divergence appears. We will briefly sketch what terms are to 
be expected in the final result.

\subsubsection{$K=E_x$, $Q=E_y$}

If we differentiate with respect to $E_x$ and $E_y$ and set $K=0=Q$ before the
integration we see that the integrand vanishes like 
$\tau^2\left(\ln{\tau}\right)^2$ for $\tau \to 0$. Thus, no divergencies 
occur in this particular case.

\subsubsection{$K=E_x$, $Q \neq E_y$}

After differentiating with respect to $E_x$ and $E_y$ and setting $K=0=Q$ we
can isolate the divergent part of the integrand:
\eq{
K^3\left(\frac{1}{\tau+i\eps}+\frac{1}{\tau-i\eps}\right)
\ln{\left(\tau+i\tilde{\eps}\right)}\ln{\left(K-\tau+i\tilde{\eps}\right)}.
}{F4n}
Infinities coming from the $\infty$ boundary cancel again completely. There is
some convergent contribution and a logarithmically divergent one.
$D_{qq'}$ acts on the result like the identity operator and $D_{kk'}$ yields 
together with the sign sum the expression
\meq{
T^{div}_a \propto \ln{(i\eps)} \left(k^3\ln{k}+{k'}^3\ln{k'} - \left(k^3+
{k'}^3\right)\ln{(k+k')+kk'(k+k')}\right),
}{F6n}
which clearly diverges logarithmically in the limit $\eps \to 0$. It does not 
vanish after the leg summation, even if the $\de$-function is used.

\subsubsection{$K \neq E_x$, $Q=E_y$} 

This case is essentially identical to the one discussed before. Both 
contributions do {\em not} cancel each other.
Thus, the symmetric part of the amplitude has a non-vanishing divergent 
contribution.

\subsection{Differentiations and summations}

Acting with $3\frac{\partial^2}{\partial E_x^2}+3\frac{\partial^2}
{\partial E_x^2}-2\frac{\partial}{\partial E_x}\frac{\partial}{\partial E_y}$ 
on (\ref{F7n}), setting $E_x=0=E_y$ and dropping all harmless terms yields
\meq{
T_a = -ci\pi\sum_{legs}D_{kk'}D_{qq'}\sum_{signs}\left[ \frac{2}{3}\left(7K^2-
2KQ+7Q^2\right)(K-Q)\ln{|K-Q|} \right. \\
\left. + (K+Q)\left(K^2+Q^2\right)\left(Li_2\left(\frac{K}{Q},0\right)-
Li_2\left(\frac{Q}{K},0\right)\right)\right].
}{F8n}
Note that the prefactor $(K-Q)$ in front of $\ln{|K-Q|}$ has survived 
differentiation, which is nontrivial and the main reason why no poles 
$(K-Q)^{-n}$ appear in the final result.

Introducing the signs $s_i = \pm$ with $i=1..4$
\eq{
K =: (1+s_1)k + (1+s_2)k', \hspace{0.5cm} Q =: (1+s_3)q + (1+s_4)q',
}{F9n}
allows an exchange of the order of sign summations and differentiations.
Clever redefinitions of the differential operators
\eqa{
&& D_{kk'} = (1+s_1)(1+s_2)kk'\frac{\partial^2}{\partial K^2} - 
K\frac{\partial}{\partial K} + 1, \label{F10na} \\
&& D_{qq'} = (1+s_3)(1+s_4)qq'\frac{\partial^2}{\partial Q^2} - 
Q\frac{\partial}{\partial Q} + 1,
}{F10n} 
further simplify the calculations considerably, provided one splits the sign 
sum into four constituents: The first one (containing 9 summands) includes all
cases where $K \neq 0 \neq Q$. Thus, it is safe to use 
(\ref{F10na}-\ref{F10n}). The second sum (containing 3 summands) includes
all cases where $K = 0$ and $Q \neq 0$. We can set $D_{kk'} = 1$ and still use
(\ref{F10n}). The third sum is the analogue of the second, with 
$K \leftrightarrow Q$. The last ``sum'' contains a single term, namely 
$K=0=Q$, and vanishes completely. 
Note that without this split problems arise in the cases where $Q=0$ or $K=0$,
because differentiation yields poles in these variables (another way of 
putting this, is to observe that the redefinitions (\ref{F10na}) and 
(\ref{F10n}) are singular for vanishing $K$ resp. $Q$).

Apart from the formerly discussed divergent contribution to $T_a$ we obtain a 
dilogarithmic one 
\meq{
T_a^{di} = 8 ci\pi\sum_{legs} \sum_{s_i = \pm}^{i=1..4}\left[\left((1+s_1)k^3+(1+s_2){k'}^3+(1+s_3)q^3+(1+s_4){q'}^3
\right) \right. \\
\left. \left(Li_2\left(\frac{K}{Q}\right)-Li_2\left(\frac{Q}{K}\right)\right)
\right],
}{F12n}
a logarithmic one, which we will not write explicitly due to its sheer size,
and a polynomial\footnote{The leg exchange symmetries allow $t(n)$ linearly 
independent polynomials of $n^{th}$ degree, where $t(n)$ is given by the 
partition of $n$ into 4 natural numbers (I am grateful to Gerd Baron for
pointing this out to me)
\meq{
t(n)=\frac{1}{864}\left( (6n^3+90n^2+405n+525) + 108 \cos(n \pi/2) + 32 
\sqrt{3} \sin(n 2\pi/3)\right. \\
+96 \cos(n 2\pi/3) + (27n+135) \cos(n \pi) {\Big )}. \nonumber
}{foot1}
It is the $n^{th}$ coefficient in the Taylor expansion of $\left((1-x)
(1-x^2)(1-x^3)(1-x^4)\right)^{-1}$ and the first 6 terms are 1,2,3,5,6 and 
9. A useful basis for third order polynomials (where according to above
we have got three independent terms) is given by the combinations
$(k+k'-q-q')^3$, $(k+k'-q-q')(k^2+{k'}^2+q^2+{q'}^2)$ and 
$(k^3+{k'}^3-q^3-{q'}^3)$. The first two vanish due to energy conservation,
which is an important simplification. Thus, only one third order polynomial can
survive all summations and energy conservation. Using the $\de$-function it 
can be represented as $(k+k')(k-q)(k'-q)$.}
\eq{
T_a^{pol} = -96ci\pi (k+k')(k-q)(k'-q).
}{F14n}

\section{The \Ix{non-symmetric part of the scattering amplitude}}
\label{nonsymmetric}

Equipped with the knowledge of certain tricks used in the derivation of the
symmetric part of the scattering amplitude, we will shorten the discussion a
little bit. We will use the same definitions as above. Our starting point is 
\meq{
T_b(q,q';k,k') = \frac{1}{2} \int\limits_0^{\infty} dr \int\limits_0^{\infty} 
dr' \frac{1}{2\pi i} \lim_{\eps \to 0} \int\limits_{\-infty}^{\infty} d\tau 
\frac{\left(e^{i\tau\left(r-r'\right)}-e^{i\tau\left(r'-r\right)}\right)}
{\tau-i\eps}  \\
\frac{2\pi \de\left(k+k'-q-q'\right)}{\sqrt{16kk'qq'}} r
\left(r^2-{r'}^2\right) \left[\tilde{I}(k,k';q,q')+\tilde{I}(k,-q';q,-k')
\right. \\
\left. +\tilde{I}(-q,k';-k,q')+\tilde{I}(-q,-q';-k,-k') \right] ,
}{F200}
with
\meq{
\tilde{I}(k,k';q,q') = \left(R_{k0}ike^{ikr}+R_{k1} (-k) e^{ikr}\right)
\left(R_{k'0}ik'e^{ik'r}+R_{k'1} (-k') e^{ik'r}\right) \\
\left(R_{q0}(-iq)e^{-iqr'}+R_{q1} q e^{-iqr'}\right)
\left(R_{q'0}(-iq')e^{-iq'r'}\right) .
}{F201}
Performing the same steps as in the previous section with the introduction
of the following new definitions
\eqa{
&& \tilde c := \frac{\de(k+k'-q-q')}{8\sqrt{kk'qq'}}, \\
&& \tilde{D}_{xy} := xy\left(\frac{\partial}{\partial x}-\frac{1}{x}\right),
\hspace{0.5cm} \tilde{D}_{xy} xf(y) = 0,  
}{F202}
we obtain a result up to $\tau$-integration
\meq{
T_b = \frac{\tilde{c}}{36}\left(\frac{\partial^3}{\partial E_x^2\partial E_y}
-\frac{\partial^3}{\partial E_x \partial E_y^2} \right) \sum_{legs} D_{kk'} 
\tilde{D}_{qq'} \lim_{\eps \to 0} \int\limits_{-\infty}^{\infty}  d\tau \\
\left(\frac{1} {\tau-i\eps}+\frac{1}{\tau+i\eps}\right)\frac{\partial}
{\partial \tau} \left(F_{kk'}(k+k'+E_x-\tau) F_{qq'}(-q-q'-E_y+\tau) \right).
}{F203}
Since we arranged our differential operators s.t. the integrand is equivalent 
to the symmetric case, we can immediately start discussing the divergent 
contributions and the action of the new differential operators.

\subsection{The \Ix{divergent part}}

The important difference comes from the differential operators. However, after
the smoke clears we obtain again a contribution of the type (\ref{F6n}). So
it is natural to ask whether these terms cancel each other. Of course, to this
end one has to track the proportionality constants for both contributions. 
This has been performed using {\em Mathematica} (for details see also 
\cite{fis01}).

\subsection{Differentiations and summations}

After energy differentiation the amplitude reads
\meq{
T_b = -\tilde{c}i\pi\sum_{legs}D_{kk'}\tilde{D}_{qq'}\sum_{signs}{\Big [} 
4 \left(K-Q\right)^2\ln{|K-Q|} \\
\left. + \left(K^2-Q^2\right)\left(Li_2\left(\frac{K}{Q},0\right)-
Li_2\left(\frac{Q}{K},0\right)\right)\right].
}{F204}
The other differentiations and summations (using the same tricks as before)
yield a non-vanishing polynomial term which cancels with the 
symmetric polynomial part, non-vanishing logarithmic terms (again too many to 
write them down) and dilogarithms which cancel precisely the symmetric 
contribution.

\section{The total \Ix{result for the scattering amplitude}}

As we have seen in the last sections, all divergent contributions vanish. 
Additionally, all dilogarithmic and all polynomial terms disappear.
Moreover, partial cancellations occur between various logarithmic terms.
Theses calculations have been performed using the pattern matching and 
simplifying algorithms of {\em Mathematica}. A documented 
\Ix{{\em Mathematica} notepad} is available at the URL \\
\myURL \\
The result (which has been confirmed in \cite{fis01}) is
\meq{
T_a+T_b = \frac{i\pi\de\left(k+k'-q-q'\right)}{\sqrt{kk'qq'}}{\Big [}\Pi 
\ln{\Pi^2} \\
\left. + \frac{1}{\Pi} \sum_{p=k,k',q,q'} p^2 \ln{p^2} \left( 3 kk'qq' - 
\frac{1}{2}\sum_{r\neq p} \sum_{s \neq r,p}\left(r^2s^2\right)\right) \right].
}{F300}
with $\Pi = (k+k')(k-q)(k'-q)$. Due to the peculiar properties of the sums, the
logarithmic scale terms cancel if all momenta are rescaled by some 
(energy-)factor $E$. Thus, the scaling behaviour of (\ref{F300}) is 
{\em monomial}.

\clearplaindoublepage

\end{appendix}


\backmatter


\lhead[\thepage]{\slshape Abbreviations}  
\rhead[\slshape Abbreviations]{\thepage}
\addcontentsline{toc}{chapter}{Abbreviations}
\chapter*{Abbreviations}\index{abbreviations}
\begin{acronym}
%

\acro{ADM}{{A}rnowitt-{D}eser-{M}isner}
\acro{AdS}{Anti-de{S}itter}
\acro{BH}{black hole}
\acro{BRST}{{Becchi}-{R}ouet-{S}tora-{T}yutin}
\acro{CGHS}{{C}allan-{G}iddings-{H}arvey-{S}trominger}
\acro{CSS}{continuous self-similarity}
\acro{DSS}{discrete self-similarity}
\acro{EBH}{{E}ternal \ac{BH}}
\acro{EF}{{E}ddington-{F}inkelstein}
\acro{EH}{{E}instein-{H}ilbert}
\acro{EMKG}{{E}instein-massless-{K}lein-{G}ordon}
\acro{EOM}{equations of motion}
\acro{GDT}{generalized dilaton theories}
\acro{JBD}{{J}ordan-{B}rans-{D}icke}
\acro{JT}{{J}ackiw-{T}eitelboim}
\acro{KV}{{K}atanaev-{V}olovich}
\acro{ODE}{ordinary differential equation}
\acro{PDE}{partial differential equation}
\acro{PSM}{{P}oisson-{$\si$}-model}
\acro{QED}{quantum electro-dynamics}
\acro{QFT}{quantum field theory}
\acro{RNBH}{{R}eissner {N}ordstr{\"o}m \ac{BH}}
\acro{RN}{{R}eissner {N}ordstr{\"o}m}
\acro{SB}{{S}achs-{B}ondi}
\acro{SRG}{spherically reduced gravity}
\acro{SSBH}{{S}chwarzschild \ac{BH}}
\acro{VBH}{virtual \ac{BH}}
\acro{dS}{de{S}itter}
\acro{l.h.s.}{left hand side}
\acro{r.h.s.}{right hand side}
\end{acronym}


\clearplaindoublepage
\lhead[\thepage]{\slshape \bibname}  
\rhead[\slshape \bibname]{\thepage}
\addcontentsline{toc}{chapter}{\bibname}
\providecommand{\href}[2]{#2}\begingroup\raggedright\endgroup


\clearplaindoublepage
\lhead[\thepage]{\slshape \indexname}  
\rhead[\slshape \indexname]{\thepage}
\addcontentsline{toc}{chapter}{\indexname}
\begin{theindex}

  \item $d$-dimensional spherical reduction, 44
  \item {\em  Mathematica} notepad, 139

  \indexspace

  \item abbreviations, 141
  \item abelianization, 97
  \item ADM mass, 20
  \item ambiguous terms, 33, 112
  \item analysis of constraints, 91
  \item asymptotics, 37

  \indexspace

  \item black hole mass, 20
  \item Bondi-mass, 20
  \item boundary conditions, 30, 31
  \item boundary terms, 48
  \item brackets, 56
  \item branch cut contribution, 133
  \item branch cuts, 57
  \item BRST-charge, 29, 66, 104
  \item BRST-differential, 65
  \item BRST-invariant, 104

  \indexspace

  \item canonical gauges, 64
  \item canonical momenta, 62
  \item canonical transformation, 101
  \item Cartan variables, 57, 59, 79
  \item Cartan's structure equations, 59
  \item conceptual remarks, 22
  \item conformal gauge, 123
  \item conformal transformations, 83
  \item conservation law, 81
  \item consistency equations, 62
  \item constraint analysis, 63
  \item coordinates, 56
  \item cosmological constant, 45
  \item covariant approach, 122
  \item critical collapse, 17, 90

  \indexspace

  \item diagonal gauges, 120
  \item dilaton action, 8
  \item dilaton field, 9, 71
  \item dilaton formulation, 9
  \item dilaton Lagrangian, 73
  \item dimensional reduction anomaly, 43
  \item divergent part, 134, 138

  \indexspace

  \item effective Lagrangian, 31
  \item effective line element, 35
  \item effective manifold, 31
  \item energetic constraints, 97
  \item energetic coordinates, 99
  \item energetic gauges, 114
  \item equations of motion, 80
  \item extended phase space approach, 104

  \indexspace

  \item fermions, 47
  \item final result, 39
  \item first class constraints, 63
  \item first order action, 9, 10, 78, 79
  \item first order formulation, 9
  \item first order Lagrangian, 79
  \item Fourier transformation, 127
  \item fundamental Poisson bracket, 62

  \indexspace

  \item gauge fields, 47
  \item gauge fixing, 102
  \item gauge fixing fermion, 29, 67, 104, 105, 109
  \item gauge fixing functions, 105
  \item gauge freedom, 63
  \item gauge transformations, 64
  \item geometry, 7
  \item ghosts, 66

  \indexspace

  \item Hamiltonian analysis, 29, 61, 91
  \item Hamiltonian formalism, 61, 91
  \item harmless terms, 133
  \item Hodge dual, 61

  \indexspace

  \item indices, 56

  \indexspace

  \item Jordan-Brans-Dicke, 46

  \indexspace

  \item Killing-horizons, 10, 30

  \indexspace

  \item Lagrange multipliers, 62
  \item Lagrangian formalism, 77
  \item Levi-Civit\'a symbol, 55
  \item loop calculations, 43
  \item lost equations, 109

  \indexspace

  \item mass-aspect function, 20, 82
  \item matter, 10
  \item matter part, 75
  \item matterless case, 25
  \item minimally coupled matter, 25
  \item multiplier gauge, 65

  \indexspace

  \item non-canonical gauges, 64
  \item non-symmetric part of the scattering amplitude, 137
  \item nonminimally coupled matter, 28

  \indexspace

  \item Painlev{\'e}-Gullstrand gauge, 121
  \item path integral, 68, 107--109
  \item path integral quantization, 31
  \item Poisson algebra of secondary constraints, 94
  \item Poisson bracket, 62
  \item Poisson-$\si$ models, 77, 95
  \item primary constraints, 63, 92

  \indexspace

  \item quantization, 67

  \indexspace

  \item reduced phase space quantization, 114
  \item relativistic gauges, 119
  \item residual gauge freedom, 111
  \item residual gauge transformations, 121
  \item result for the scattering amplitude, 139

  \indexspace

  \item SB gauge, 30, 120
  \item scattering amplitudes, 38
  \item Schwarzschild term, 45
  \item second order action, 79
  \item secondary constraints, 93
  \item signature, 55
  \item signs, 55
  \item simplifications in $d=2$, 60
  \item spherical reduction, 71
  \item spherically reduced Lagrangian, 9, 76
  \item spin connection, 79
  \item static solutions, 84
  \item symmetric part of the scattering amplitude, 129

  \indexspace

  \item target space coordinates, 79
  \item temporal gauge, 30, 106
  \item torsion terms, 45
  \item trivial gauge transformations, 64
  \item twodimensional models, 7, 60, 79

  \indexspace

  \item units, 57

  \indexspace

  \item vertices, 34
  \item vielbein, 57
  \item Virasoro algebra, 96
  \item virtual black hole, 36

\end{theindex}




\clearemptydoublepage
\addcontentsline{toc}{chapter}{Lebenslauf}
\begin{samepage}

\thispagestyle{empty}

\enlargethispage*{5cm}
\setlength{\parindent}{0pt}

\selectlanguage{austrian}
\frenchspacing


\vspace*{-1truecm}

{\Huge\textbf{Lebenslauf}}

\bigskip\bigskip\bigskip

{\Large\textbf{Dipl.-Ing.\ Daniel Martin Lukas Grumiller}}

\vspace*{1truecm}

{\large\textbf{Persönliche Daten}}
\begin{description} \setlength{\itemsep}{-0.8ex}
\item[Geburtsdatum:] 4.~Mai 1973 in Wien
\item[Eltern:] Ingo Grumiller und Helga Weule 
\item[Staatsbürgerschaft:] Österreich
\item[Familienstand:] Okt. 1997 Heirat mit Wiltraud Grumiller
\item[Kinder:] Laurin (13.~Nov. 1997) und Armin Grumiller (2.~Okt. 1999)
\end{description}

\vspace*{0.5truecm}

{\large\textbf{Bildungsgang}}
\begin{description} \setlength{\itemsep}{-0.8ex}
\item[1979--1983:] Volksschule in Klagenfurt und Wien
\item[1983--1989:] BRG Rahlgasse, Wien 6
\item[1989--1990:] Bachillerato Col\'egio Siracusa, Edo. de M\'exico\\
  einjähriger Schulbesuch in Mexiko mit AFS
\item[1990--1991:] BRG Rahlgasse, Wien 6\\
    Reifeprüfung am 12.~Juni 1991 mit Auszeichnung bestanden
\item[1991:] Teilnahme an der österreichischen Physikolympiade
\item[1991--1997:] Studium der Technischen Physik an der TU-Wien \\
  Diplomprüfung am 27.~März\ 1997 mit Auszeichnung bestanden
\item[1998--2001:] Doktoratsstudium der Technischen Physik an der TU-Wien
\item[1999--2001:] Teilnahme an der 38.-40. Winterschule für Kern- und 
Teilchenphysik in Schladming
\item[2000:] Teilnahme an der SIGRAV Graduate School in Contemporary 
Relativity and Gravitational Physics, Como (Italien)
\end{description}

\vspace*{0.5truecm}

{\large\textbf{Berufspraxis}}
\begin{description} \setlength{\itemsep}{-0.8ex}
\item[Okt. 1993--Feb. 1997:] Fachtutor an den Instituten für 
Experimentalphysik und Theoretische Physik (2-3 Wochenstunden/Semester)
\item[Nov. 1993--Dez. 1996:] Arbeit im Vorstand von AFS Wien
\item[seit Mai 1997:] regelmäßige Jonglierauftritte im Rahmen von ArtisTick
\item[Jul.-Aug. 1997:] Technische und informatische Arbeit am 
CERN im Rahmen des CP-Verletzungsexperimentes NA48
\item[Okt. 1997--Jun. 1999:] Arbeit an der VHS Stöbergasse als 
Kursleiter
\item[Okt. 1997--Sep. 1998:] Zivildienst im Männerwohnheim der Heilsarmee
\item[seit Sep. 1997:] Wiss. Mitarb. am Institut für Theoretische Physik der 
TU Wien, Arbeitsgruppe Univ.-Prof. Dr. Wolfgang Kummer
\item[seit Aug. 2000:] Wartung und Weiterentwicklung von 
{\URL{http://www.teilchen.at}} für den Fachausschuss Kern- und Teilchenphysik
der ÖPG
\end{description}

\vspace*{0.5truecm}

\leftline{Wien, am 4. Mai 2001}

\end{samepage}
\nonfrenchspacing


\end{document}